\documentclass[10pt,aps,prd,nofootinbib,superscriptaddress,twocolumn,preprintnumbers,balancelastpage]{revtex4}
\usepackage{amssymb,amsmath,latexsym,graphics, graphicx,epsfig,multirow,comment,hyperref,appendix,feyn,slashed,color,afterpage,multirow,makecell}

\newcommand{\abs}[1]{\left\lvert #1 \right\rvert}

\newcommand {\be} {\begin {equation}}
\newcommand {\ee} {\end {equation}}

\newcommand {\bes} {\begin {equation*}}
\newcommand {\ees} {\end {equation*}}

\newcommand{\es}[2] {\begin{equation} \label{#1} \begin{split} #2 \end{split} \end{equation}}

\newcommand\Tstrut{\rule{0pt}{2.6ex}}         
\newcommand\Bstrut{\rule[-0.9ex]{0pt}{0pt}}   

\newcommand{\beq}{\begin{equation}}
\newcommand{\eeq}{\end{equation}}

\begin{document}

\title{
Evidence for Unresolved Gamma-Ray Point Sources in the Inner Galaxy}
\preprint{
MIT-CTP-4684}

\author{Samuel K. Lee}
\affiliation{Princeton Center for Theoretical Science, Princeton University, Princeton, NJ 08544 }
\affiliation{Broad Institute, Cambridge, MA 02142}

\author{Mariangela Lisanti}
\affiliation{Department of Physics, Princeton University, Princeton, NJ 08544}

\author{Benjamin R. Safdi}
\affiliation{Center for Theoretical Physics, Massachusetts Institute of Technology, Cambridge, MA 02139}

\author{Tracy R. Slatyer}
\affiliation{Center for Theoretical Physics, Massachusetts Institute of Technology, Cambridge, MA 02139}

\author{Wei Xue}
\affiliation{Center for Theoretical Physics, Massachusetts Institute of Technology, Cambridge, MA 02139}

\date{\today}

\begin{abstract}
We present a new method to characterize unresolved point sources (PSs),  generalizing traditional template fits to account for non-Poissonian photon statistics.  We apply this method to \emph{Fermi} Large Area Telescope gamma-ray data to characterize PS populations at high latitudes and in the Inner Galaxy.  We find that PSs (resolved and unresolved) account for $\sim$50\% of the total extragalactic gamma-ray background in the energy range $\sim$1.9 to 11.9 GeV.  Within 10$^\circ$ of the Galactic Center with $|b| \geq 2^\circ$, we find that $\sim$5--10\% of the flux can be accounted for by a population of unresolved PSs, distributed consistently with the observed $\sim$GeV gamma-ray excess in this region. The excess is fully absorbed by such a population, in preference to dark-matter annihilation. The inferred source population is dominated by near-threshold sources, which may be detectable in future searches.
\end{abstract}
\maketitle
Dark-matter (DM) annihilation in the Galactic halo can contribute to the flux of high-energy gamma rays detected by experiments such as the \emph{Fermi} Large Area Telescope~\cite{Atwood:2009ez}.  Currently, an excess of $\sim$GeV gamma rays has been observed by \emph{Fermi} near the Galactic Center (GC)~\cite{0910.2998, 1010.2752, 1012.5839, 1110.0006, 1207.6047, 1302.6589, 1306.5725, 1307.6862, 1312.6671, 1402.4090, 1402.6703, 1406.6948, 1409.0042, 1410.6168, Murgia2014}.  The signal extends $\sim$10$^\circ$ off the plane, is approximately spherically symmetric, and has an intensity profile that falls as $r^{-2\gamma}$ with $\gamma \approx 1.1$--$1.4$~\cite{1402.6703,1409.0042}.  The morphology and energy spectrum of the signal is consistent with DM annihilation. There is some possible tension between the DM interpretation and other searches, especially in dwarf galaxies~\cite{Ackermann:2015zua}; alternate explanations include a new population of millisecond pulsars (MSPs)~\cite{1011.4275, 1305.0830, 1309.3428, 1402.4090, 1404.2318, Calore:2014oga, 1407.5625, 1411.2980, 1411.4363,OLeary:2015gfa} or cosmic-ray injection~\cite{Carlson:2014cwa, Petrovic:2014uda}.

This Letter addresses the potential contribution of unresolved point sources (PSs) to the excess through the use of a new statistical method, called a non-Poissonian template fit (NPTF).  Our approach is model-independent, in that we remain agnostic about the nature of the PSs.  To verify the method, we use it to characterize unresolved gamma-ray PSs at high Galactic latitudes.  These findings represent one of the most precise measurements of the contribution of PSs to the extragalactic gamma-ray background (EGB) and have important implications for characterizing its source components.

The main focus of this Letter is to use the NPTF to search for a population of unresolved gamma-ray PSs in the Inner Galaxy (IG) with a morphology consistent with that of the excess.  We find that the NPTF strongly prefers a PS origin for the excess over a DM-like (smooth diffuse) origin.  The Supplementary Material provides further details on the method, as well as additional cross-checks that support these conclusions.

This study analyzes the Extended Pass~7 Reprocessed \emph{Fermi} data from $\sim$August 4, 2008 to $\sim$December 5, 2013 made available by~\cite{1406.0507}.  A HEALPix~\cite{astro-ph/0409513} pixelization of the data with \mbox{$nside = 128$} is used, corresponding to pixels $\sim$0.5$^\circ$ to a side.  We emphasize that our study focuses on data in a single energy bin from 1.893--11.943~GeV and does not rely on or extract spectral information for the excess.  The choice of this energy range keeps the signal-to-background ratio in the region of interest (ROI) high, maintains a sufficiently large number of photons over the full sky, and keeps the point-spread function relatively small and energy-independent.

The analysis utilizes the photon-count probability distribution in each pixel.  In general, a given model for the gamma-ray flux, with parameters $\theta$,  predicts a probability $p_{k}^{(p)}(\theta)$ of observing $k$~photons in a pixel~$p$.  Several source components, each modeled by a spatial template, can contribute photons in a pixel.  To date, analyses using templates have assumed Poisson statistics for the photon-count distribution---specifically, that $p_k^{(p)}(\theta)$ is the Poisson probability to draw $k$ counts with mean given by the sum of the template components in pixel $p$.

To account for unresolved PSs, the standard template-fitting procedure must be generalized to include non-Poissonian photon counts.  In the NPTF procedure, $p^{(p)}_{k}(\theta)$ depends on a potentially pixel-dependent PS source-count function $d N_p / dF$.  The source-count function determines the average number of PSs within pixel $p$ that contribute photon flux between $F$ and $F + dF$.  In this work, the source-count function is assumed to follow a broken power law, $d N_p / dF \propto A_p F^{-n}$, with pixel-dependent normalization $A_p$ and indices $n_1$ ($n_2$) above (below) the break $F_b$ that are constant between pixels.  For isotropically distributed PSs, $A_p$ is constant between pixels.  To model a population of PSs that mimics a DM annihilation signal, $A_p$ must instead follow the  DM annihilation template.  Semi-analytic methods for calculating the $p^{(p)}_{k}(\theta)$ with a broken power law source-count function were developed in ~\cite{1104.0010,Lee:2014mza}.

\begin{figure}[t]
	\leavevmode
	\begin{center}$
	\begin{array}{c}
	\scalebox{0.45}{\includegraphics{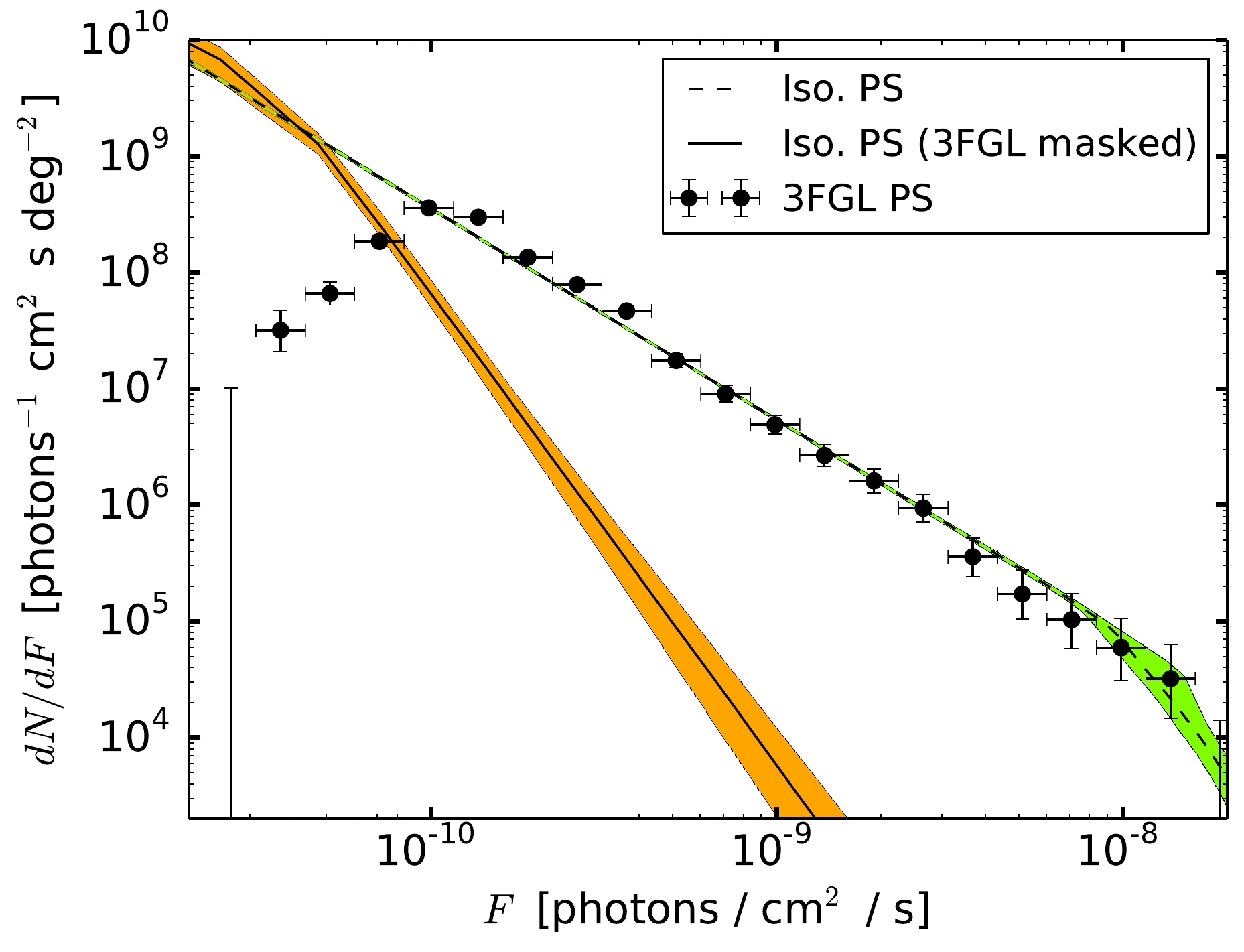}} 
	\end{array}$
	\end{center}
	\vspace{-.50cm}
	\caption{The source-count function for high-latitude point sources, derived from applying non-Poissonian template fits to data with 3FGL sources~\cite{TheFermi-LAT:2015hja} unmasked (green band) and masked (orange band).  The colored bands indicate 68\% confidence intervals, which are computed pointwise in $F$ from the posteriors for the source-count--function parameters, while the solid and dashed black lines show the median source-count functions.  The black points show the source-count function of the detected point sources in the 3FGL catalog.  The vertical error bars indicate 68\% confidence intervals; the horizontal bars denote bins in $F$.}      
	\vspace{-0.15in}
	\label{Fig: AGN}
\end{figure}

We include templates for up to seven different components in the NPTF analysis:  \emph{(1)} diffuse background, assuming the \emph{Fermi} \texttt{p6v11} diffuse model;  \emph{(2)} \emph{Fermi} bubbles, assuming uniform emission within the bubbles~\cite{Su:2010qj};  \emph{(3)} isotropic background;  \emph{(4)} annihilation of Navarro-Frenk-White (NFW)~\cite{Navarro:1995iw,astro-ph/9611107}-distributed DM, assuming no substructure; \emph{(5)}  isotropic PSs;  \emph{(6)} disk-correlated PSs, and \emph{(7)} NFW-distributed PSs.\footnote{More specifically, the 2-D spatial distribution of flux from the PSs, projected along the line-of-sight, follows the flux distribution of gamma-rays from DM annihilation,  assuming the DM is distributed with a generalized NFW distribution. }  Templates \emph{(1)} through \emph{(4)} are specified by a single normalization parameter each.  Templates \emph{(4)} and \emph{(7)} assume a generalized NFW distribution with inner slope $\gamma = 1.25$. Template \emph{(6)} corresponds to a doubly exponential thin-disk source distribution with scale height 0.3~kpc and radius 5~kpc.
The PS templates each have four parameters describing their respective source-count functions. 
  
Bayesian methods (implemented with \texttt{MultiNest}~\cite{Feroz:2008xx,Buchner:2014nha}) are used to extract posterior distributions for the model parameters.  The prior distributions of all parameters are flat, except for those of the DM and PS normalization factors, which are log flat. Unless otherwise stated, the prior ranges of all parameters are sufficiently large so that the posterior distribution is well-converged.  

We begin by applying the NPTF to data at high Galactic latitudes ($|b| \geq 30^\circ$).  \emph{Fermi}'s third source catalog (3FGL)~\cite{TheFermi-LAT:2015hja} identifies 1307 gamma-ray PSs in this region of the sky, with $\sim$55\% associated with Active Galactic Nuclei and $\sim$24\% associated with pulsars, supernova remnants and other known gamma-ray sources.  The remaining $\sim$21\% are unassociated.   
Figure~\ref{Fig: AGN} shows the source-count function $d N /d F$ in terms of the flux of the 3FGL PSs in our energy bin (black points).  The observed source-count function is suppressed below \mbox{$F \sim 10^{-10}$~photons$/ \text{cm}^2 /\text{s}$}, where it is hard to detect PSs over the diffuse background with the current exposure. 

The NPTF is performed in this high-latitude region, including templates for the diffuse background, \emph{Fermi} bubbles, isotropic emission, and isotropic PSs.  The best-fit source-count function values are given in Tab.~\ref{tab:bestfit}.\footnote{The exposure map in~\cite{1406.0507}, with  average exposure $\epsilon_\text{avg} = 4.45\times10^{10}$ $\text{cm}^2 \, \text{s}$, is used to translate between the number of photons $S$ and the flux $F$.}  The pointwise $68$\% confidence interval for the source-count function is shown in Fig.~\ref{Fig: AGN}, shaded green.  The source-count function matches the 3FGL data well above $F \sim 10^{-10}$~photons$/ \text{cm}^2 /\text{s}$.  

The best-fit intensities obtained from the NPTF can be compared to those obtained from a standard template fit that neglects PSs.  The diffuse-background and \emph{Fermi}-bubbles intensities (averaged over the ROI) are consistent between both procedures.  When the PS template is included, the isotropic-background intensity is $I_\text{iso} = 1.38_{-0.07}^{+0.07} \times 10^{-7}$ photons/cm$^2$/s/sr and the isotropic PS intensity is $I_\text{PS}^\text{iso} = 1.67_{-0.09}^{+0.08} \times 10^{-7}$ photons/cm$^2$/s/sr.  With no PS template, the isotropic-background intensity is over twice as high, $I_\text{iso} = 3.00_{-0.03}^{+0.03} \times 10^{-7}$ photons/cm$^2$/s/sr.  Thus, the PS intensity is absorbed by the isotropic-background template in the standard procedure.

\begin{table*}
\renewcommand{\arraystretch}{1.7}
\setlength{\tabcolsep}{5.2pt}
\begin{center}
\begin{tabular}{ c c  c | c  c  c | c  c | c  }
ROI & Template & 3FGL & $n_1$ & $n_2$ & $F_b$ & Bayes Factor & Bayes Factor & NFW DM\\ 
& & & & & [photons/cm$^2$/s] & (Data) & (Simulation) & (95\% confidence) \\
\hline 
\hline
\multirow{ 2}{*}{HL}  & \multirow{ 2}{*}{Iso. PS}  & unmasked  & $3.98_{-0.72}^{+2.72}$ & $1.82_{-0.02}^{+0.01}$ & $9.05_{-2.06}^{+5.68}\times10^{-9}$   & \multirow{ 2}{*}{---} &  \multirow{ 2}{*}{---} & \multirow{ 2}{*}{---} \\ 
& & masked  & $4.06_{-0.29}^{+0.40}$ & $1.56_{-0.16}^{+0.12}$ & $3.72_{-0.71}^{+1.40}\times10^{-12}$  &  & & \\
\hline
\multirow{ 2}{*}{IG}  &   \multirow{2}{*}{\makecell{NFW PS \vspace{-3pt}  \\ \tiny{+} \\ \vspace{-2pt}Disk PS}} & \multirow{ 2}{*}{unmasked}& $18.2_{-7.91}^{+8.44}$ & $-0.66_{-0.90}^{+0.98}$ & $1.76_{-0.35}^{+0.44}\times10^{-10}$ & \multirow{2}{*}{$\sim$10$^6$} & \multirow{2}{*}{$\sim$10$^5$} &   \multirow{ 2}{*}{$< 0.44$ \% }\\
& & & $17.5_{-8.40}^{+8.19}$ & $1.40_{-0.15}^{+0.12}$ & $6.80_{-1.25}^{+1.92}\times10^{-9}$ &  & & \\
\hline
\multirow{ 2}{*}{IG}  &   \multirow{2}{*}{\makecell{NFW PS \vspace{-3pt}  \\ \tiny{+} \\ \vspace{-2pt}Disk PS}} & \multirow{ 2}{*}{masked}& $18.5_{-8.09}^{+7.78}$ & $-0.73_{-0.83}^{+1.07}$ & $1.62_{-0.32}^{+0.45}\times10^{-10}$ & \multirow{2}{*}{$\sim$10$^2$} & \multirow{2}{*}{$\sim$10$^2$}  & \multirow{ 2}{*}{$< 0.48$ \%}\\
& & & $17.0_{-8.68}^{+8.85}$ & $-0.21_{-1.18}^{+1.24}$ & $6.58_{-4.36}^{+9.24}\times 10^{-10}$  &  & & \\
\end{tabular}
\end{center}
\caption{Best-fit values (16$^\text{th}$, 50$^\text{th}$, and 84$^\text{th}$ percentiles) for the source-count functions associated with the PS templates in the High Latitude (HL) and Inner Galaxy (IG) ROIs.  The source-count function is fit by a broken power-law, where $n_{1(2)}$ is the slope above (below) the break in $dN/dF$, given by $F_b$.  The source-count function for the isotropic PS component in the IG is not included, as its flux fraction is subdominant.  Depending on the analysis, the \emph{Fermi} 3FGL sources may either be masked or unmasked.  Where appropriate, we provide the Bayes factor in preference for including the NFW PS component, in both the real data and in simulations, as well as the constraint on the flux fraction (calculated as in Fig.~\ref{Fig: IG_dnds_unmasked}) attributed to NFW DM. }
\label{tab:bestfit}
\end{table*}

The averaged intensity of the observed 3FGL PSs is $\sim$9.32$\times10^{-8}$ photons/cm$^2$/s/sr at high latitudes.  Using the result of the NPTF described above and neglecting systematic uncertainties in modeling the 3FGL PSs, we predict that the intensity of unresolved PSs is $7.38_{-0.85}^{+0.83}$$\times10^{-8}$ photons/cm$^2$/s/sr.  This can be checked explicitly by repeating the NPTF with all 3FGL PSs masked (at a $\sim$1$^\circ$ radius).
The results of this fit are given in Tab.~\ref{tab:bestfit} and illustrated by the orange band in Fig.~\ref{Fig: AGN}.  The source-count function for the unresolved PSs agrees with that computed from the unmasked sky at low flux.  This suggests that the NPTF is sensitive to contributions from unresolved PSs below \emph{Fermi}'s detection threshold.  The intensity of the isotropic background is  $I_\text{iso} = 1.55_{-0.07}^{+0.07} \times 10^{-7}$ photons/cm$^2$/s/sr, which agrees with that from the 3FGL-unmasked NPTF, within uncertainties.  The intensity of the isotropic PSs is $I_\text{PS}^\text{iso} = 4.61_{-0.88}^{+0.72} \times 10^{-8}$ photons/cm$^2$/s/sr, which is slightly lower than the value inferred from the 3FGL-unmasked NPTF.  

$I_\text{iso}$ corresponds to the intensity of the isotropic gamma-ray background (IGRB), while $I_\text{iso} + I_\text{PS}^\text{iso}$ gives the intensity of the EGB.  While the PS template does absorb some contribution from Galactic PSs, extragalactic PSs are expected to dominate.  From the 3FGL-unmasked NPTF, we infer that $55_{-2}^{+2}$\% of the EGB in this energy range is associated with both resolved and unresolved PS emission; from the 3FGL-masked NPTF and using the intensities of the 3FGL PSs, we find that $47_{-2}^{+2}$\% of the EGB is due to PS emission.  These estimates appear to be consistent with those in~\cite{2010ApJ...720..435A,1104.0010}, though a direct comparison is made difficult by the fact that these analyses cover a different energy range and only use the first $\sim$11 months of \emph{Fermi} data. 
Our estimates for $I_\text{iso}$ agree with the most recently published results from \emph{Fermi}~\cite{Ackermann:2014usa}.

\begin{figure*}[t]
	\leavevmode
	\begin{center}$
	\begin{array}{cc}
	\scalebox{0.515}{\includegraphics{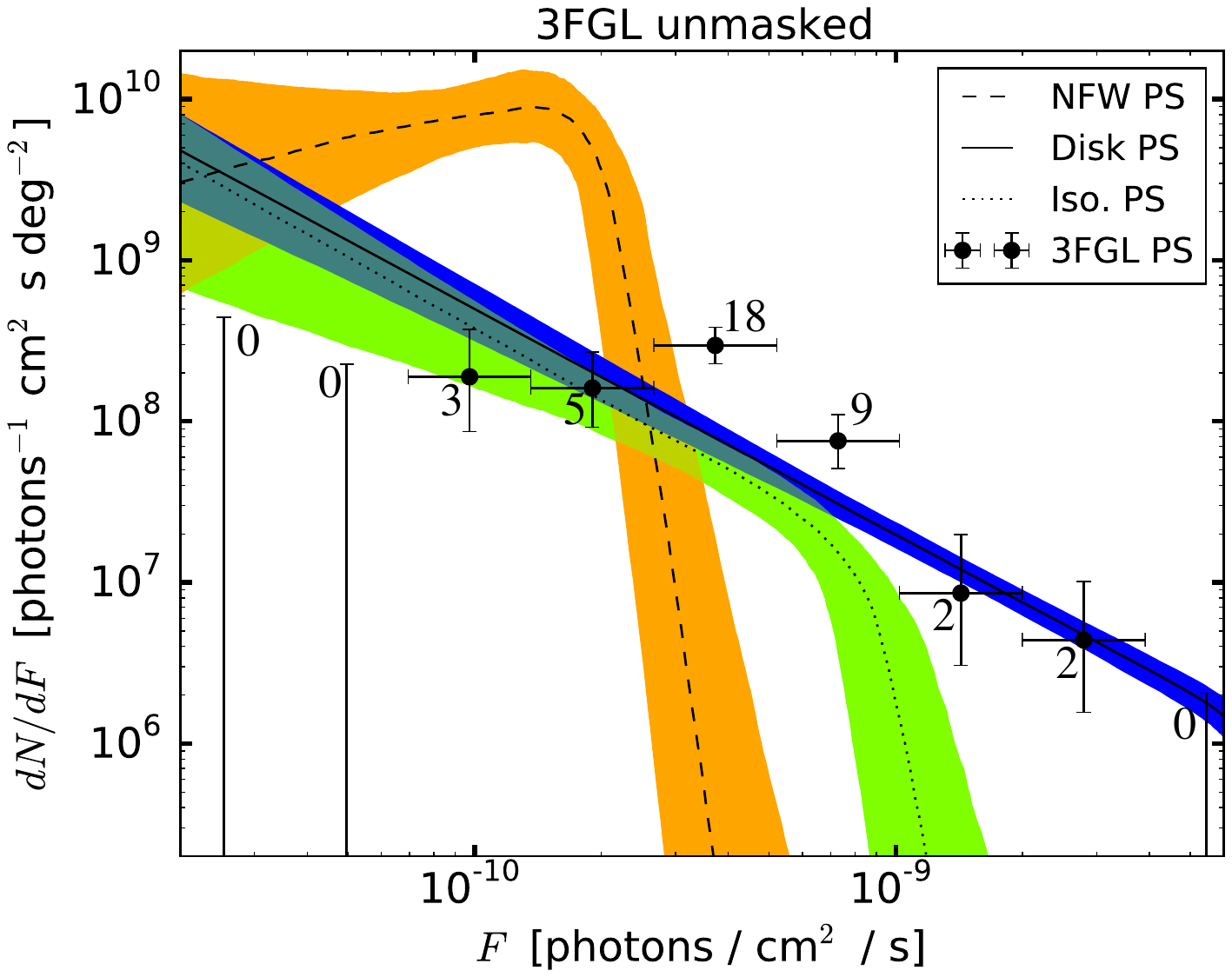}} &\scalebox{0.4}{\includegraphics{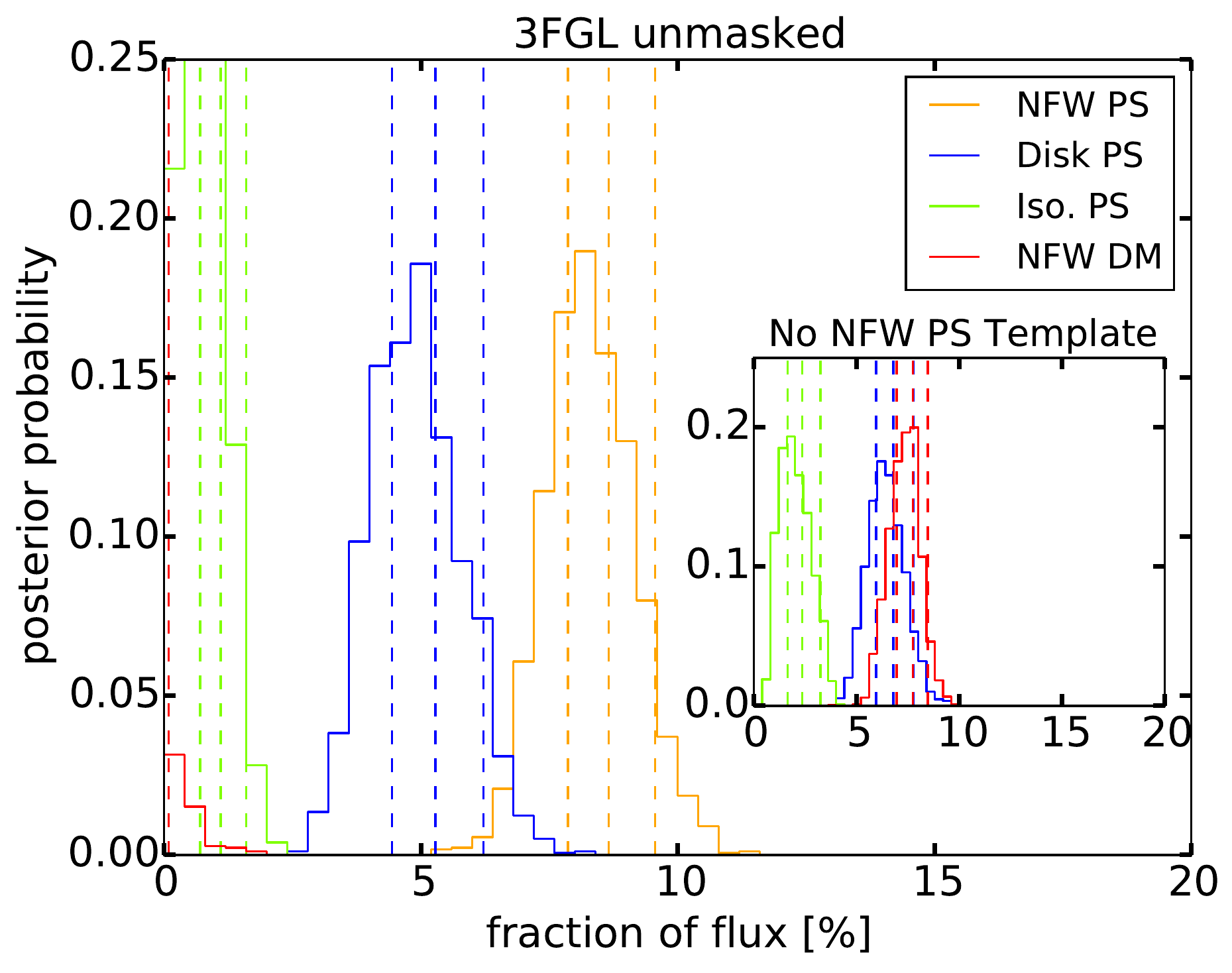}} 
	\end{array}$
	\end{center}
	\vspace{-.50cm}
	\caption{ (Left) Best-fit source-count functions within $10^\circ$ of the GC and $|b| \geq 2^\circ$, with the 3FGL sources unmasked.  The median and 68\% confidence intervals are shown for each of the following PS components: NFW (dashed, orange), thin-disk (solid, blue), and isotropic (dotted, green).  The number of observed 3FGL sources in each bin is indicated.  The normalization for the diffuse emission in the fit is consistent with that at high latitudes, as desired.  (Right)  Posteriors for the flux fraction within $10^\circ$ of the GC with $|b| \geq 2^\circ$ arising from the separate PS components, with 3FGL sources unmasked.  The inset shows the result of removing the NFW PS template from the fit.  Dashed vertical lines indicate the $16^\text{th}$, $50^\text{th}$, and $84^\text{th}$ percentiles.}
	\vspace{-0.15in}
	\label{Fig: IG_dnds_unmasked}
\end{figure*}

\begin{figure*}[t]
	\leavevmode
	\begin{center}$
	\begin{array}{cc}
	\scalebox{0.40}{\includegraphics{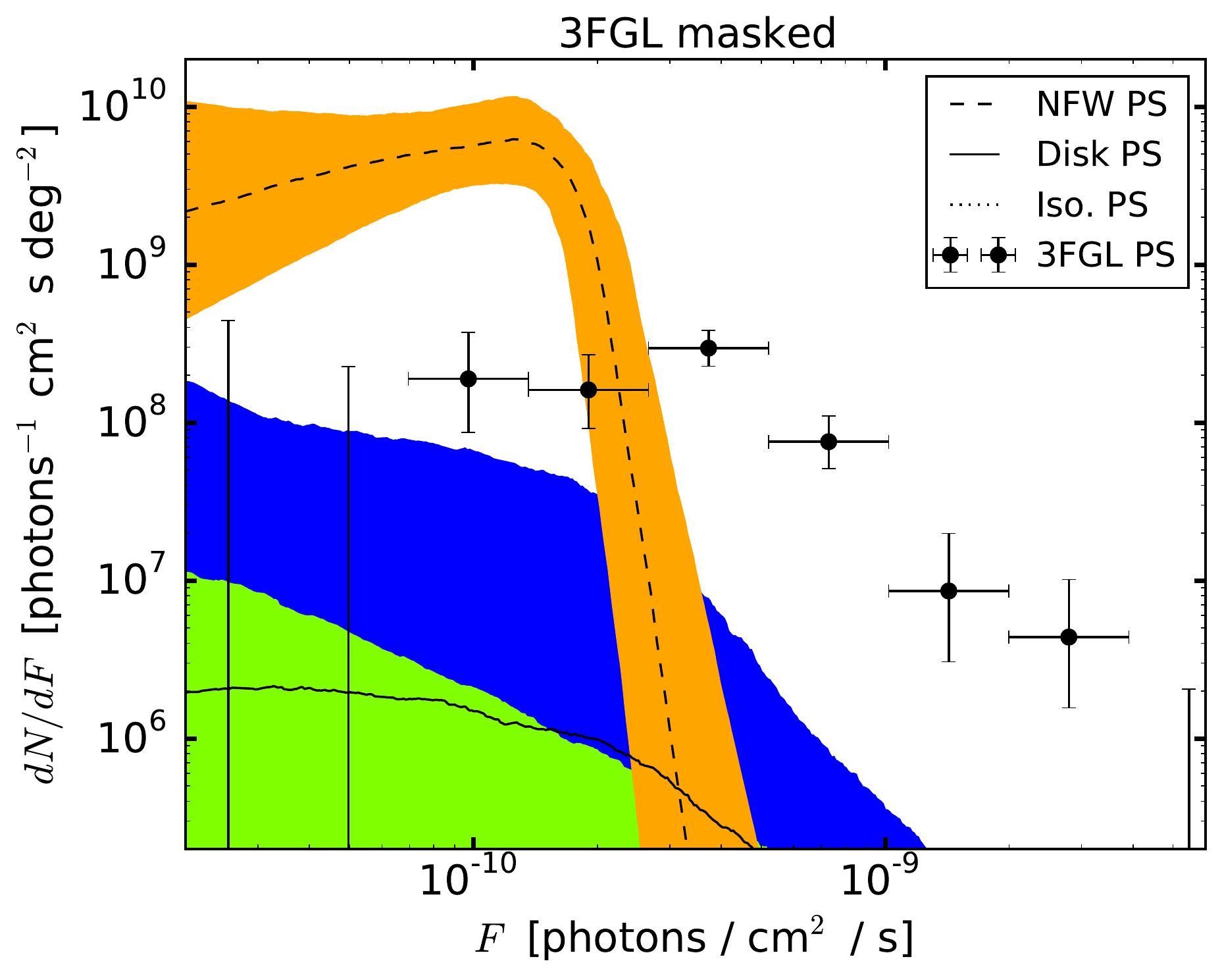}} &\scalebox{0.40}{\includegraphics{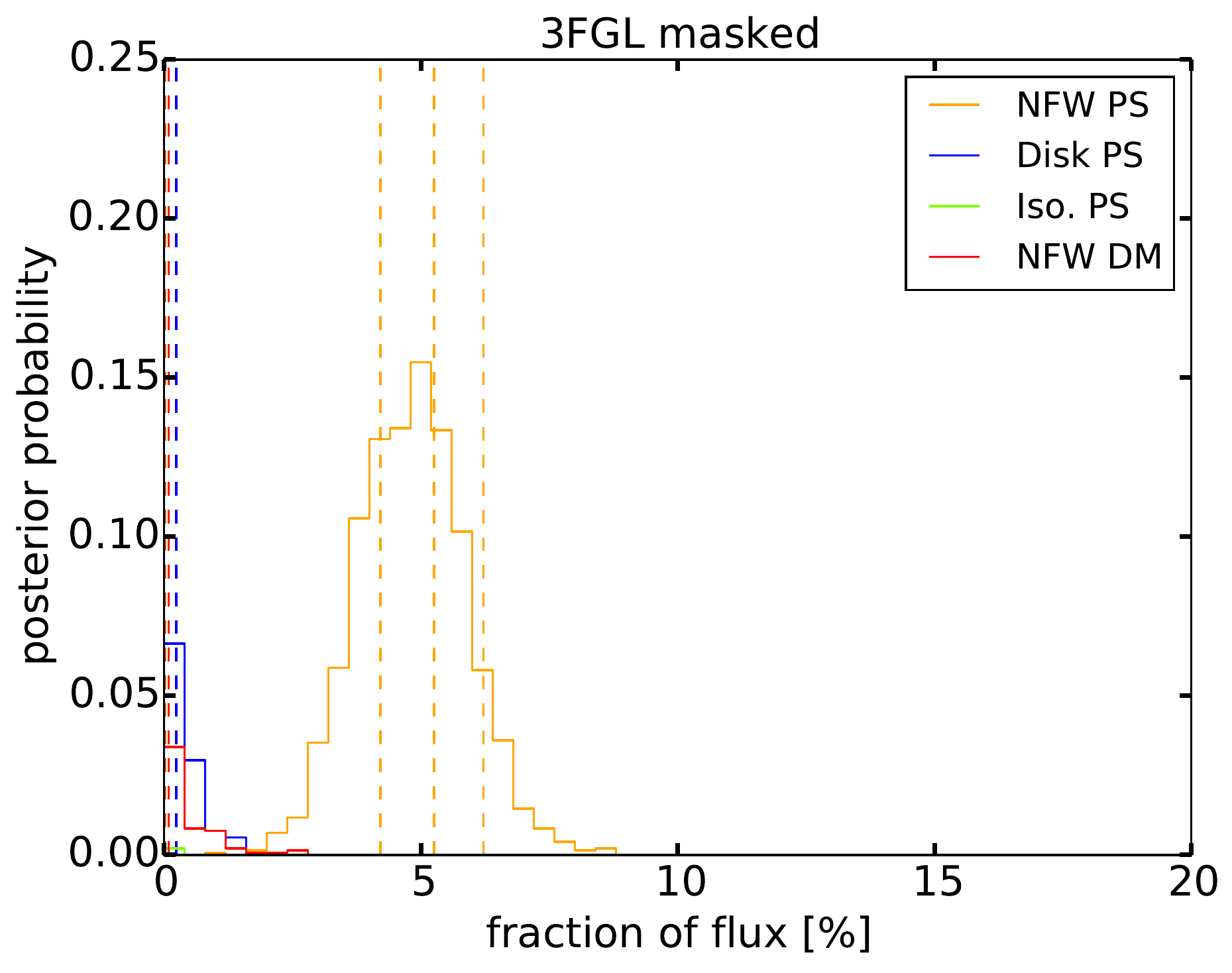}} 
	\end{array}$
	\end{center}
	\vspace{-.50cm}
	\caption{ Same as Fig.~\ref{Fig: IG_dnds_unmasked}, except with 3FGL sources masked.  } 
	\vspace{-0.15in}
	\label{Fig: IG_dnds_masked}
\end{figure*}

Next, we use the NPTF procedure to determine the fraction of flux from unresolved PSs in the IG.  These analyses include templates for the diffuse background, the \emph{Fermi} bubbles, isotropic background, and NFW-distributed DM, in addition to isotropic, disk-correlated, and NFW-distributed PSs.
While the prior ranges for the isotropic, isotropic PS, \emph{Fermi} bubbles, and diffuse background template parameters are not constrained by the high-latitude fit, restricting these parameters to their high-latitude values does not significantly affect the results.\footnote{In particular, allowing the isotropic and isotropic PS template parameters to float allows the isotropic components to partly compensate for flaws in the other templates. Mismodeling that is roughly uniform across the relatively small IG ROI can be absorbed in this way. For example, if our disk-correlated PS template is more sharply peaked toward the Galactic plane than the true disk PS population, the isotropic PS template can pick up an additional positive contribution that absorbs the higher-latitude disk PSs. If our disk PS template is broader in latitude than the true disk PS population, a negative contribution to the isotropic PS template can help account for this. Thus the ``isotropic'' templates in the IG may in principle be either brighter or fainter than their high-latitude counterparts.}

The ROI consists of all pixels within $30^\circ$ of the GC  with $|b| \geq 2^\circ$, masking out the plane.  As above, we perform two analyses, one on the full ROI and another with all 3FGL PSs masked.  For both cases, the source-count functions and flux fractions are quoted with respect to the region within $10^\circ$ of the GC and $|b| \geq 2^\circ$, with no PSs masked.  The source-count function of the Galactic and unassociated 3FGL PSs in the IG is given by the black points in the left panel of Fig.~\ref{Fig: IG_dnds_unmasked}, with the number of PSs in each bin indicated.  The majority ($\sim$90\%) of these PSs are unassociated.  

Consider, first, the case where the 3FGL sources in the IG are unmasked.  The left panel of Fig.~\ref{Fig: IG_dnds_unmasked} shows the best-fit source-count function for the NFW PS (dashed, orange), isotropic PS (dotted, green), and disk PS (solid, blue) populations. 
The disk-correlated PS template accounts for the high-flux 3FGL sources.  
Below $F\sim2\times10^{-10}$~photons/cm$^2$/s, the NFW PS template accounts for nearly all the PS emission;   its source-count function has a steep cutoff just below the source sensitivity threshold.  It is worth noting that there is no externally imposed threshold for the PS population in this case, as the 3FGL sources are not masked.  
\begin{figure*}[t]
	\leavevmode
	\begin{center}$
	\begin{array}{cc}
\scalebox{0.40}{\includegraphics{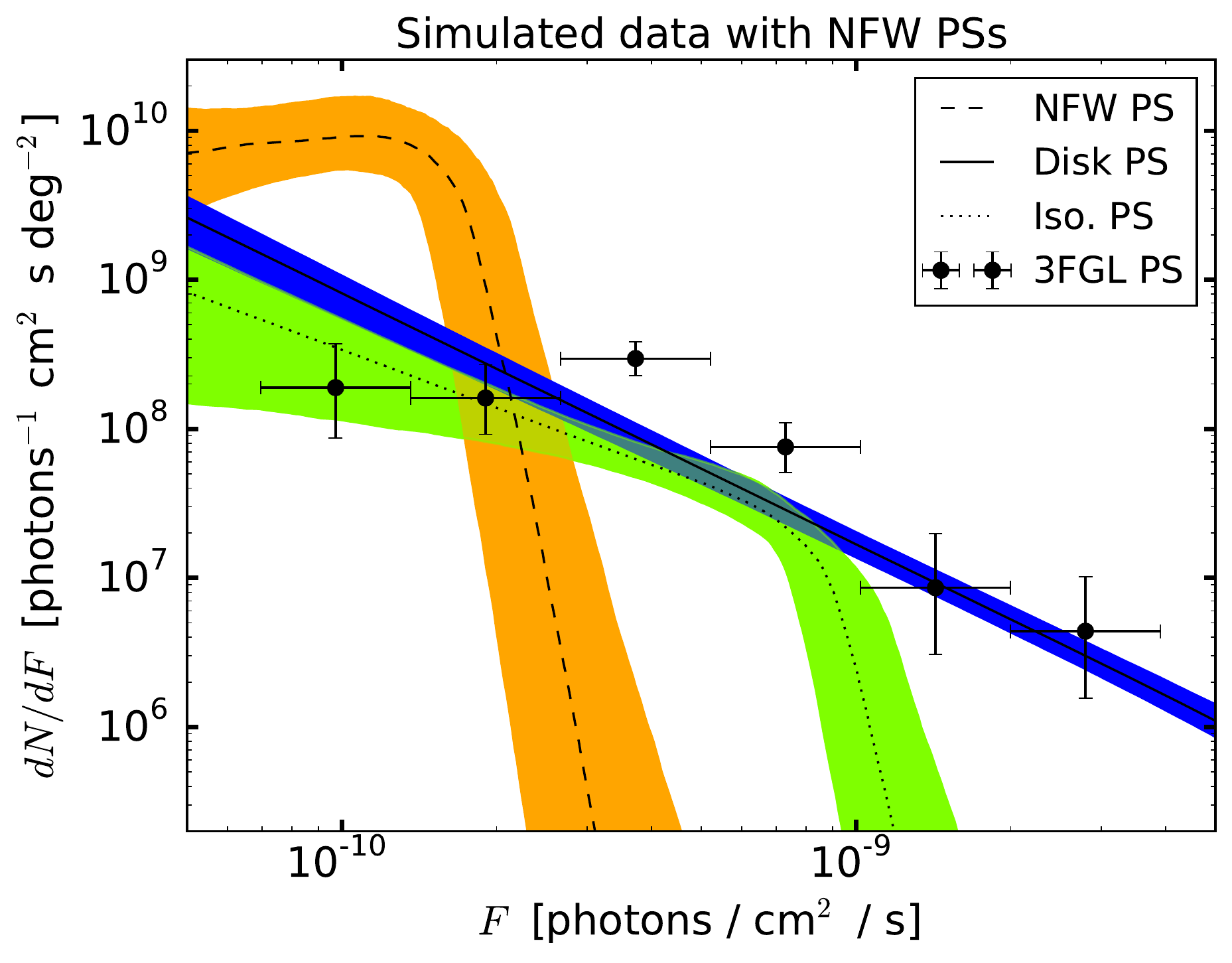}}  &	\scalebox{0.40}{\includegraphics{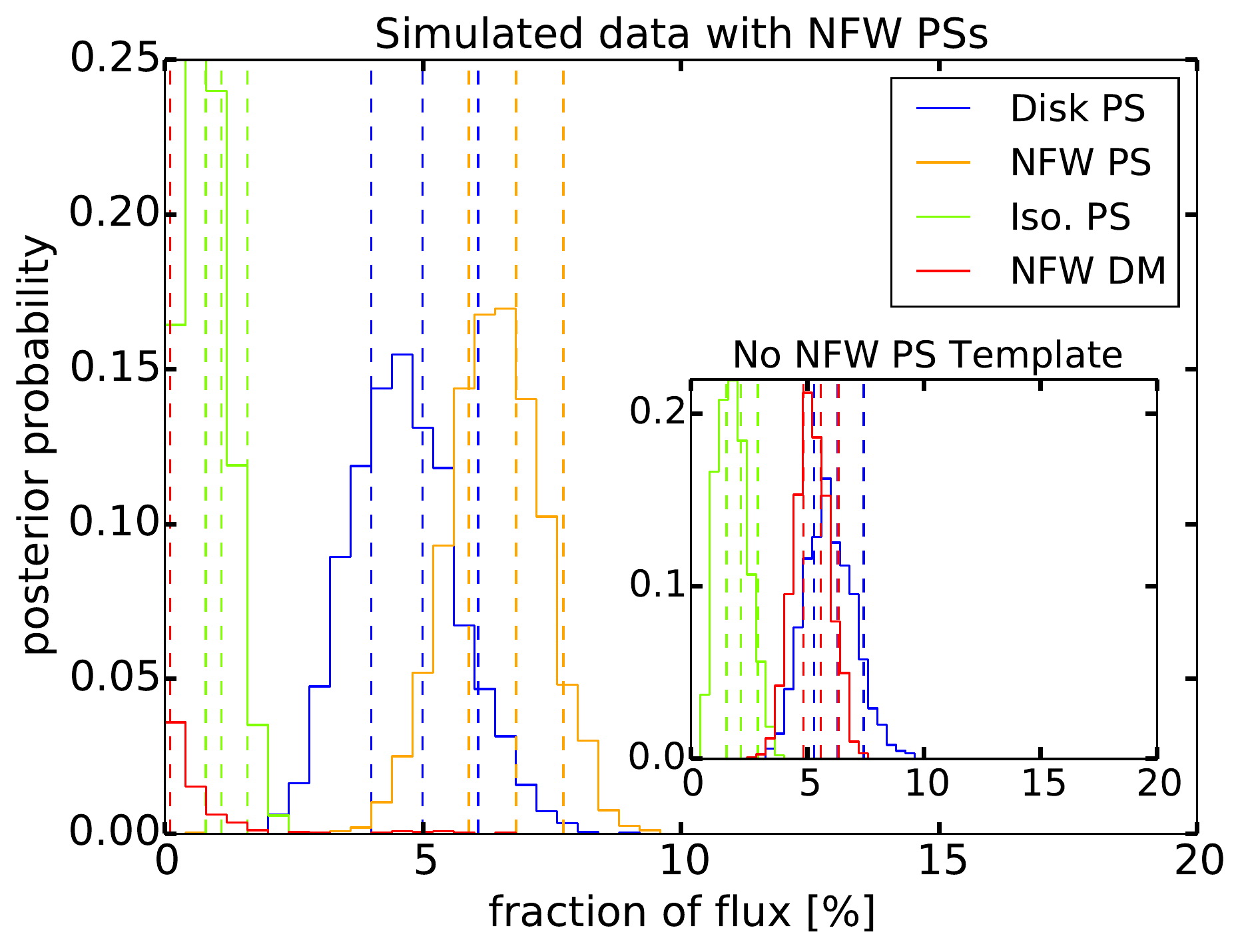}}  \\
\scalebox{0.40}{\includegraphics{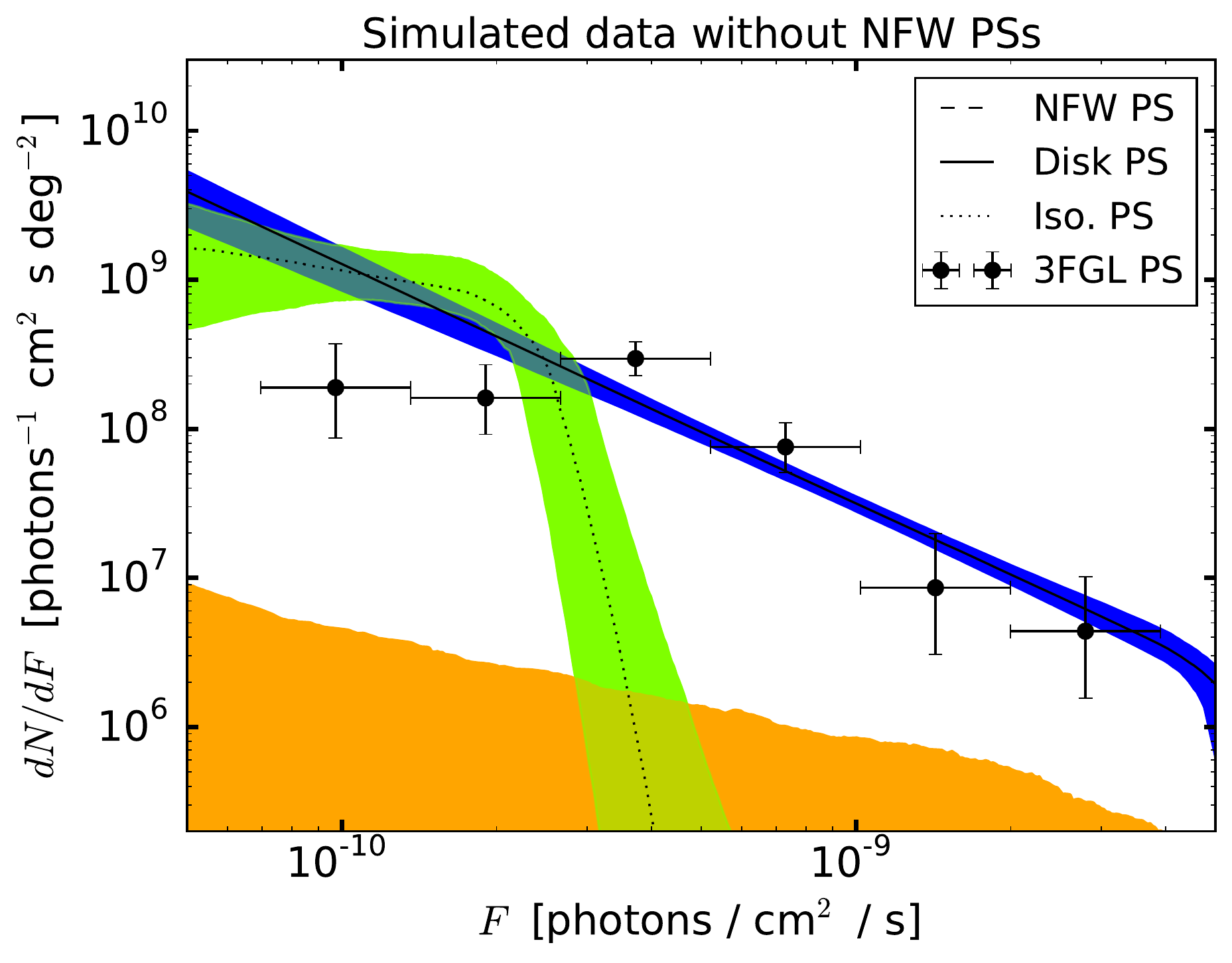}}  &	\scalebox{0.40}{\includegraphics{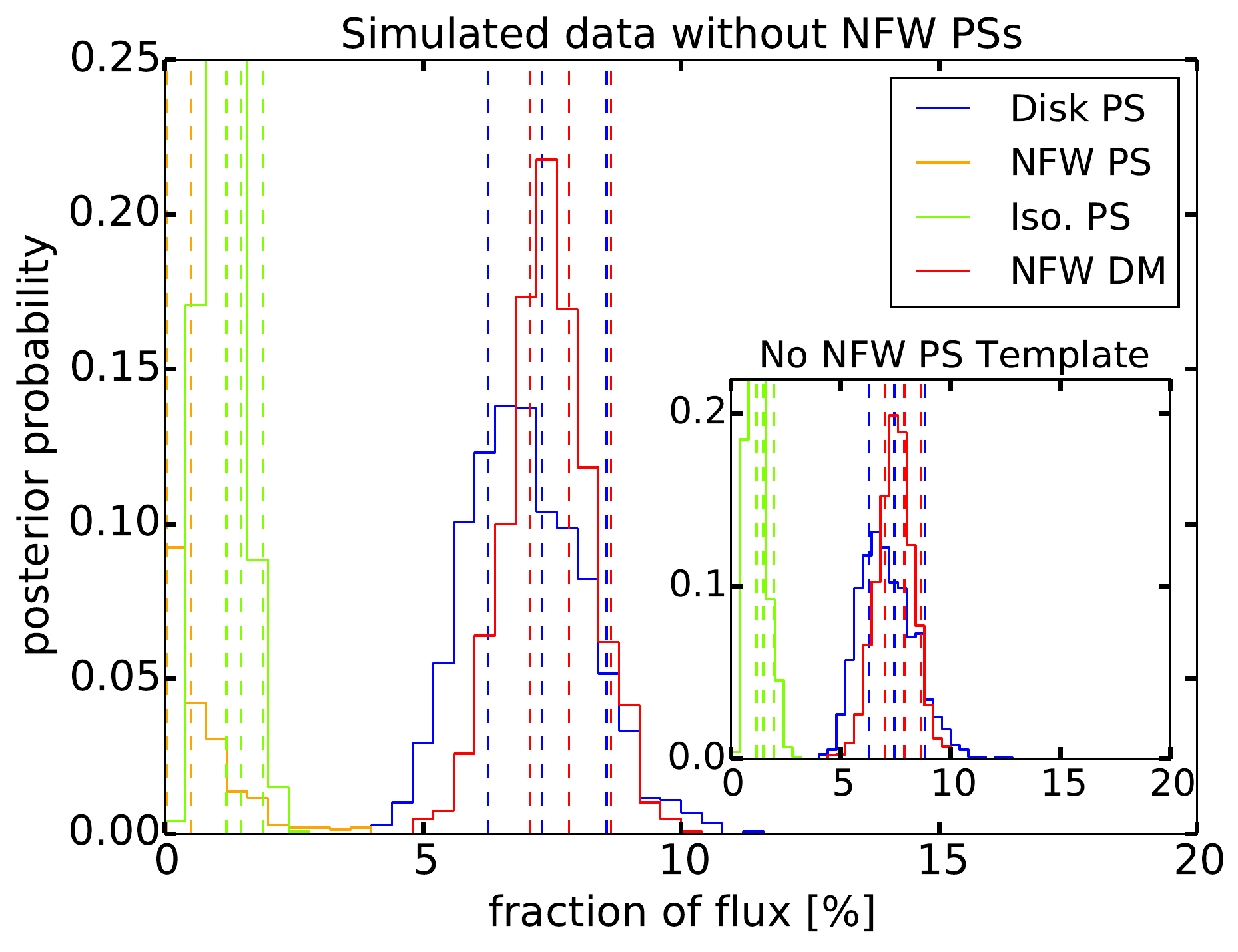}}  \\
	\end{array}$
	\end{center}
	\vspace{-.50cm}
	\caption{
	Results obtained by applying the NPTF to simulated data.  (Left column) The source-count functions for the PS templates in the fit when NFW PSs are included in the simulated data (top) or not (bottom).  Note that when NFW PSs are not simulated, an NFW DM component is instead.  (Right column)  The associated posteriors for the fraction of flux absorbed by the different templates in the fit.  The inset plots show the results of analyzing the simulated data without an NFW PS template in the fit.    All plots are relative to the region within 10$^\circ$ of the GC with $|b| \geq 2^\circ$ and 3FGL sources unmasked.  For the flux-fraction plots, the fractions are computed relative to the total number of counts observed in the real data. }
	\vspace{-0.15in}
	\label{Fig: sim_data_1}
\end{figure*}

The most pressing question to address is whether the excess flux in the IG is better absorbed by the NFW PS or NFW DM template.  The right panel of Fig.~\ref{Fig: IG_dnds_unmasked} shows the respective flux fractions, computed relative to the total photon count in the inner $10^\circ$ region with $|b| \geq 2^\circ$ and the 3FGL sources unmasked.  The disk and isotropic PSs contribute $5.3_{-0.9}^{+0.9}\%$ and $1.1_{-0.4}^{+0.5}\%$ of the flux, respectively.  In contrast, the best-fit flux fraction for the NFW PS component is $8.7_{-0.8}^{+0.9}\%$, while the best-fit DM flux fraction is consistent with zero.  The normalization of the diffuse model remains consistent (to within 1\%) with expectations from the fit at high latitudes, suggesting that the NFW PS template is absorbing the excess, and only the excess, and corresponds to a source population distinct from the more disk-like population of resolved sources.  When the NFW PS template is omitted (inset), the fraction of flux absorbed by the disk PS population is essentially unchanged at $6.8_{-0.9}^{+0.7}\%$, and the DM template absorbs $7.7_{-0.8}^{+0.7}\%$ of the flux.  The DM flux obtained in absence of an NFW PS template is consistent with other estimates in the literature~\cite{1402.6703,1409.0042}.  The model including the NFW PS contribution is preferred over that without by a Bayes factor $\sim$$10^6$.\footnote{For reference, this corresponds to test statistic $2 \Delta \ln \mathcal{L} \approx 36$. }

When the 3FGL sources are masked, the NPTF procedure yields a best-fit source-count function given by the orange band in the left panel of Fig.~\ref{Fig: IG_dnds_masked}.  Below the break, the source-count function agrees well with that found by the unmasked fit.  In this case, the contributions from the isotropic and disk-correlated PS templates are negligible.  The flux fraction attributed to the NFW PS component is $5.3_{-1.1}^{+1.0}\%$, while the NFW DM template absorbs no significant flux.  

In the masked analysis, the Bayes factor for a model that contains an NFW PS component, relative to one that does not, is $\sim$10$^2$, substantially reduced relative to the result for the unmasked case. Masking the 3FGL sources removes most of the ROI within $\sim$$5^\circ$ of the GC, reducing photon statistics markedly, especially for any signal peaked at the GC. Furthermore, in the masked ROI, non-NFW PS templates can absorb a substantial fraction of the excess. For example, if only disk and isotropic PS templates are added, the flux fraction attributed to the disk template is $2.5_{-0.62 }^{+0.70}$\%, while that attributed to NFW DM is $2.2_{-2.2 }^{+1.6}$\% (the flux attributed to isotropic PSs is negligible).  When no PS templates are included in the fit, the NFW DM template absorbs $4.1_{-1.2}^{+1.1}$\% of the total flux. As we will discuss later, this behavior agrees with expectations from simulated data. In this statistics-limited regime, the fit does not distinguish different models for the PS distribution at high significance,\footnote{However, if we repeat the analysis using Pass 8 data up to June 3, 2015, corresponding to an average exposure increase of $\sim$$30\%$ and a slight improvement in angular resolution, the Bayes factor in favor of NFW PSs increases from $\sim$$10^2$ to $\sim$$10^4$ in the 3FGL masked analysis; in the 3FGL unmasked analysis, the Bayes factor in favor of NFW PSs increases from $\sim$$10^6$ to $\sim$$ 10^9$.} but there is still a strong preference for unresolved PSs. The Bayes factor for a model with disk and isotropic PSs, relative to one with no PSs, is $\sim$10$^6$, while the Bayes factor for a model with NFW, disk and isotropic PSs, relative to one with no PSs, is $\sim$10$^8$.  The Bayes factors, best-fit source-count function parameters, and DM flux fractions for the 3FGL masked and unmasked analyses are summarized in Tab.~\ref{tab:bestfit}.

To validate the analysis procedure, we generate simulated data using the best-fit parameters from the unmasked IG analysis; we include isotropic, \emph{Fermi} bubbles, and \emph{Fermi} \texttt{p6v11} diffuse emission, as well as isotropic, disk and NFW-distributed PSs.  The simulated data is then passed through the 3FGL-unmasked IG analysis pipeline described above.  Details for how we perform the simulations may be found in the Supplementary Material.

  The top row of Fig.~\ref{Fig: sim_data_1} shows the source-count functions that are recovered from the NPTF (left), as well as the posterior distributions for the flux fractions of the separate components of the fit (right).   The fitting procedure attributes the correct fraction of flux to NFW-distributed PSs, within uncertainties, and finds no evidence for NFW DM.  When no NFW PS template is included in the fit (inset, top right), the NFW DM template absorbs the excess.  Both the source-count functions and the flux fractions are consistent with the results obtained using real data.  Additionally, we recover a Bayes factor of $\sim$10$^5$ in preference for NFW PSs when using the simulated data, which is similar to what we found for the actual analysis. 
  
   For comparison, the bottom row of Fig.~\ref{Fig: sim_data_1} shows the result of running the NPTF on a simulated data set that does not include NFW-distributed PSs but does include NFW DM.  The model parameters used to generate the simulated data are taken from the best-fit values of the analysis without NFW PSs on the real data.  In this case, the fitting procedure finds no evidence for NFW PSs, as it should, and the Bayes factor in preference for NFW PSs is much less than 1.  The source-count functions recovered for disk-correlated and isotropic PSs are consistent with those used to generate the simulated data.

The source-count function that we recover for NFW PSs in the IG differs at low flux from those previously considered in the literature, which were motivated by population models and/or data for disk MSPs~\cite{1407.5583,1407.5625,1411.2980,1305.0830}.  In particular, our source-count function seems to prefer an increasing $dN/d\log F$ below the break, implying most sources lie close to the cutoff luminosity, while previously-considered source-count functions tend to be flatter or falling in $dN/d\log F$.  If confirmed,  this may suggest novel features of the source population; however, our results are also consistent with a flat or falling  $dN/d\log F$ within uncertainties.

The results of the NPTF analyses presented here predict a new population of PSs directly below the PS-detection threshold in the IG.  We estimate from the 3FGL unmasked (masked) analysis that half of the excess within $10^\circ$ of the GC with $|b| \geq 2^\circ$ may be explained by a population of $132_{-25}^{+31}$ ($86_{-25}^{+32}$) unresolved PSs, with flux above \mbox{$1.51_{-0.25}^{+0.30} \times 10^{-10}$} (\mbox{$1.40_{-0.27}^{+0.29} \times 10^{-10}$})~photons$/ \text{cm}^2 /\text{s}$.  The entire excess within this region could be explained by $402_{-91}^{+159}$ ($258_{-83}^{+135}$) PSs, although this estimate relies on extrapolating the  source-count function to very low flux, where systematic uncertainties are large.  
Detecting members of this PS population, which appears to lie just below the current \emph{Fermi} PS-detection threshold, would be convincing evidence in favor of the PS explanation of the $\sim$GeV excess.

{\it Acknowledgements.}--- We thank S. Ando, K. Blum, D. Caprioli, I. Cholis, C. Dvorkin, D. Finkbeiner, \mbox{D. Hooper}, M. Kaplinghat, T. Linden, S. Murgia, \mbox{N. Rodd}, J. Thaler, and N. Weiner for useful discussions. We thank the \emph{Fermi} Collaboration for the use of \emph{Fermi} public data and the Fermi Science Tools.  B.R.S was supported by a Pappalardo Fellowship in Physics at MIT.  This work is supported by the U.S. Department of Energy under grant Contract Numbers DE-SC00012567, DE-SC0013999 and DE-SC0007968.   B.R.S. and T.R.S.
thank the Aspen Center for Physics, which is supported
by National Science Foundation grant PHY-1066293, for
support during the completion of this work.

{\it Note added in proof.}---Recently, we became aware of Ref.~\cite{Bartels:2015aea}, which studied the distribution of sub-threshold PSs in the inner Galaxy using a wavelet technique.

\twocolumngrid
\def\bibsection{} 
\bibliographystyle{apsrev}
\bibliography{pulsars_real_data}


\clearpage
\newpage
\maketitle
\onecolumngrid
\begin{center}
\textbf{\large Evidence for Unresolved Gamma-Ray Point Sources in the Inner Galaxy} \\ 
\vspace{0.05in}
{ \it \large Supplementary Material}\\ 
\vspace{0.05in}
{ Samuel K. Lee, Mariangela Lisanti, Benjamin R. Safdi, Tracy R. Slatyer, and Wei Xue}
\end{center}
\onecolumngrid
\setcounter{equation}{0}
\setcounter{figure}{0}
\setcounter{table}{0}
\setcounter{section}{0}
\setcounter{page}{1}
\makeatletter
\renewcommand{\theequation}{S\arabic{equation}}
\renewcommand{\thefigure}{S\arabic{figure}}
\renewcommand{\thetable}{S\arabic{table}}

The supplementary material is organized as follows.  The analysis methods are described in Sec.~\ref{sec: methods}.  Section \ref{sec: thresh} estimates the \emph{Fermi} PS-detection threshold.  Extended results for the analyses described in the main text are then given in Sec.~\ref{sec: extended}.  Section~\ref{sec: systematic} investigates a variety of systematic issues that may affect the Inner Galaxy analysis.  In Sec.~\ref{sec: comp}, we make a detailed comparison to previous work on the luminosity function for PSs in the Inner Galaxy. 
Section~\ref{Sec: SF} describes an  additional test of the unresolved PS models utilizing the survival function.  Lastly, Sec.~\ref{sec: sim} describes how the NPTF is validated using simulated data in the Inner Galaxy.

\section{Methods}
\label{sec: methods}

This section describes in detail the analysis framework for the NPTF and the dataset used in this work. 

\subsection{Analysis framework}
In the standard template-fitting procedure, the photon-count probability distribution of observing $k$ photons in a given energy bin and pixel $p$ is assumed to be Poissonian,
\es{eq: template}{
p_k^{(p)}= \frac{\left(\mu_p\right)^{k} \, e^{-\mu_p}}{k!} \, ,
}
where $\mu_p$ is the expected number of photons in the energy bin, summed over all the templates that contribute in that pixel.  In particular,
\es{eq: template2}{
\mu_p = \sum_{\ell} \alpha_{\ell} \, \mu_{p, \ell} \,,
}
where $\alpha_{\ell}$ is the best-fit normalization of the $\ell^\text{th}$ template in the given energy bin, and $\mu_{p,\ell}$ specifies the spatial dependence of the template.  That is, $\alpha_{\ell}\,\mu_{p,\ell}$ gives the expected number of events in the energy bin and pixel $p$ (accounting for an energy-dependent exposure, which may vary across the sky) for that template.  Past studies of the GeV excess have included templates that account for the spatial dependence of the diffuse and isotropic backgrounds, the \emph{Fermi} bubbles, and an NFW-distributed DM component.  We have repeated the standard template-fitting procedure and verified that our analysis pipeline reproduces results from those studies (see Sec.~\ref{SubSec: IG} for details).    

Unlike the standard template-fitting procedure, the NPTF does not assume that the photon counts are Poissonian.  We compute the non-Poissonian photon-count probabilities using the method of generating functions.  In particular, the total generating function for the photon-count probability distribution in each pixel is
\es{eq: genfun}{
\mathcal{P}^{(p)}(t) = \mathcal{D}^{(p)}(t) \cdot \mathcal{G}^{(p)} (t) \, ,
}
where $t$ is an auxiliary variable, and $\mathcal{D}^{(p)} (t)$ and $\mathcal{G}^{(p)} (t)$ are the pixel-dependent generating functions for the diffuse and non-Poissonian PS contributions, respectively.  The probability of observing $k$ photons in pixel $p$ is then
\es{eq: genfunP}{
p_k^{(p)} = \frac{1}{k!} \frac{d^k \mathcal{P}^{(p)}}{d t^k} \Big|_{t=0}  \, .
}
If $\mathcal{P}^{(p)}(t) = \exp[\mu_p(t-1)]$, then~\eqref{eq: genfunP} reduces to the simple Poissonian form found in~\eqref{eq: template}.  Explicit analytic formulae for these generating functions are derived in~\cite{1104.0010,Lee:2014mza}.

The benefit of the NPTF procedure is that it allows one to include additional spatial templates in the fitting procedure that are not properly characterized by Poissonian statistics---in particular, it allows for PS templates.  In this work, we include up to seven different templates, with up to 16 free parameters.  These parameters are listed in Tab.~\ref{tab:priors}, along with the prior ranges over which they are scanned.  The Bayesian analysis results in posterior probability distributions for each of these parameters.  We now explain each of the templates and the corresponding parameters in detail.

\begin{itemize}
\item \emph{Isotropic Background}\\
The isotropic-background template, which is uniform over the sky before correcting for exposure, is specified by the normalization parameter $A_\text{iso}$ in the NPTF.  We define $A_\text{iso}$ such that the number of counts arising from the isotropic background in pixel $p$ is given by $A_\text{iso}\, \alpha_{\text{iso}} \, \mu_{p, \text{iso}}$; that is, $A_\text{iso} = 1$ when the NPTF and the standard template fit agree exactly.\footnote{We neglect energy dependence in the NPTF and use only a single wide-energy bin, necessitating only one normalization parameter $A_l$ for each template.  However, we do account for energy dependence when performing the standard template fit, yielding individual values for the normalization parameters $\alpha_l$ for each of the narrower energy bins contained within the wide bin.  Implicit in this expression and those that follow is the appropriate sum over these bins.}  The prior for $A_\text{iso}$ is taken to be linear flat; the posterior is well-converged and is typically peaked around unity.

The best-fit pixel-averaged intensity for the isotropic background  in a given ROI is 
\es{eq: fluxnorm}{
I_\text{iso} = \frac{A_\text{iso} \, \alpha_{\text{iso}} \, }{\Delta\!\Omega} \, \Big\langle\frac{\mu_{p, \text{iso}}}{\epsilon^{(p)}} \Big\rangle_\text{ROI}\, ,
}
where $\Delta\!\Omega$ is the pixel solid angle, $\epsilon^{(p)}$ is the exposure in pixel $p$, and the angle brackets denote an average over the pixels in the ROI.  Note that $\mu_{p, \text{iso}} \propto \epsilon^{(p)}$ for the isotropic-background template.

\begin{table}
\renewcommand{\arraystretch}{1.5}
\setlength{\tabcolsep}{5.2pt}
\begin{center}
\begin{tabular}{| c | c | c |}
\hline
\multirow{2}{*}{Parameter} 	& \multicolumn{2}{c|}{Prior Ranges} 		 \Tstrut\Bstrut		\\   
\cline{2-3}
& High Latitude & Inner Galaxy	\Tstrut\Bstrut \\
\hline 
\hline
$A_\text{iso}$  & $[0,10]$ & $[-2, 2]$ \Tstrut\Bstrut \\ 
$A_\text{diff}$  & $[0, 10]$  & $[0, 2]$ \Tstrut\Bstrut \\
$A_\text{bub}$  & $[0, 10]$ & $[0,2]$ \Tstrut\Bstrut \\ 
$\log_{10} A_\text{NFW}$ 	     & $[-6, 6]$  &  $[-6, 6]$  \Tstrut\Bstrut	\\
$\log_{10}A_\text{PS}$ 		     	 & $[-6, 6]$ & $[-6, 6]$ \Tstrut\\
$S_b$ 	    & $[0, k_\text{max}]$   & $[0.05, k_\text{max}]$  \\
$n_1$ 	    & $[2.05, 50]$   & $[2.05, 30]$ \\
$n_2$     & $[-2, 1.95]$ & $[-2, 1.95]$\Bstrut\\
\hline
\end{tabular}
\end{center}
\caption{Parameters and associated prior ranges for the high-latitude and IG analyses.  Note that there can be up to three copies of the PS parameters $(A_\text{PS}, S_b, n_1, n_2)$ when isotropic, disk-correlated, and NFW-distributed PS templates are all included in the fit.  The break $S_b$ is scanned up to $k_\text{max}$, the maximum number of photons observed in a given pixel within the ROI; in general, $k_\text{max} > 50$ photons.}
\label{tab:priors}
\end{table}

\item \emph{Diffuse Background}\\
The spatial dependence $\mu_{p, \text{diff}}$ of the diffuse-background template is based on the \emph{Fermi} \texttt{p6v11} model for the majority of our analyses; however, we do perform a number of cross-checks with other diffuse models.  As with the isotropic-background template, the diffuse-background template is specified by a single normalization parameter, $A_{p, \text{diff}}$, also defined such that $A_{p, \text{diff}} = 1$ when the NPTF agrees exactly with the standard template fit.\footnote{In Sec.~\ref{sec:varydiffuse}, we consider a class of diffuse models for which the $\pi^0$, bremsstrahlung, and ICS components are allowed to float independently; in these cases, there are three normalization parameters for the diffuse background. }
We assume a linear-flat prior, and the resulting best-fit value is typically close to unity.  The best-fit pixel-averaged intensity $I_\text{diff}$ is computed as in \eqref{eq: fluxnorm}.
\item \emph{Fermi Bubbles}\\ 
The \emph{Fermi}-bubbles template assumes uniform emission within the bubbles~\cite{Su:2010qj} and is also specified by a single normalization parameter, $A_\text{bub}$, defined analogously as above.  We similarly assume the prior for $A_\text{bub}$ is linear flat; the best-fit value is again typically close to unity.  The best-fit pixel-averaged intensity $I_\text{bub}$ is computed as in \eqref{eq: fluxnorm}.

\item \emph{NFW Dark Matter}\\
 The spatial dependence of the DM template is determined by $\Phi_p$, which gives the integrated photon-count intensity at the center of pixel $p$ from DM annihilation in the Galactic halo.  The intensity $\Phi_p$ is computed by the line-of-sight integral
\es{eq:DM flux}{
\Phi(\psi) \propto \int_\text{los} \! ds \, \rho\left[ r(s, \psi) \right]^2  \, ,
}
where $\rho$ is the DM density profile and $r$ gives the distance from the GC.  This work assumes a generalized Navarro-Frenk-White (NFW) density profile~\cite{Navarro:1995iw,astro-ph/9611107}
\begin{equation} \label{eq: gNFW}
\rho(r) \propto \frac{(r/r_s)^{-\gamma}}{(1+r/r_s)^{3-\gamma}} \,,
\end{equation}
where $r_s = 20$~kpc is the scale radius.  We take $\gamma = 1.25$ for the majority of the analyses.

With the spatial dependence of $\mu_{p, \text{NFW}}$ fixed by the NFW intensity profile and exposure map, the DM template can be specified by a single normalization parameter,  $A_\text{NFW}$, defined analogously as above.  We assume a prior that is log flat and covers a broad range (see Tab.~\ref{tab:priors}).  The normalization of $A_\text{NFW}$ is such that it is equal to unity for $\sim$35~GeV DM annihilating into $b\bar{b}$ with cross section $\langle \sigma v \rangle_0 \approx 1.7\times 10^{-26}$~cm$^3$/s, and a local DM density at the solar circle of 0.3 GeV/cm$^3$.

\item \emph{NFW Point Sources}\\
Previous standard template fits have included the above Poissonian templates.  However, in the NPTF, PS templates have non-Poissonian statistics, which can be derived from the source-count function. In the majority of this work, the PS source-count function in a given pixel $p$ is modeled as a broken power law
\es{dNdS}{
{d N_p(S) \over dS} = A_{p} \left\{ 
\begin{array}{cc}
\left( {S \over S_b} \right)^{-n_1} & S \geq S_b \\
\left( {S \over S_b} \right)^{-n_2} & S < S_b  \, ,
\end{array}
\right.
}
where $S_b$ is the break, $n_{1,2}$ are the slopes above and below the break, and $A_p$ is a pixel-dependent normalization.  We require $n_1>2$ and $n_2<2$ so that the total number of photons contributed from PSs is finite.  The priors for the parameters $S_b$, $n_1$, and $n_2$ are all linear flat.

For the NFW PS template, the pixel-dependent normalization of the source-count function, $A_p$, is assumed to be proportional to $\mu_{p, \text{NFW}}$, as would be needed for the PSs to mimic a DM signal.\footnote{We fold the spatial dependence arising from both the NFW intensity profile and the exposure into the pixel-dependent normalization $A_p$ as an approximation.  Strictly speaking, the exposure correction should be modeled by a pixel-dependent break $S_{b,p}$.  However, this increases the computational complexity required to perform the NPTF.  We use the same approximation to incorporate variations in the exposure map in the high-latitude analysis.  While this approximation is probably valid in the IG, since the exposure map does not vary significantly within $\sim$10$^\circ$ of the GC, this may have an important effect on the isotropic PS population in the high-latitude analysis.}  We can thus specify the NFW PS template by the three source-count function parameters and an overall normalization parameter, $A_\text{PS}$, defined such that
\es{eq: PSnorm}{
A_p = A_\text{PS} \frac{\mu_{p, \text{NFW}}}{\langle\mu_{p, \text{NFW}}\rangle_\text{ROI}}\,.
}
As with $A_\text{NFW}$, the prior for $A_\text{PS}$ is log flat and covers a broad range.

We also quote the pixel-averaged intensity $I_\text{PS}$, which is given by 
\es{IPS}{
I_\text{PS} =   \frac{1 }{\Delta\!\Omega} \, \Big\langle\frac{\mu_p^\text{PS}}{\epsilon^{(p)}} \Big\rangle_\text{ROI}\,, 
}
with $\mu_p^\text{PS}$ the expected number of counts in pixel $p$ arising from the PS population:
\es{SPS}{
\mu_p^\text{PS} = \int \!dS\, S\, {d N_p \over dS} = A_p \, S_b^2 \left( {1 \over n_1 - 2} + {1 \over 2 - n_2} \right)   \,.
}

\item \emph{Isotropic Point Sources}\\
The source-count function for the isotropic PS template is also modeled as a broken power law, as in~\eqref{dNdS}, except with $A_p \propto \mu_{p, \text{iso}} \propto \epsilon^{(p)}$.  As above, the isotropic PS template can be specified by either its overall normalization parameter $A_\text{PS}^\text{iso}$ or the pixel-averaged intensity $I_\text{PS}^\text{iso}$.

\item \emph{Disk-correlated Point Sources}\\
The thin-disk template is the projection along the line-of-sight of the source spatial distribution given by:
\es{eq: diskymodel}{
n(z, R) \propto \exp\left[ \frac{-R}{5~\text{kpc}} \right] \, \exp \left[\frac{-|z|}{0.3~\text{kpc}}\right] \, ,
}
where $R$, $z$ are cylindrical polar coordinates.  The source-count function associated with this template is modeled as a broken power law, as in~\eqref{dNdS}, except with $A_p \propto \mu_{p, \text{disk}}$.

\end{itemize}

Given these templates and their associated generating functions, the overall photon-count probability distribution $p_{k}^{(p)} (\theta)$ can be written as a function of the 16 parameters
\es{eq: parameters}{
\theta = \{A_\text{iso}, A_\text{diff}, A_\text{bub}, A_\text{NFW}, A_\text{PS}, S_b, n_1, n_2, A_\text{PS}^\text{iso}, S_b^\text{iso}, n_1^\text{iso}, n_2^\text{iso},A_\text{PS}^\text{disk}, S_b^\text{disk}, n_1^\text{disk}, n_2^\text{disk}\} \,.
}
Then, for a data set $d$ consisting of the set of $\{n_p\}$ photon counts in each pixel $p$, the likelihood function for observing a particular photon-count distribution over all pixels in the ROI is
\es{llf}{
p(d | \theta, {\cal M}) =  \prod_p p_{n_p}^{(p)} (\theta) \, .
}
With the priors specified above, this likelihood function can be used in the standard framework of Bayesian inference to compute both the posteriors and the evidence for models $\mathcal{M}$ that include various subsets of the parameters $\theta$.  We use the \texttt{MultiNest} package for the Bayesian calculations~\cite{Feroz:2008xx,Buchner:2014nha}.  All \texttt{MultiNest} runs use 500 live points with importance nested sampling disabled, constant efficiency mode set to false, and sampling efficiency set for model-evidence evaluation; the typical number of posterior samples generated for each run is $\sim$10$^4$.

\subsection{Data Selection Criteria}
The NPTF analysis was performed using the Extended Pass~7 Reprocessed \emph{Fermi} data from $\sim$August 4, 2008 to $\sim$December 5, 2013 made available by~\cite{1406.0507}.  \emph{Ultraclean} front-converting events with zenith angle less than 100$^\circ$ and ``\texttt{DATA\_QUAL==1 \&\& LAT.CONFIG==1 \&\& ABS(ROCK.ANGLE) < 52}'' are selected, and a Q2 cut on the CTBCORE parameter is used to remove events with poor directional reconstruction.  
 We have used the original CTBCORE-cut dataset throughout this work~\cite{1406.0507}, but we have tested the effect of including \emph{Fermi} data up to March 8, 2015 (mission week 353) and the results are consistent with those presented here.  Additionally, we have rerun the analysis with Pass 8 \emph{Fermi} data up to June 3, 2015, using the new \emph{Ultracleanveto} class and the top quartile of events ranked by PSF (denoted PSF3).  We adopt the recommended data quality cuts for this analysis,\footnote{http://fermi.gsfc.nasa.gov/ssc/data/analysis/documentation/Cicerone/Cicerone\_Data\_Exploration/Data\_preparation.html} which are slightly different to those in our main analysis (\emph{e.g.}, the zenith angle cut has been reduced from $100^\circ$ to $90^\circ$ and the rocking angle cut has been removed).  Again, the results are consistent with those presented in this work.

\begin{figure}[b]
	\leavevmode
	\begin{center}$
	\begin{array}{cc}
	\scalebox{0.40}{\includegraphics{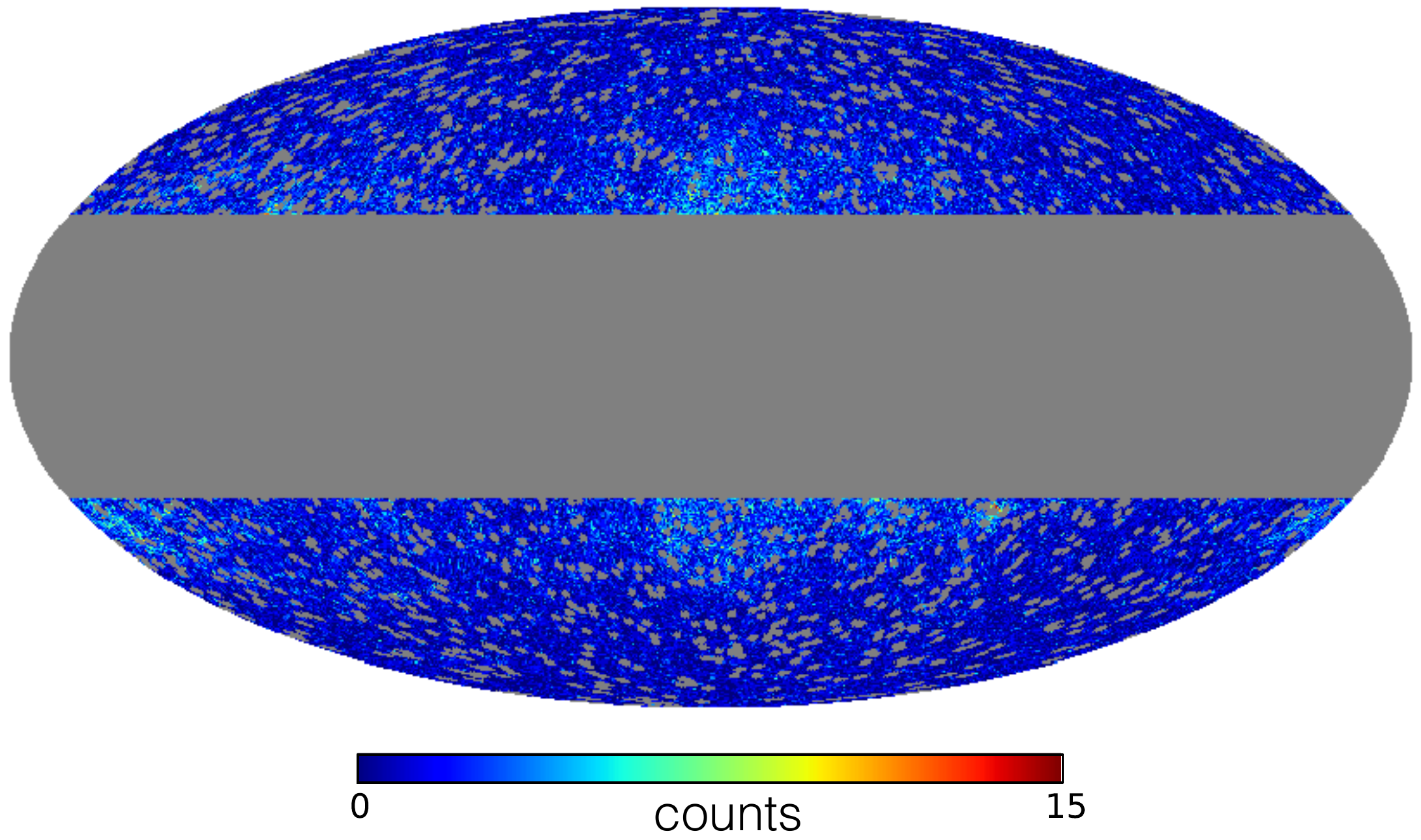}} & \scalebox{0.45}{\includegraphics{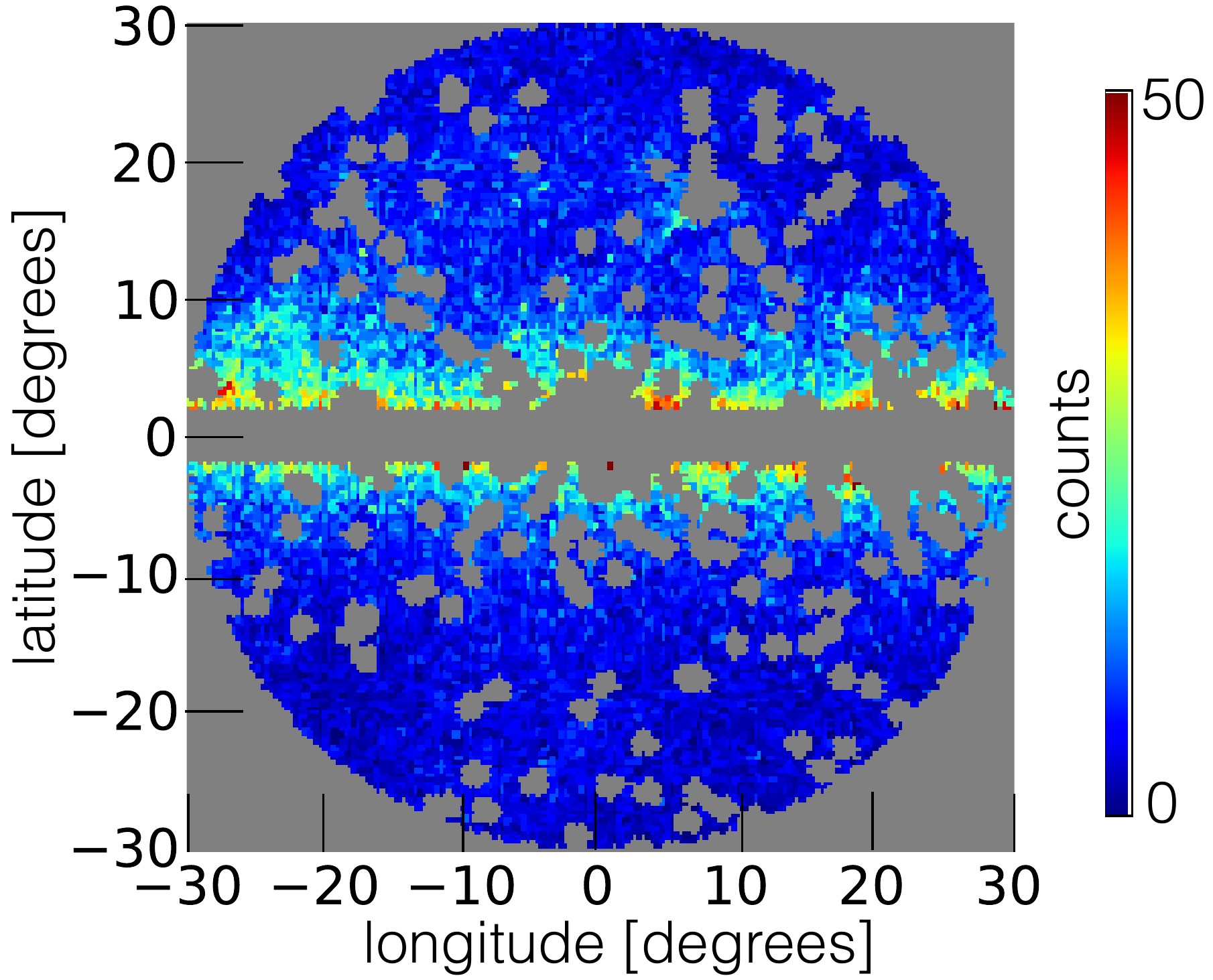}} 
	\end{array}$
	\end{center}
	\vspace{-.50cm}
	\caption{The counts map for the high-latitude ($|b| \geq 30^\circ$) analysis (clipped at 15 counts) is shown in the left panel.  The IG analysis focuses on the region within $30^\circ$ of the GC, with $|b| \geq 2^\circ$.  The associated counts map (clipped at 50 counts) is shown in the right panel.  All pixels within $\sim$1$^\circ$ of known \emph{Fermi} 3FGL sources are masked.   }      
	\vspace{-0.15in}
	\label{Fig: counts maps}
\end{figure}

The main body of the Letter focused primarily on two regions of interest: a high-latitude analysis with $|b| \geq 30^\circ$ and an IG analysis that included all pixels within $30^\circ$ of the GC, with $|b| \geq 2^\circ$.  These regions are shown in Fig.~\ref{Fig: counts maps}.  When masking identified PSs from the \emph{Fermi} 3FGL catalog~\cite{TheFermi-LAT:2015hja}, all pixels within $5\times0.198^\circ$ of the source are excluded.  This mask is sufficiently large to completely contain the flux from the majority of the PSs; the results do not qualitatively change as the mask size is varied, for example, to $7\times0.198^\circ$.  

If the detector's point spread function (PSF) is comparable to or larger than the pixel size, photons from a PS in a given pixel can leak into neighboring pixels.  As a result, the PSF must be properly accounted for in the calculation of the photon-count probability distributions for the PS templates~\cite{1104.0010,Lee:2014mza}.  To model the PSF for the CTBCORE-cut data, we use the instrument response functions made available by~\cite{1406.0507} to approximate the PSF as a two-dimensional Gaussian with energy-dependent width, $\sigma$ (see Sec.~\ref{Sec: PSF} for details).\footnote{The Gaussian PSF is used to smear the NFW DM map.  In contrast, the diffuse-background model is smeared with the exact \emph{Fermi} PSF using the Fermi Science Tools to construct the diffuse-background template.}  The NPTF analysis includes data from a single energy bin from $\sim$1.893--11.943~GeV.  Small modifications to the choice of energy bin do not affect the conclusions.  For these energies, the PSF parameter $\sigma$ varies from $\sim$$0.198^\circ$ at low energies to $\sim$$0.0492^\circ$ at high energies.  In calculating the photon-count probability distributions for the PS components, we use the energy-averaged value for $\sigma$ assuming the spectrum of the excess (high-latitude 3FGL PSs) for all PS templates in the IG (high-latitude) analysis.     We further explore systematic uncertainties arising from the PSF in Sec. \ref{Sec: PSF}.

\section{The \emph{Fermi}-LAT Point-Source Detection Threshold}
\label{sec: thresh}

The analysis presented in this Letter does not rely on any prescription for the sensitivity of \emph{Fermi} to PSs, except implicitly via masking out the known sources.  However, it is still worth considering whether the PS population we infer should already have been resolved as individual sources, given the nominal sensitivity of \emph{Fermi}.

To some degree, the detection threshold can be read off directly from figures such as Fig. \ref{Fig: IG_dnds_unmasked}; the threshold may be responsible for the turnover in the histogram of known sources between $\sim$2--4$\times 10^{-10}$ photons/cm$^2$/s. However, our analysis is restricted to a single high-energy bin; the ``faintest'' resolved sources in our sample  include sources that are not intrinsically faint but have very soft spectra, and so have very few photons at high energy. The analysis by which the 3FGL catalog was created used a wider range of energies, and so it is entirely possible that a source might be detected by its bright emission at low energies, yet be too faint to be independently detected in our 1.893--11.943~GeV energy range, thus giving the false appearance that very faint sources can be detected. The spectrum of the excess is rather hard relative to the known 3FGL sources in our ROI, so PSs comprising the excess will tend to become less detectable relative to the resolved 3FGL sources in lower energy bands.

Estimates for the PS sensitivity in the literature generally assume a specific source spectrum and are based on fewer years of \emph{Fermi} data than our current analysis (although they may have similar statistics, because we are using front-converting \emph{Ultraclean} events and have applied further cuts on the CTBCORE parameter). The \emph{Fermi} Collaboration has presented longitude-averaged sensitivity estimates based on the integral flux from 0.1--100 GeV with three years of data, for sources with a pulsar-like spectrum \cite{1305.4385}; for Galactic latitudes with $2^\circ < |b| < 10^\circ$, the mean sensitivity varies from $\sim$$5.3$--$10 \times 10^{-12}$ erg/cm$^2$/s, with the $10\%$ percentile sensitivity ranging from $\sim$$3.6$--$5.4 \times 10^{-12}$ erg/cm$^2$/s. The average over longitude probably leads to a somewhat lower threshold than is accurate in the neighborhood of the GC, but on the other hand, the 3FGL catalog is based on four years of data rather than three.

Let us assume the spectrum of the excess, and all sources comprising it, is approximately given by the best-fit broken power-law spectrum of \cite{1409.0042}, with indices $\sim$1.42 and $\sim$2.63 below and above the break, respectively, and a break energy of $2.06$ GeV. Then, a luminosity of $10^{-12}$ erg/cm$^2$/s in the 0.1--100 GeV band corresponds to $\sim$$7 \times 10^{-11}$ photons/cm$^2$/s in our energy band. Thus, a threshold of $3.6 \times 10^{-12}$ erg/cm$^2$/s in integrated energy flux translates to a photon-flux threshold $\sim$$2.5 \times 10^{-10}$ photons/cm$^2$/s in the relevant energy range, for the $10\%$ percentile sensitivity 10$^\circ$ from the Galactic plane. This is directly above the break in our inferred source-count functions.

\section{Extended results}
\label{sec: extended}
This section includes extended discussion of the results presented in the main text.  

\subsection{High latitudes}
\begin{figure}[tb]
	\leavevmode
	\begin{center}$
	\begin{array}{cc}
	\scalebox{0.40}{\includegraphics{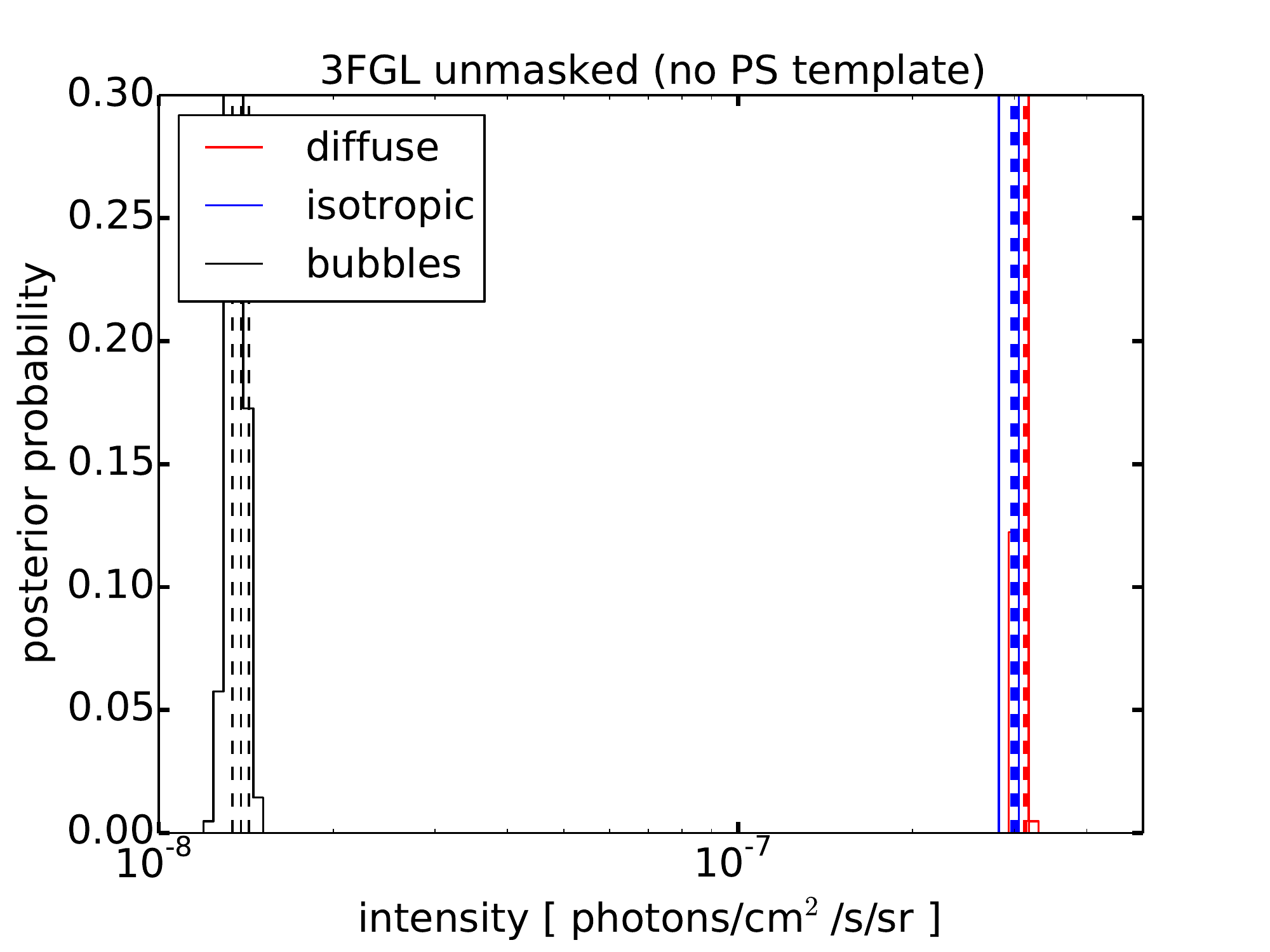}} & \scalebox{0.40}{\includegraphics{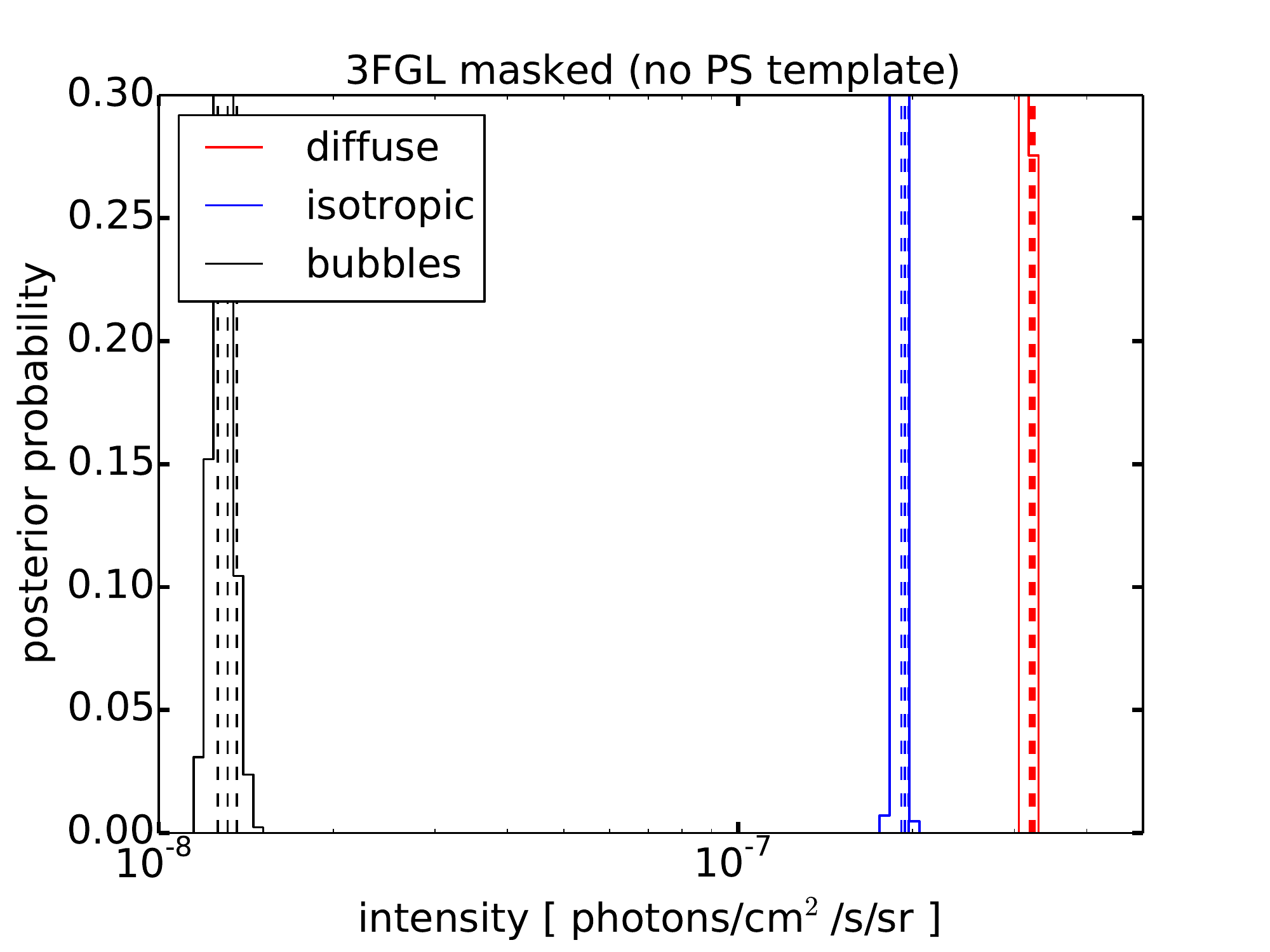}}  \\
	\scalebox{0.40}{\includegraphics{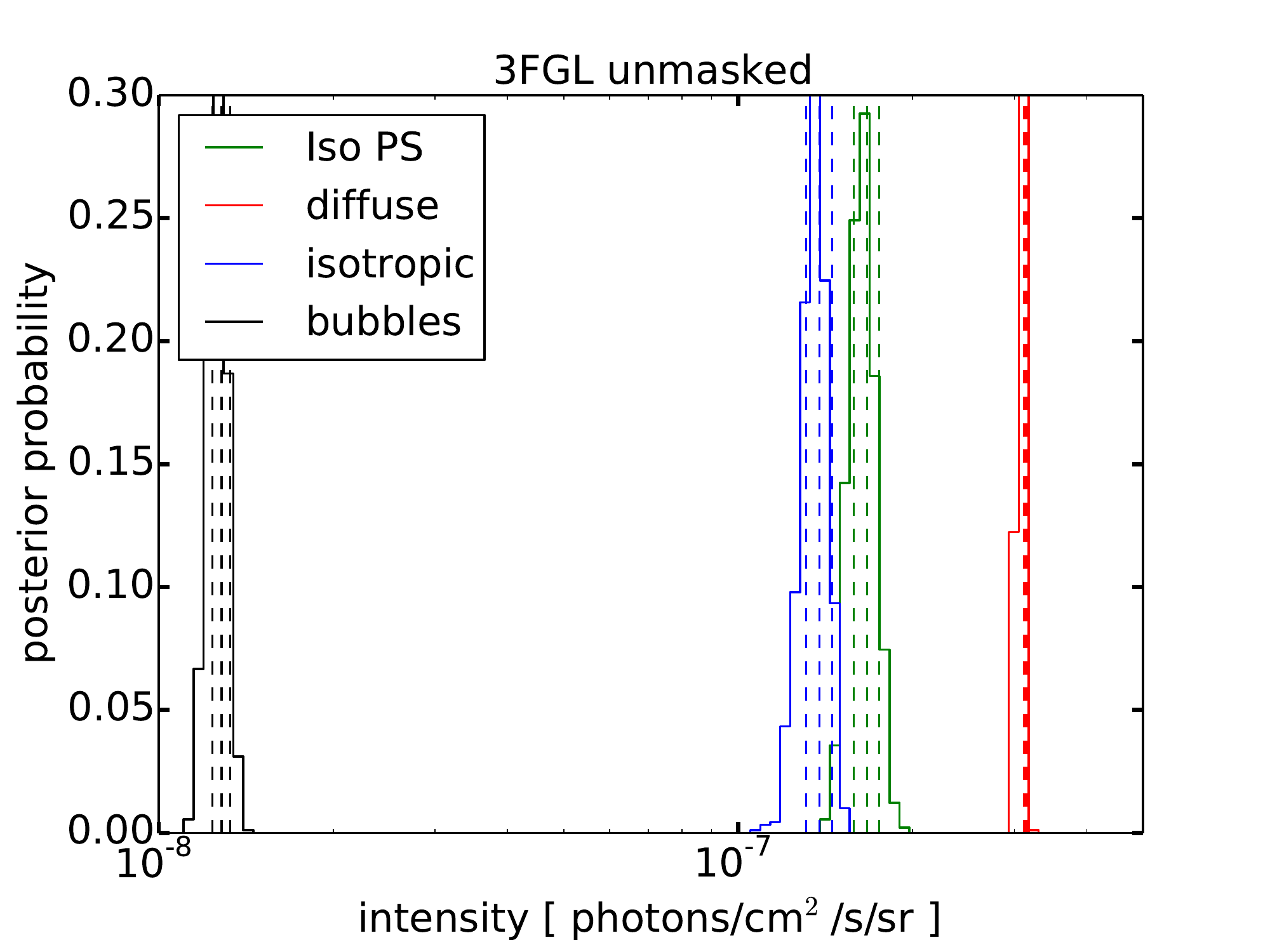}} & \scalebox{0.40}{\includegraphics{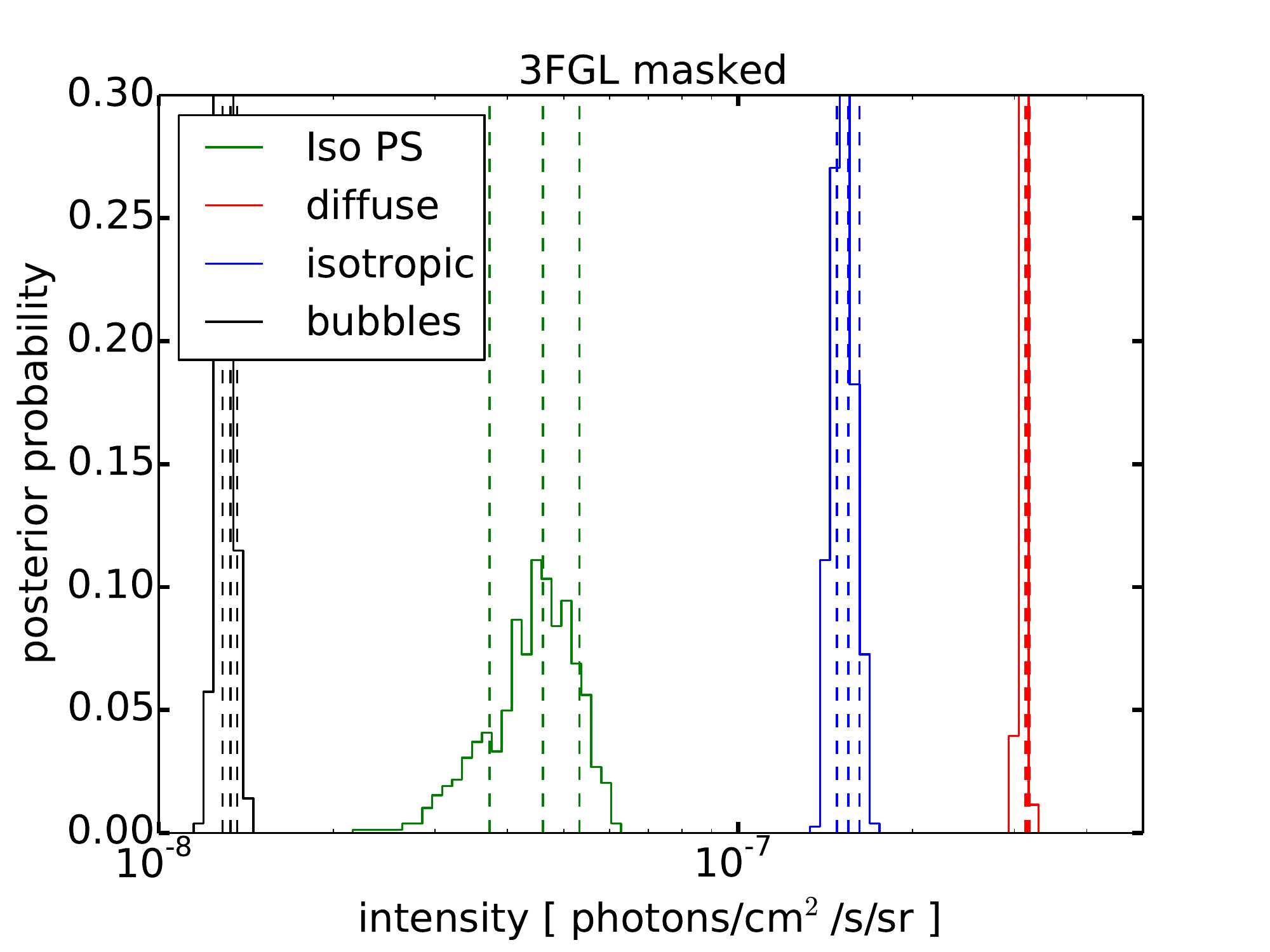}}  
	\end{array}$
	\end{center}
	\vspace{-.50cm}
	\caption{The posterior probabilities for the photon intensity associated with a given template in the high-latitude analysis with 3FGL sources unmasked (left column) and masked (right column).  The top row shows the result without a PS template (\emph{i.e.}, the standard template fit), while the bottom row shows the result of the NPTF with an isotropic PS template included.  Whether or not the 3FGL sources are masked, the isotropic template in the standard template fit clearly absorbs intensity that is assigned to the isotropic PS template in the NPTF. Dashed vertical lines indicate the $16^\text{th}$, $50^\text{th}$, and $84^\text{th}$ percentiles.}   
	\vspace{-0.15in}
	\label{Fig: highL photon flux}
\end{figure}

Fig.~\ref{Fig: highL photon flux} shows the posterior probabilities for the photon intensity associated with each template in the high-latitude analysis ($|b| \geq 30^\circ$), with 3FGL sources unmasked and masked (left and right columns, respectively).  The top row shows the results when only the isotropic, diffuse, and bubbles templates are included in the fit.  The best-fit intensity values for the diffuse and bubbles components are essentially constant, regardless of whether or not the 3FGL sources are masked.  However, the isotropic template is sensitive to the presence of the 3FGL sources, with its median best-fit intensity increasing from $\sim$$1.9\times10^{-7}$ to $\sim$$3.0\times10^{-7}$~photons/cm$^2$/s/sr when going from the right to left panels.  

When the isotropic PS template is included in the NPTF at high-latitudes (bottom row in Fig.~\ref{Fig: highL photon flux}), there is no effect on the flux absorbed by the diffuse and bubbles template.  In this case, the isotropic template is insensitive (within uncertainties) to the masking of the 3FGL sources, with a photon intensity that remains unchanged between the left and right panels.  Instead, the isotropic PS template accounts for the resolved PSs when the 3FGL sources are unmasked, and picks up flux (presumably from unresolved PS emission) when the 3FGL sources are masked.  

The intensity of the IGRB that is predicted by the NPTF can be compared with published values from the \emph{Fermi} collaboration~\cite{Ackermann:2014usa}.  Applying a standard template analysis to 50 months of \emph{Fermi} LAT data,~\cite{Ackermann:2014usa} determined the IGRB spectrum from 100~MeV to 820~GeV.  The brightest PSs from the 2FGL catalog (the predecessor of the 3FGL catalog) were fitted individually, while a standard template was used to model the other identified sources.  Three foreground models---labeled A, B, C---were studied to better understand the effects of variations in the diffuse emission.  For each foreground model, the spectrum of the IGRB intensity was fit to a broken power law with an exponential cutoff.  According to our estimates, the results in~\cite{Ackermann:2014usa} predict IGRB intensities of $I_\text{iso}^{A} =  1.33 \times 10^{-7}$, $I_\text{iso}^{B} =  1.80 \times 10^{-7}$, $I_\text{iso}^{C} =  1.35 \times 10^{-7}$~photons/cm$^2$/s/sr in the energy range from 1.893--11.943~GeV for the three models, with systematic uncertainties on the order of $10\%$ in each case.  The intensity of the IGRB computed using the NPTF (\emph{e.g.}, $1.55_{-0.07}^{+0.07}\times10^{-7}$~photons/cm$^2$/s/sr for the 3FGL-masked fit) is generally in agreement with the results of~\cite{Ackermann:2014usa}.
  However, a direct comparison is difficult to make because, for example, the analyses treat the known PSs differently, use different PS catalogs, use different data sets, and use different foreground models. 
It would be interesting to repeat the analysis in~\cite{Ackermann:2014usa} including a non-Poissonian PS template to more accurately extract the contribution of unresolved PSs to the total EGB.  

The central analyses in this work assume the source-count function parameterization in~\eqref{dNdS}, however we also consider the effect of forcing the source-count function to have a sharp cutoff at high flux, while still allowing a break at lower flux.  For example, we can impose a high-flux cutoff consistent with the flux of the brightest 3FGL PS in the ROI: \mbox{$F_\text{cut} \approx 1.45 \times 10^{-8} $ } photons$/\text{cm}^2/\text{s}$.  In this case, the NPTF gives the following best-fit values: $n_1 = 1.97_{-0.03}^{+0.05}$, $n_2 = 1.61_{-0.14}^{+0.05}$, and $F_b = 1.02_{-0.72}^{+1.4} \times 10^{-10} $ photons$/\text{cm}^2/\text{s}$.  Additionally, $I_\text{PS}^\text{iso} =  1.41_{-0.09}^{+0.09}\times10^{-7}$ photons/cm$^2$/s/sr, which---subtracting the intensity of the identified 3FGL PSs---predicts that the unresolved PSs have an intensity $4.73_{-0.93}^{+0.89} \times 10^{-8}$  photons/cm$^2$/s/sr.  This is in good agreement with the estimate obtained from the 3FGL-masked NPTF described in the main body of the Letter (\emph{e.g.}, $I_\text{PS}^\text{iso} = 4.61_{-0.88}^{+0.72} \times 10^{-8}$ photons/cm$^2$/s/sr).  Indeed, the agreement is better than that obtained from estimating the intensity of the unresolved PS population using the source-count function without the high-flux cutoff (\emph{e.g.}, $7.38_{-0.85}^{+0.83}$$\times10^{-8}$ photons/cm$^2$/s/sr).

\subsection{Inner Galaxy} \label{SubSec: IG}

\afterpage{%
\begin{figure}[b]
	\leavevmode
	\begin{center}$
	\begin{array}{c}
	\scalebox{0.2}{\includegraphics{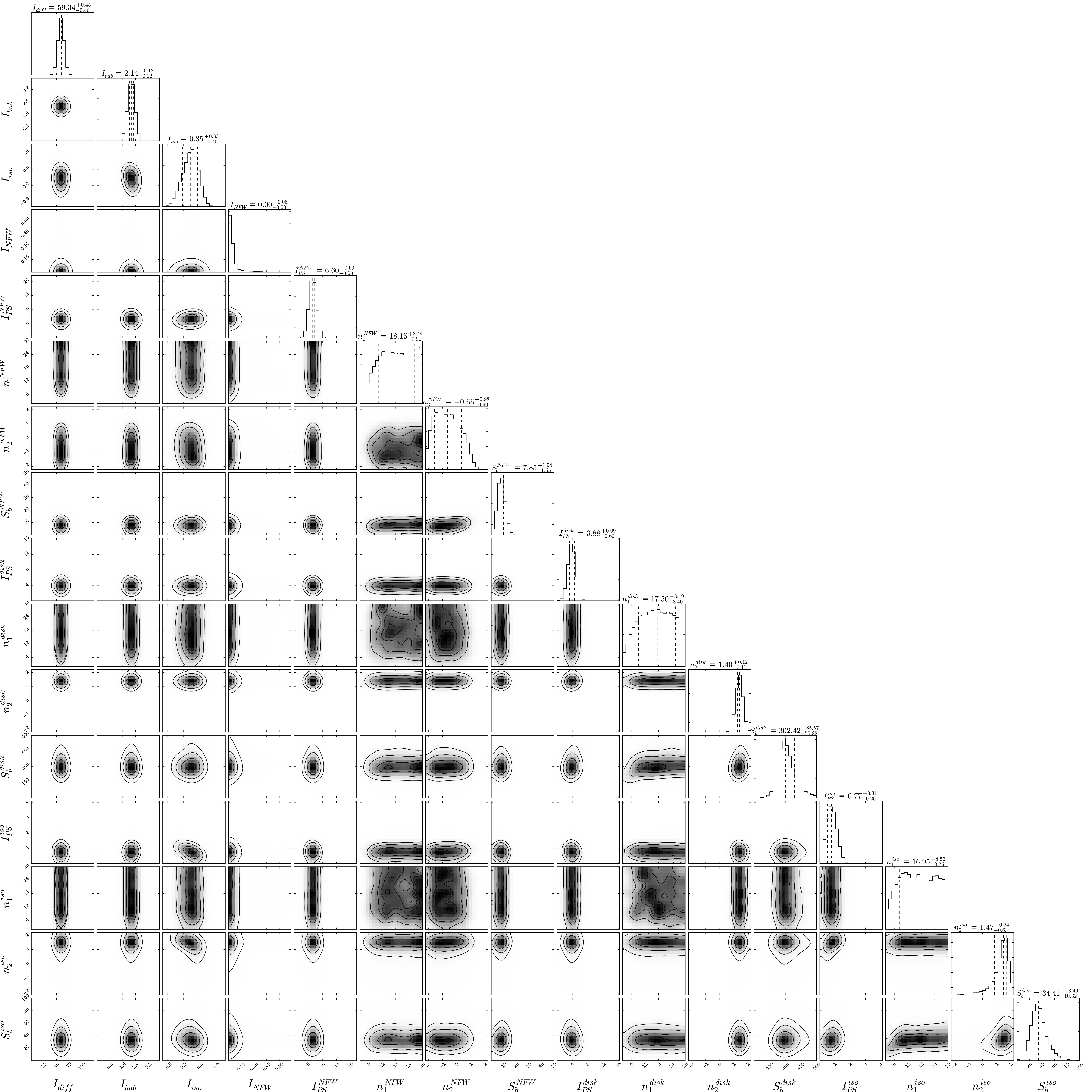}} 
	\end{array}$
	\end{center}
	\vspace{-.50cm}
	\caption{Triangle plot for the IG analysis with the 3FGL sources unmasked. Intensities are in units of $10^{-7}$~photons/cm$^2$/s/sr and are calculated with respect to the region within $10^\circ$ of the GC, with $\abs{b} \geq 2^\circ$.}     
	\vspace{-0.15in}
	\label{Fig: IG triangle nomask}
\end{figure}
\clearpage
}

\afterpage{%
\begin{figure}[b]
	\leavevmode
	\begin{center}$
	\begin{array}{c}
	\scalebox{0.2}{\includegraphics{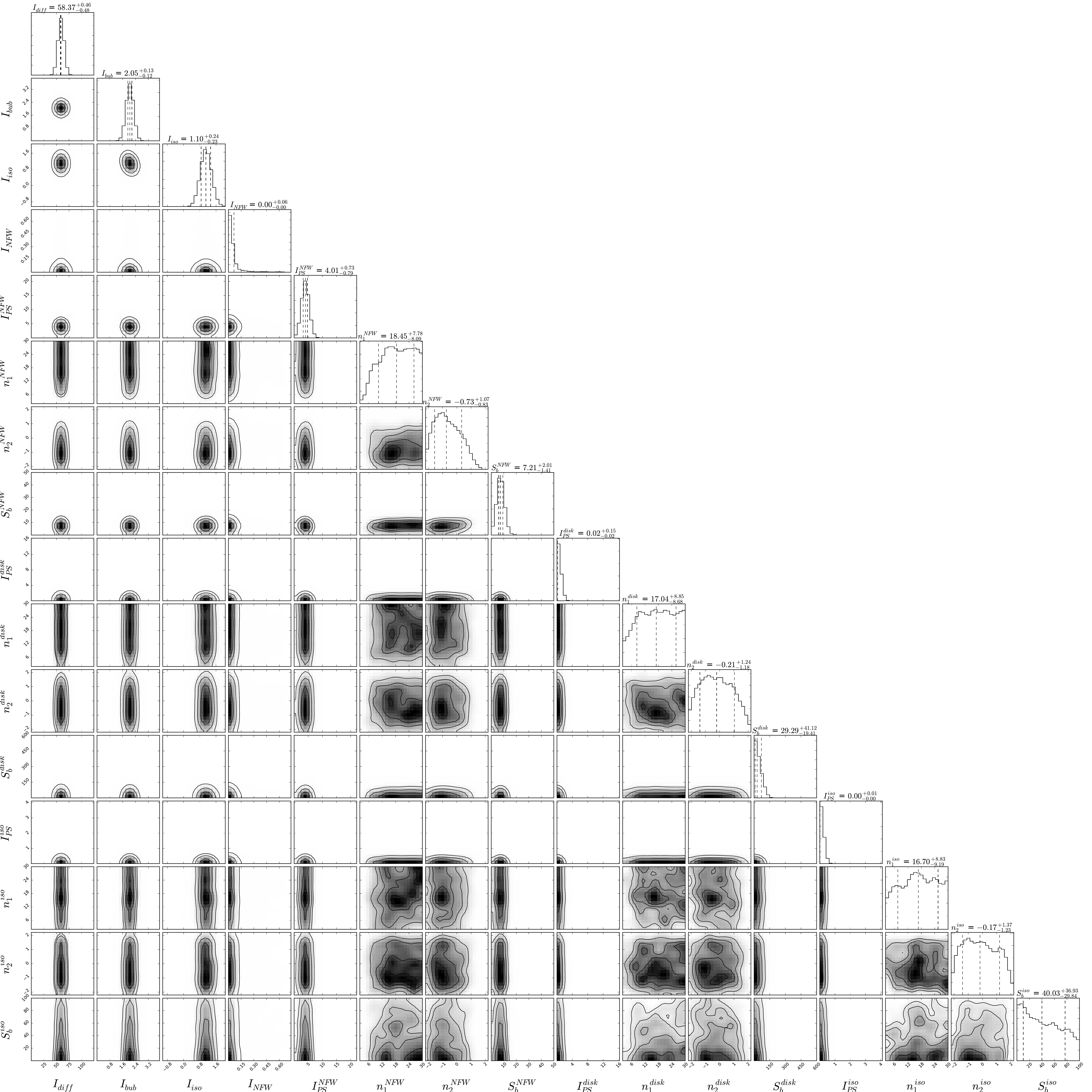}} 
	\end{array}$
	\end{center}
	\vspace{-.50cm}
	\caption{Triangle plot for the IG analysis with the 3FGL sources masked.  Intensities are in units of $10^{-7}$~photons/cm$^2$/s/sr and are calculated with respect to the region within $10^\circ$ of the GC, with $\abs{b} \geq 2^\circ$.}     
	\vspace{-0.15in}
	\label{Fig: IG triangle mask5}
\end{figure}
\clearpage
}

Figs.~\ref{Fig: IG triangle nomask} and~\ref{Fig: IG triangle mask5} show the posterior probabilities for the IG analysis (within $30^\circ$ of the GC, with $|b| \geq 2^\circ$), with 3FGL sources unmasked and masked, respectively.  In both cases, the parameters are all well-constrained within the prior ranges, with one exception.\footnote{
 While the parameter $n_2$ is peaked within the prior range, it is also not very well-constrained by the fit because it mostly affects the low-flux part of the distribution.
  Widening the prior range for $n_2$ to values well below $-2$ may add a tail to the posterior distribution for $n_2$ in some analyses, which skews the posterior for this parameter towards negative values.  This has the effect of  widening the source-count function confidence interval at low flux.  With that said, we find that the mean source-count function is not affected by broadening this prior.  One reason we present results with the prior constraint $n_2 > -2$ is that the mean and median source-count functions, computed point-wise in $F$, are in good agreement in all analyses at low flux in this case.  Similarly, in the PS-masked fit, the median source-count function may significantly differ from the mean at high flux, above the PS-detection threshold, as the source-count function falls steeply.  However, it should be understood that all our results are subject to considerable systematic uncertainties well below the PS-detection threshold.  In the 3FGL-masked fit, the source-count function falls steeply above the PS-detection threshold, and here as well systematic uncertainties are expected to be large.  See Sec.~\ref{sec:binned} for more details.
  }  Namely, the slope $n_1$ above the break of the PS source-count functions may have large error bars.  This tends to happen in the masked analyses, where the source-count functions fall off steeply near the detection threshold.
 
As shown in Figs.~\ref{Fig: IG triangle nomask} and~\ref{Fig: IG triangle mask5}, the NPTF finds that the average intensity of the diffuse emission in this region is $I_\text{diff} = 59.34_{-0.46}^{+0.45} \times10^{-7}$ ($58.37_{-0.48}^{+0.46} \times 10^{-7}$)~photons/cm$^2$/s/sr with 3FGL sources unmasked (masked); the predicted intensities are similar in both scenarios, as desired.  Additionally, these intensities change by less than $\sim$$1\%$ from the respective values obtained by the standard template procedure (\emph{i.e.}, when only an NFW DM template is included).  The fact that the intensity of the diffuse emission is essentially constant between the NPTF and standard template analysis highlights that the addition of the PS templates does not affect the flux absorbed by the diffuse background template in the IG region.

One cross-check of the results is to compare the predicted fractions of flux from DM in the template fits that do not include PSs---where the excess appears to be absorbed by the NFW DM template---to results found in previous template studies.  Ref.~\cite{1402.6703} performed two analyses, with different ROIs, that are relevant for this comparison.\footnote{We only present a detailed check of our flux fractions against the results of~\cite{1402.6703} because the results of Ref.~\cite{1409.0042} are consistent with those of~\cite{1402.6703}, within estimated systematic uncertainties.}  The ROI for the first analysis was a $40^\circ \times 40^\circ$ region around the GC, while the second analysis used the full sky.  In both cases, the plane was masked ($\abs{b} \geq 1^\circ$) in addition to the 300 brightest and most variable PSs in the 2FGL PS catalog, using estimated 95\% containment masks.  A spectral and spatial model was also included for the remaining 2FGL sources, based on their positions and spectra in the catalog.  Ref.~\cite{1402.6703} used the same Poissonian templates as we do, so we can compare their results to our fits that do not include NFW PSs.  

Using the results of~\cite{1402.6703}, we can compute the predicted fraction of flux from DM relative to the total number of observed counts within $10^\circ$ of the GC, with $\abs{b} \geq 2^\circ$, no PSs masked, and within the energy range considered in this Letter.  For the restricted (full-sky) ROI, we find that~\cite{1402.6703} predicts the fraction of flux from DM to be $9.2 \pm 0.5\%$ ($6.3 \pm 0.4$\%).  The difference between these values gives a sense for the systematic uncertainty that comes from changing ROIs.  Given the fact that our analyses use different ROIs from~\cite{1402.6703}, we consider our results to be consistent with theirs within systematic uncertainties.  But importantly, regardless of the ROI, we find that when we include an NFW PS template in addition to an NFW DM template, the excess flux is preferentially absorbed by the PS template.           

For the results presented in this work, we have used a double-exponential thin-disk template with scale height and radius of 0.3 and 5~kpc, respectively.  However, we find that our results are not sensitive to variations in the disk template.  For example, we have repeated the masked and unmasked analyses using a thick disk with scale height and radius of 1 and 5~kpc, respectively.  This thick-disk distribution has been shown to be a good model for the distribution of MSPs identified by \emph{Fermi}~\cite{1305.0830,1407.5625,Gregoire:2013yta}.  We have also considered the case of a thin disk with scale height and radius of 0.3 and 10~kpc, respectively.  In both these cases, the results of our analyses remain essentially the same.

One substantial difference between our work and previous studies of the excess is that we include no energy information, using only a single large energy bin. In future work, it would be very useful to incorporate energy dependence into the NPTF likelihood function in order to extract a spectrum for the PS population.  It would also be useful to understand whether DM substructure could give rise to all or part of the excess.  In this Letter, we have modeled DM emission as smooth emission, but it may be the case that the DM emission is more PS-like because of, for example, DM subhalos.

\section{Systematic Uncertainties in the NPTF} 
\label{sec: systematic}

There are various systematic uncertainties that may influence the IG analysis, which we examine in some detail in this section.\footnote{We have additionally performed multiple tests that we do not show here for brevity.  For example, we have performed analyses with isotropic and isotropic PS emission constrained by the results of the high latitude analyses.  In general, constraining the priors for these templates is found to increase the evidence in favor of the model with NFW PSs.  Thus, to be conservative, we leave the priors for these templates unconstrained throughout the Letter. } 

\subsection{Broadening the ROI}
\label{sec: broad}

As a first cross check of our results, we perform the NPTF on the full sky with $|b| \geq 2^\circ$. 
This analysis can be used to test the North-South symmetry of the excess by masking each hemisphere in turn.  The results for the full-sky and hemisphere analyses are similar, so only the latter are presented here.  Additionally, we only show results for the 3FGL-unmasked analysis, since, when when masking 3FGL sources, the fraction of the sky masked near the GC is different in the North versus the South.  

\begin{figure}[tb]
	\leavevmode
	\begin{center}$
	\begin{array}{cc}
	\scalebox{0.51}{\includegraphics{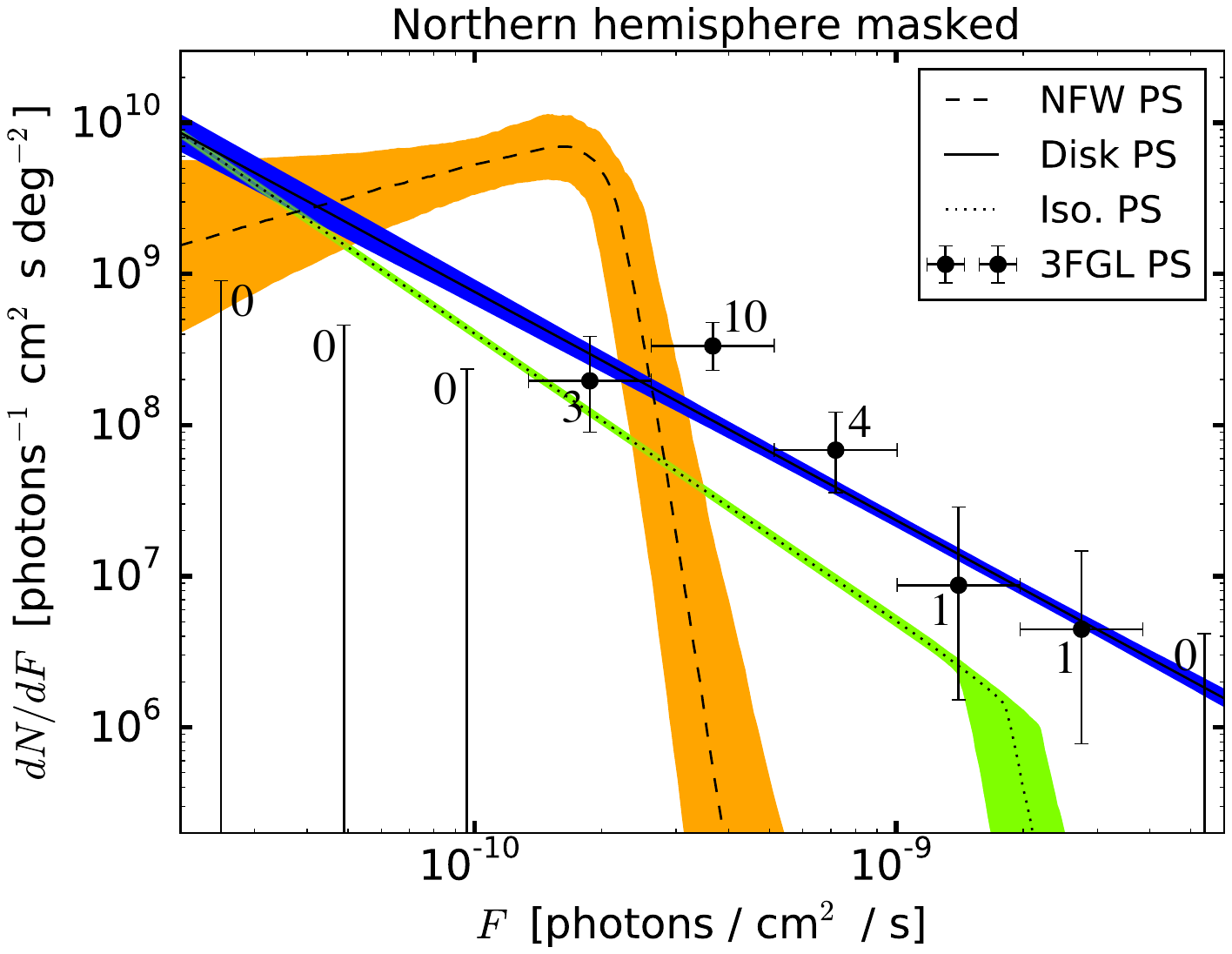}} & \scalebox{0.40}{\includegraphics{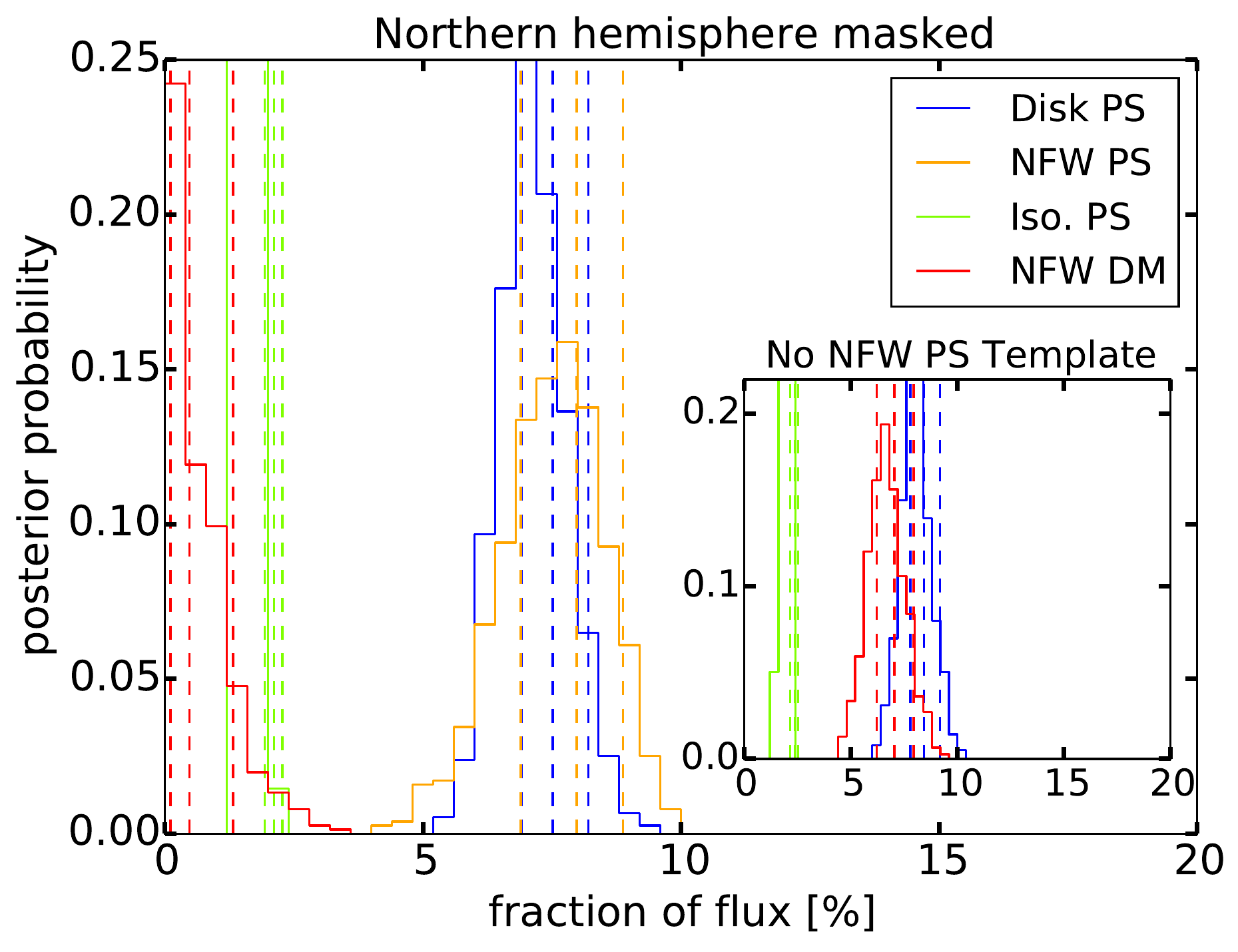}}  \\
		\scalebox{0.51}{\includegraphics{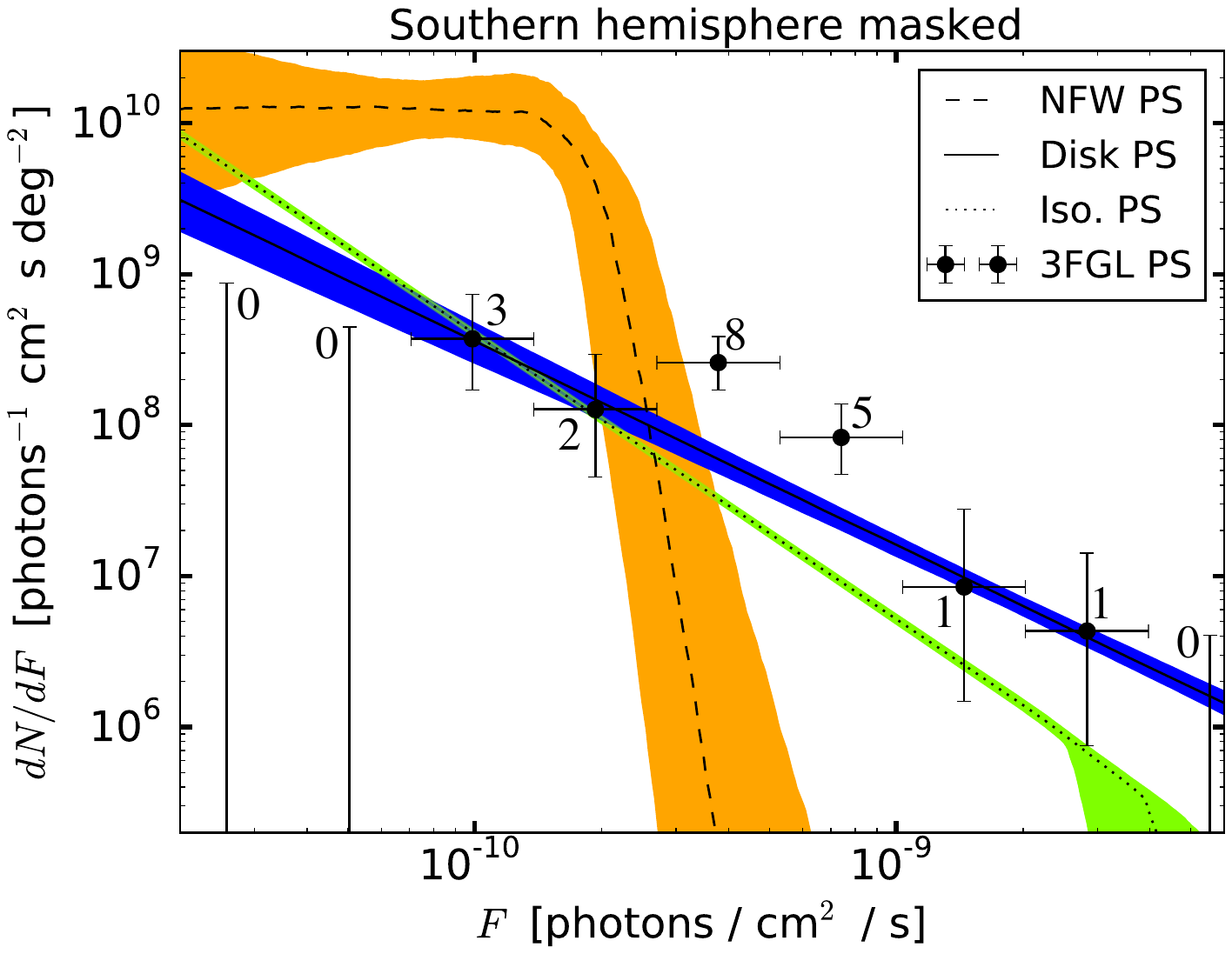}} & \scalebox{0.40}{\includegraphics{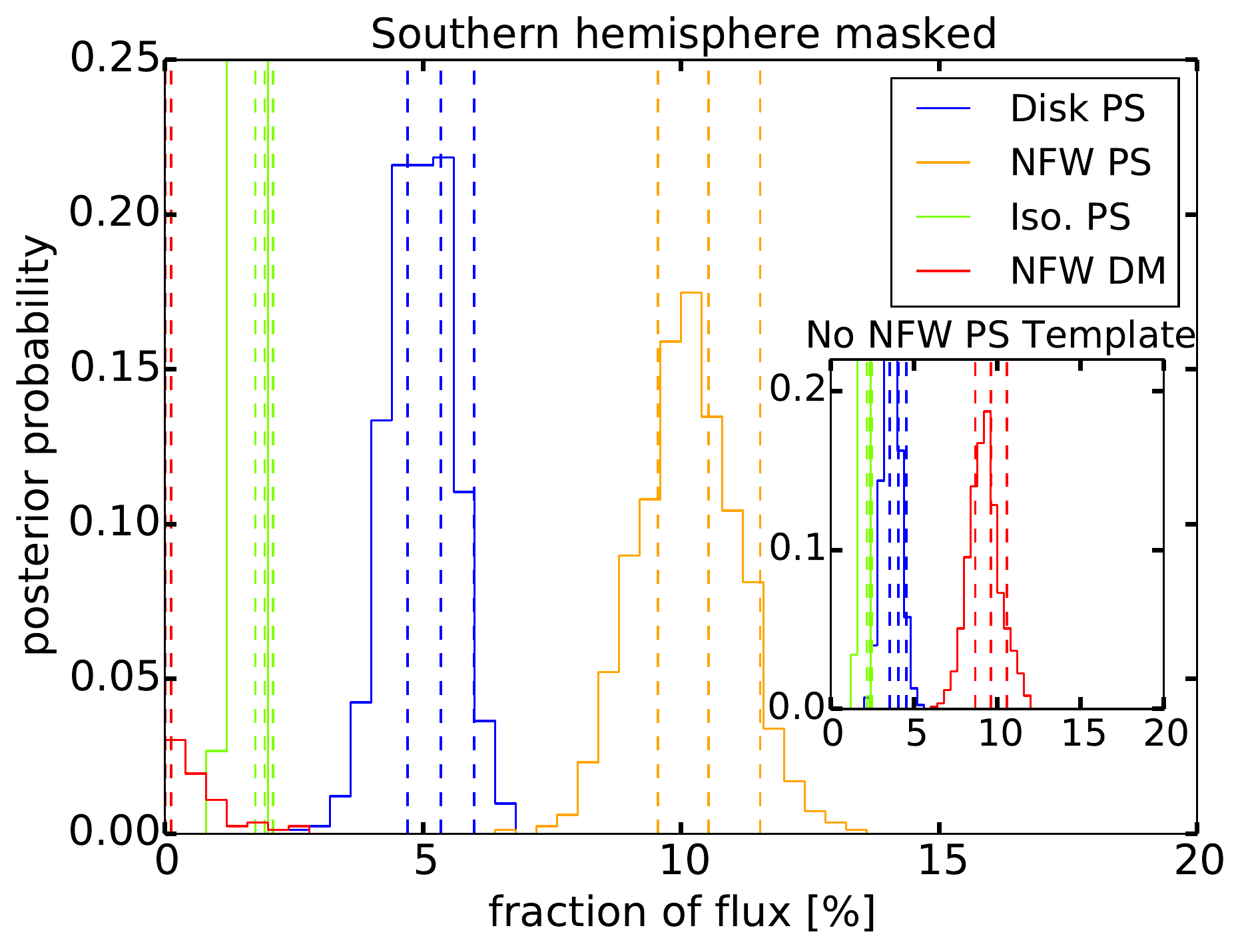}} 
	\end{array}$
	\end{center}
	\vspace{-.50cm}
	\caption{Results obtained by applying the NPTF to a full-sky map with either the Northern (top row) or Southern (bottom row) hemispheres masked.  The templates included are: isotropic, diffuse, bubbles, NFW DM, NFW PSs, disk PSs, and isotropic PSs.  The best-fit source-count functions (with 68\% confidence intervals shaded) are shown in the left column, and the posterior probabilities for the flux fractions are shown in the right column.  The source-count functions are plotted with respect to the region within 10$^\circ$ of the GC with $b \geq 2^\circ$ ($b \leq -2^\circ$) for the Northern (Southern) analysis.  The number of 3FGL sources in these regions is indicated.  
The flux-fraction plots are calculated for the region within 10$^\circ$ of the GC with $|b| \geq 2^\circ$ in both cases.  
The inset shows the posterior probabilities for the flux fractions when the NFW PS template is not included.}      
	\vspace{-0.15in}
	\label{Fig: hemispheres}
\end{figure}

Fig.~\ref{Fig: hemispheres} shows the best-fit source-count function and flux fractions for the Northern and Southern hemispheres in the top and bottom rows, respectively.  The source-count functions for the two regions are consistent within statistical uncertainties.  In both cases, the orange band cuts off steeply around \mbox{$F\sim2\times10^{-10}$} photons/cm$^{2}$/s.
The inferred flux fraction of NFW PSs in the Northern (Southern) hemisphere analysis is $7.98_{-1.08}^{+0.90}$\% ($10.54_{-0.98}^{+1.00}$\%) in the region within 10$^\circ$ of the GC, with $\abs{b} \geq 2^\circ$, while    
 the Bayes factor in favor of the model with NFW, disk, and isotropic PSs over the model without NFW PSs is $\sim$$10^3$ ($\sim$$10^4$) in the Northern (Southern) hemisphere analysis.  
We conclude that there are no significant asymmetries in the inferred 
NFW PS population between the Northern and Southern hemispheres.

\subsection{Varying the Diffuse Model}
\label{sec:varydiffuse}

\begin{figure}[t]
	\leavevmode
	\begin{center}$
	\begin{array}{cc}
	\scalebox{0.40}{\includegraphics{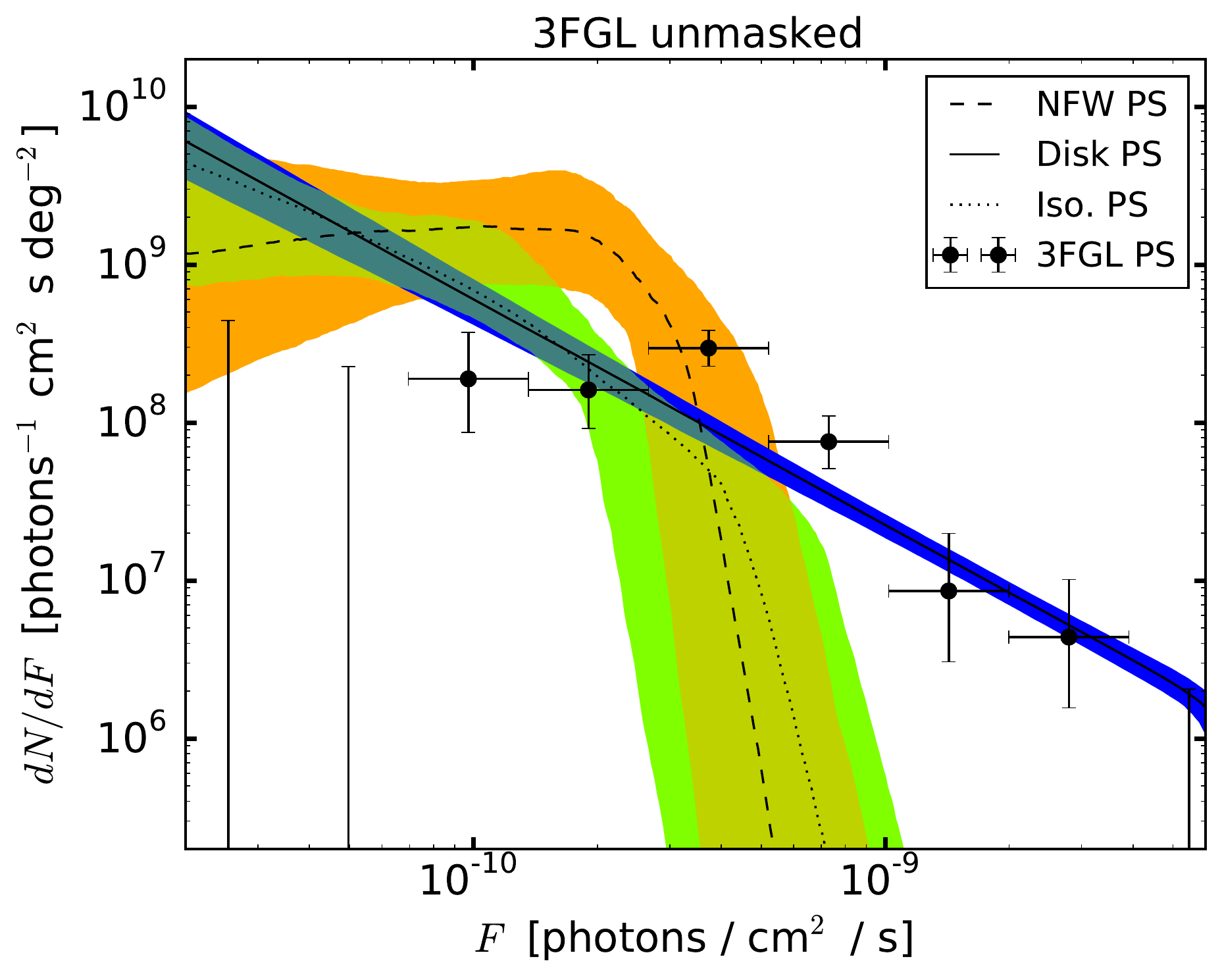}} & \scalebox{0.40}{\includegraphics{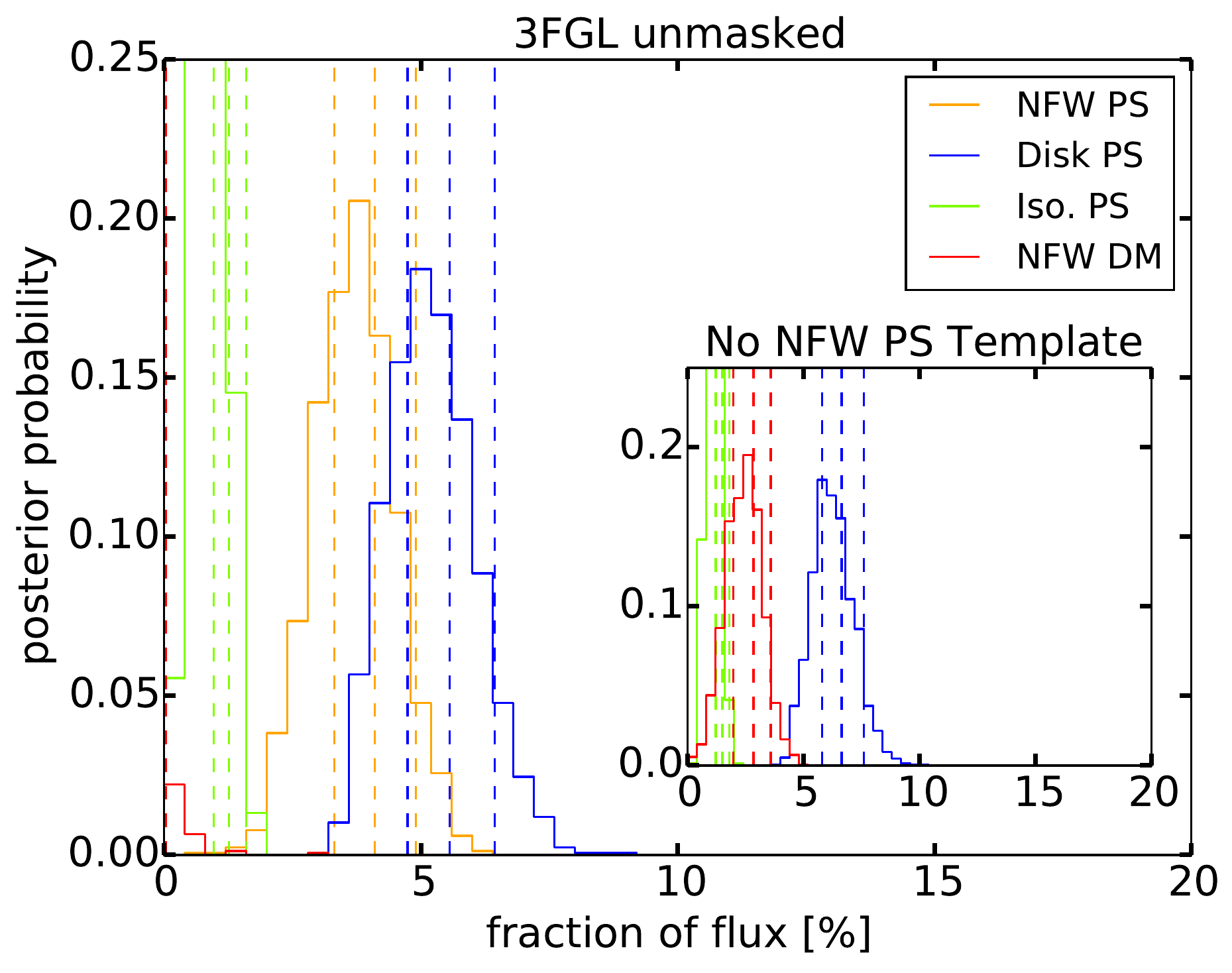}}
		\end{array}$
	\end{center}
	\vspace{-.50cm}
	\caption{Same as Fig.~\ref{Fig: IG_dnds_unmasked}, except using the \emph{Fermi} \texttt{p7v6} diffuse background model.}
	\vspace{-0.15in}
	\label{Fig: p7dnds_unmasked}
\end{figure}

A potentially serious source of systematic uncertainty is due to  limitations in modeling the diffuse gamma-ray background arising from the propagation of high-energy cosmic rays in the Galaxy.  The primary contributions come from bremsstrahlung of high-energy cosmic rays passing through the interstellar gas, inverse Compton scattering of Cosmic Microwave Background, interstellar, and infrared radiation off high-energy electrons, and the decay of boosted pions produced in inelastic proton collisions with the interstellar gas.  Modeling the cosmic-ray emission depends on many factors, including the location and spectrum of source injection, the gas distribution, magnetic fields, and the interstellar radiation field, as well as the diffusion parameters.  Repeating the NPTF with different diffuse-background models can help to quantify the effects of mis-modeling the diffuse background.

The primary results presented in this work used the \emph{Fermi} \texttt{p6v11} diffuse model.  This model is not the most recent to be released by \emph{Fermi}.  The more recent \emph{Fermi} \texttt{p7v6} diffuse model includes contributions from large-scale diffuse substructures such as Loop 1 and the bubbles.  As found in~\cite{1402.6703}, repeating the template analysis with the \texttt{p7v6} model does not qualitatively affect the results for the GeV excess, except for the fact that the overall flux absorbed by the DM template is reduced.  This may be due to the largely data-driven \texttt{p7v6} diffuse model having absorbed part of the GeV excess.
\begin{figure}[b]
	\leavevmode
	\begin{center}$
	\begin{array}{cc}
	\scalebox{0.40}{\includegraphics{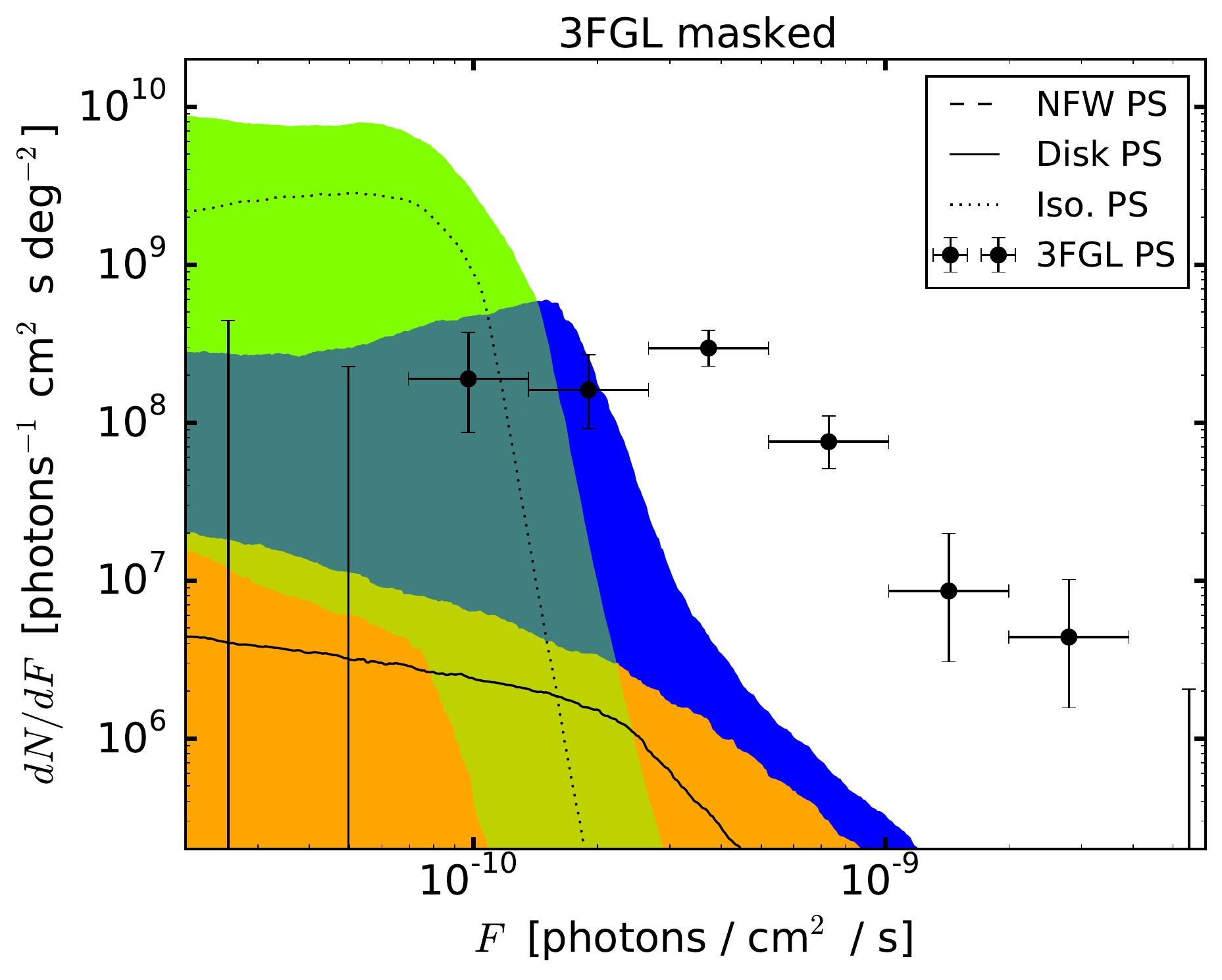}} & \scalebox{0.40}{\includegraphics{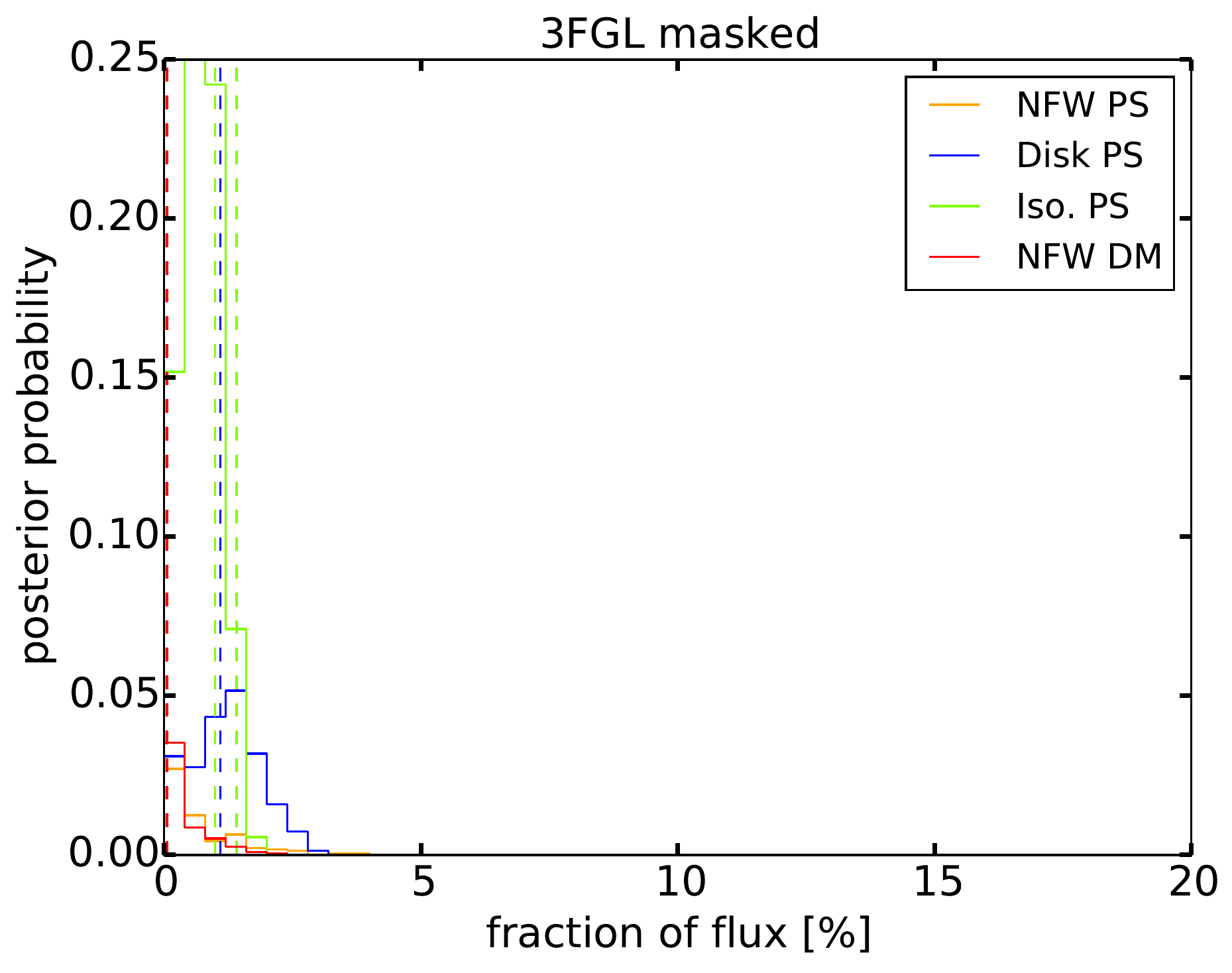}} 
	\end{array}$
	\end{center}
	\vspace{-.50cm}
	\caption{Same as Fig.~\ref{Fig: IG_dnds_masked}, except  using the \emph{Fermi} \texttt{p7v6} diffuse background model.}     
	\vspace{-0.15in}
	\label{Fig: p7dnds_masked}
\end{figure}

\begin{figure}[t]
	\leavevmode
	\begin{center}$
	\begin{array}{cc}
	\scalebox{0.40}{\includegraphics{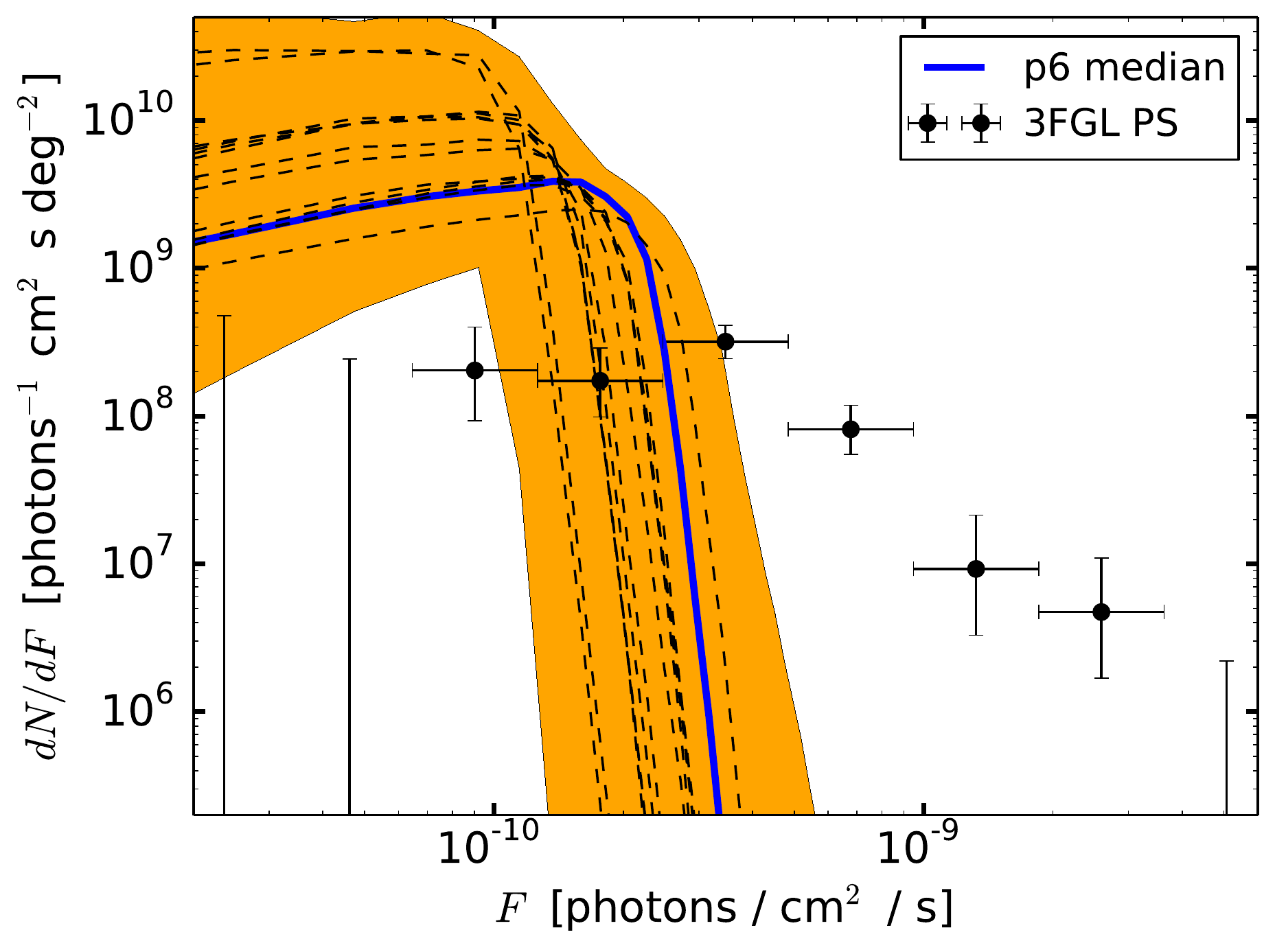}}
	\end{array}$
	\end{center}
	\vspace{-.50cm}
	\caption{Same as Fig.~\ref{Fig: IG_dnds_masked}, except showing the median source-count functions for the additional thirteen diffuse models described in the text in dashed black; the median function for the \texttt{p6v11} model is shown in solid blue, for comparison.  The orange band is the overlap of the 68\% confidence bands for all thirteen models.  For computational reasons, the only PS template included in these tests is that for NFW PSs.}
	\vspace{-0.15in}
	\label{Fig: p7sourcecount}
\end{figure}

The left panel of Fig.~\ref{Fig: p7dnds_unmasked} shows the best-fit source-count functions obtained for the IG study  using the \texttt{p7v6} diffuse model, with all other templates the same.  Comparing to Fig.~\ref{Fig: IG_dnds_unmasked}, we see that the overall features are recovered. The right panel of Fig.~\ref{Fig: p7dnds_unmasked} shows the corresponding flux fraction plot.  In this case, the flux fraction absorbed by the NFW (disk) PS template is $\sim$$4\%$ $(6\%)$; when the NFW PS template is removed from the fit (inset, right panel), the majority of its corresponding flux is absorbed by the NFW DM template instead.  
The main difference with the \texttt{p6v11} results is that the flux associated to NFW PSs is lower with \texttt{p7v6}, which is consistent with previous findings of a smaller flux fraction for the GeV excess in \texttt{p7v6} studies.  As a result, the Bayes factor in preference for the NFW PS template is reduced to $\sim$$10^2$ in this case.\footnote{This Bayes factor increases to $\sim$$10^3$ with Pass 8 data.}

Figure~\ref{Fig: p7dnds_masked} shows the corresponding results for \texttt{p7v6} when the 3FGL sources are masked.  Now, the NPTF finds no excess flux in the ROI; the flux fractions for the PS templates, as well as the NFW DM template, are all consistent with 0\%.  The fact that no flux is picked up by the NFW DM template is different from what was previously observed in~\cite{1402.6703}.  We have verified that this is due to the larger PS mask implemented here, which removes a considerable region close to the Galactic Center; repeating the analysis with the PS masking of~\cite{1402.6703} recovers their result.

 In addition to studying the \texttt{p7v6} diffuse model, we also consider thirteen other diffuse models from~\cite{1202.4039}: $^\text{S}$L$^\text{Z}$6$^\text{R}$20$^\text{T}$100000$^\text{C}$5, $^\text{S}$L$^\text{Z}$6$^\text{R}$20$^\text{T}$150$^\text{C}$2, $^\text{S}$L$^\text{Z}$6$^\text{R}$20$^\text{T}$150$^\text{C}$5, $^\text{S}$L$^\text{Z}$6$^\text{R}$20$^\text{T}$100000$^\text{C}$2, $^\text{S}$L$^\text{Z}$10$^\text{R}$20$^\text{T}$150$^\text{C}$5, $^\text{S}$O$^\text{Z}$10$^\text{R}$20$^\text{T}$150$^\text{C}$5, $^\text{S}$Y$^\text{Z}$10$^\text{R}$20$^\text{T}$150$^\text{C}$5, $^\text{S}$S$^\text{Z}$10$^\text{R}$20$^\text{T}$150$^\text{C}$5, $^\text{S}$S$^\text{Z}$8$^\text{R}$20$^\text{T}$150$^\text{C}$5, $^\text{S}$S$^\text{Z}$8$^\text{R}$20$^\text{T}$150$^\text{C}$5, $^\text{S}$S$^\text{Z}$4$^\text{R}$20$^\text{T}$150$^\text{C}$5, $^\text{S}$O$^\text{Z}$10$^\text{R}$30$^\text{T}$150$^\text{C}$5, $^\text{S}$L$^\text{Z}$10$^\text{R}$30$^\text{T}$150$^\text{C}$5.  These models are chosen to span variations in the source distributions, the diffusion scale height and radius, the gas spin temperature, and cuts on the magnitude of E(B-V).   For these models, we include separate templates for the $\pi^0$, bremsstrahlung, and ICS components.  The prior ranges for each of these three templates is dealt with in the same way as the diffuse-model prior ranges for the \texttt{p6v11} and \texttt{p7v6} diffuse models; in all cases, the three component templates are well-converged within the prior ranges. 
 For brevity, we only show results for the 3FGL-masked analyses.    
 Additionally, for computational reasons, we only include a non-Poissonian template for NFW PSs. When only this PS template is included, the Bayes factor in preference for the model with NFW PSs relative to that without is $\sim$$10^7$ ($\sim$$10^3$) for the \texttt{p6v11} (\texttt{p7v6}) diffuse model. 
  
Fig.~\ref{Fig: p7sourcecount} summarizes the results, showing the spread in the median source-count functions (dashed black lines) obtained by running the NPTF in the IG with each of these diffuse models.  The orange band indicates the combination of the thirteen 68\% confidence intervals.  While there is some spread in the predicted source-count function, the NPTF consistently finds a non-zero flux for the NFW PS contribution in all cases, ranging from $\sim$3\% to $\sim$9\%; also, the DM flux is always consistent with zero when the NFW PS template is included in the fit.  Specifically, the model including an NFW PS template is preferred over the model without such a template by Bayes factors in the range $\sim$$10^6$--$10^9$ for all diffuse emission scenarios considered.  In this sense, these thirteen diffuse models appear more similar to \texttt{p6v11} than to \texttt{p7v6}.
  
For the model $^\text{S}$L$^\text{Z}$6$^\text{R}$20$^\text{T}$100000$^\text{C}$5, which provided the formal best fit in a previous analysis of the IG \cite{1409.0042}, we also test the effect of adding an independent diffuse template with free normalization, corresponding to the predicted gas-correlated emission (pion decay and bremsstrahlung) within the innermost Galactocentric ring. The addition of this template does not significantly alter the source-count function or flux fraction attributed to the NFW PSs.

\subsection{Scan Along the Galactic Plane}
\label{sec:planescan}

\begin{figure}[b]
	\leavevmode
	\begin{center}$
	\begin{array}{cc}
	\scalebox{0.515}{\includegraphics{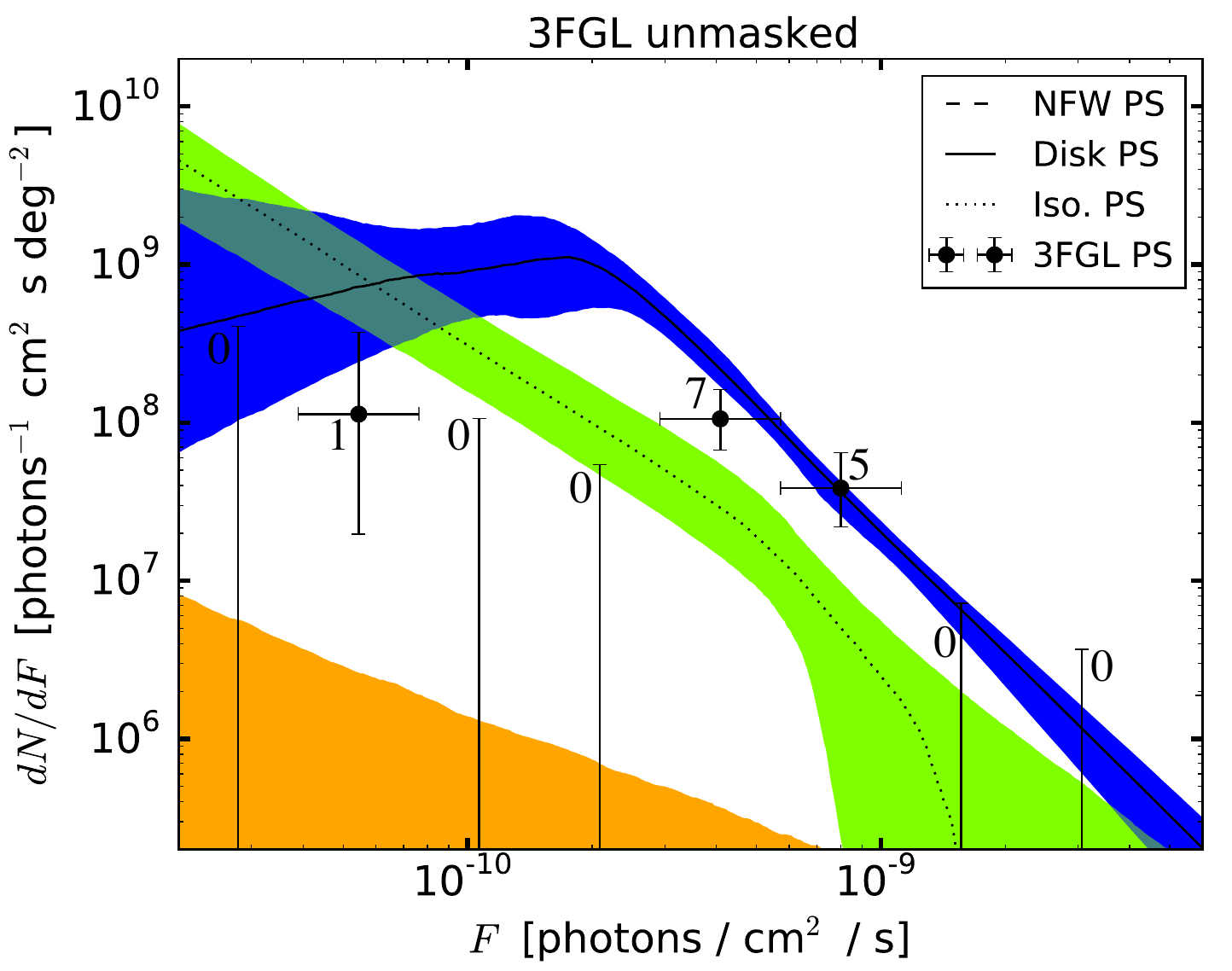}} & 	\scalebox{0.4}{\includegraphics{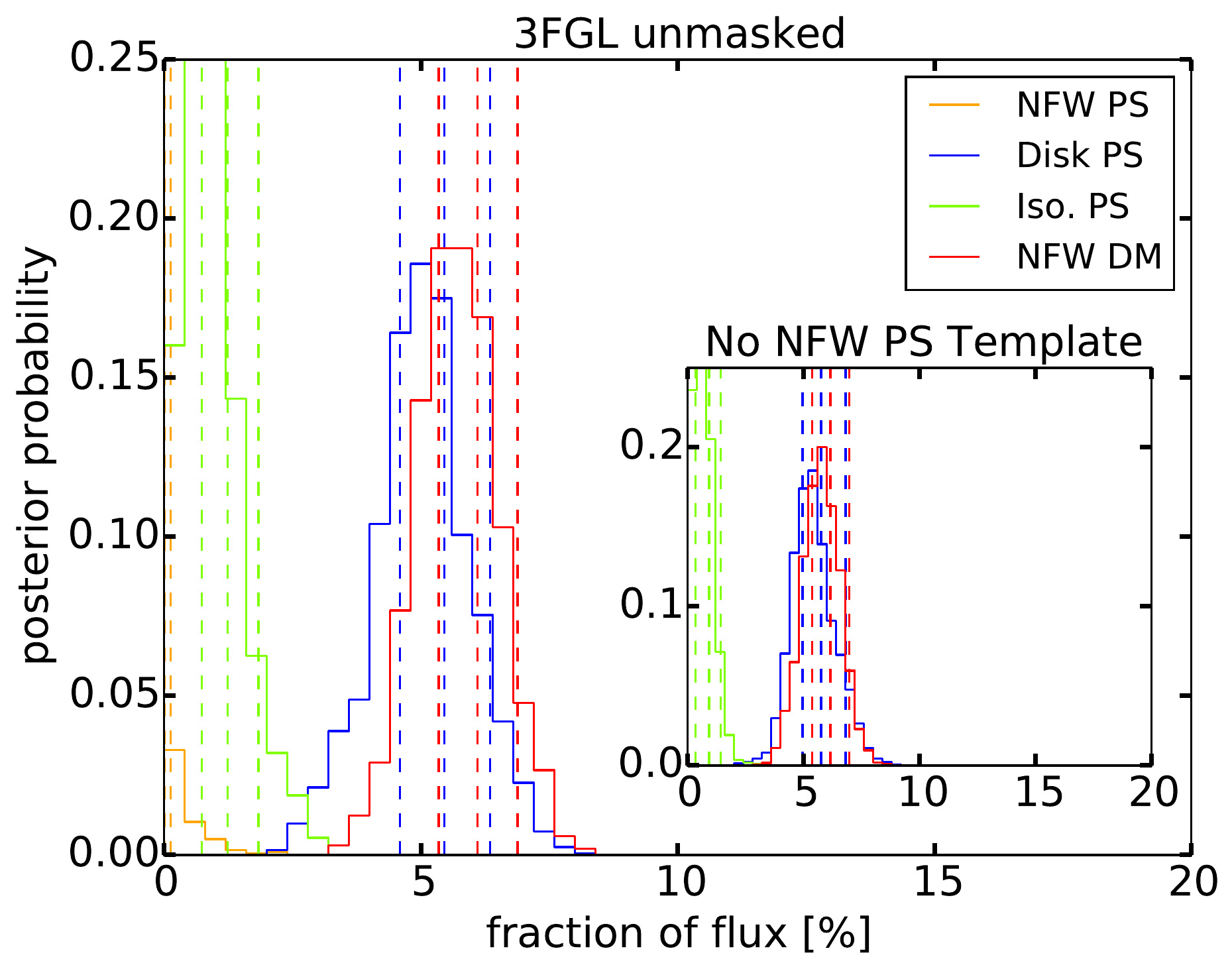}} \\
	\end{array}$
	\end{center}
	\vspace{-.50cm}
	\caption{Same as Fig.~\ref{Fig: IG_dnds_unmasked}, except repeating the NPTF in the region within $30^\circ$ of $(l,b) = (30^\circ, 0^\circ)$, with $|b| \geq 2^\circ$.}   
	\vspace{-0.15in}
	\label{Fig: lon30sourcecount}
\end{figure}

To address the concern that the PS templates may be erroneously absorbing contributions from the diffuse background, we repeat the NPTF in regions centered at different points along the Galactic plane.  Previous studies have reported additional bright excesses along the plane, with the most significant at $l=30^\circ$~\cite{1402.6703,1409.0042}.  The residual emission in this region is roughly similar to that at the GC; however, its energy spectrum is softer, suggesting a different origin.  Repeating the NPTF on another region of sky---with an excess that is unlikely to arise from PSs---allows us to test whether the fitting procedure can adequately distinguish between extra diffuse emission on top of a mis-modeled diffuse background and extra PS emission.

We apply the NPTF to regions of the sky within $30^\circ$ of the central points $(l,b) = (30n^\circ, 0^\circ)$, where $n = -3, -2, \ldots, 3$, requiring $|b| \geq 2^\circ$ throughout.  The same templates are included as in the IG analysis, except the NFW templates are centered at the middle of each ROI.  The most significant excess is found for the region centered at $(l,b) = (30^\circ, 0^\circ)$, consistent with previous findings.  Fig.~\ref{Fig: lon30sourcecount} shows the best-fit source-count functions for the different PS templates in the 10$^\circ$ region centered at this point. The disk template (solid, blue) successfully recovers the known PSs in this region of sky (numbering $\sim$13).  Most notably, the error band on the source-count function for the NFW PS population (dashed, orange) is consistent with \emph{no} unresolved PSs in this ROI.

The right panel of Fig.~\ref{Fig: lon30sourcecount} shows that the flux fractions of the NFW DM and disk PS components are comparable in this region, while the isotropic and NFW PS templates pick up negligible flux.  Excluding the NFW PS template (inset) does not significantly affect the DM and disk PS flux fractions.  The Bayes factor in favor of the model with NFW PSs relative to that without is $\sim$0.1.  We note that repeating the analysis with the exact masking criteria and NFW inner-slope value used by~\cite{1402.6703}, we reproduce the (slightly smaller) flux fraction of the excess found by those authors.  

\subsection{Point Spread Function}
\label{Sec: PSF}

\begin{figure}[b]
	\leavevmode
	\begin{center}$
	\begin{array}{cc}
	\scalebox{0.40}{\includegraphics{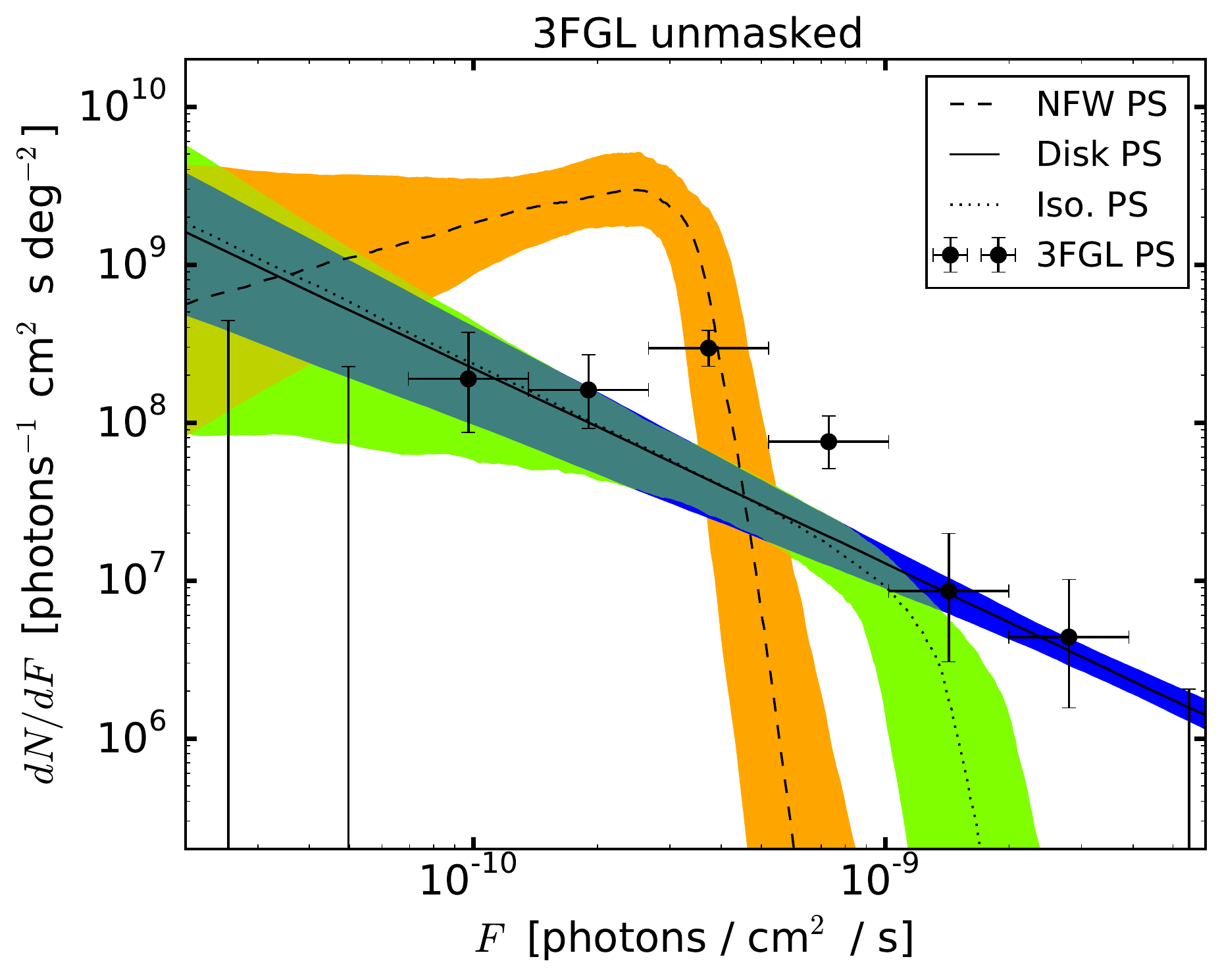}} & \scalebox{0.40}{\includegraphics{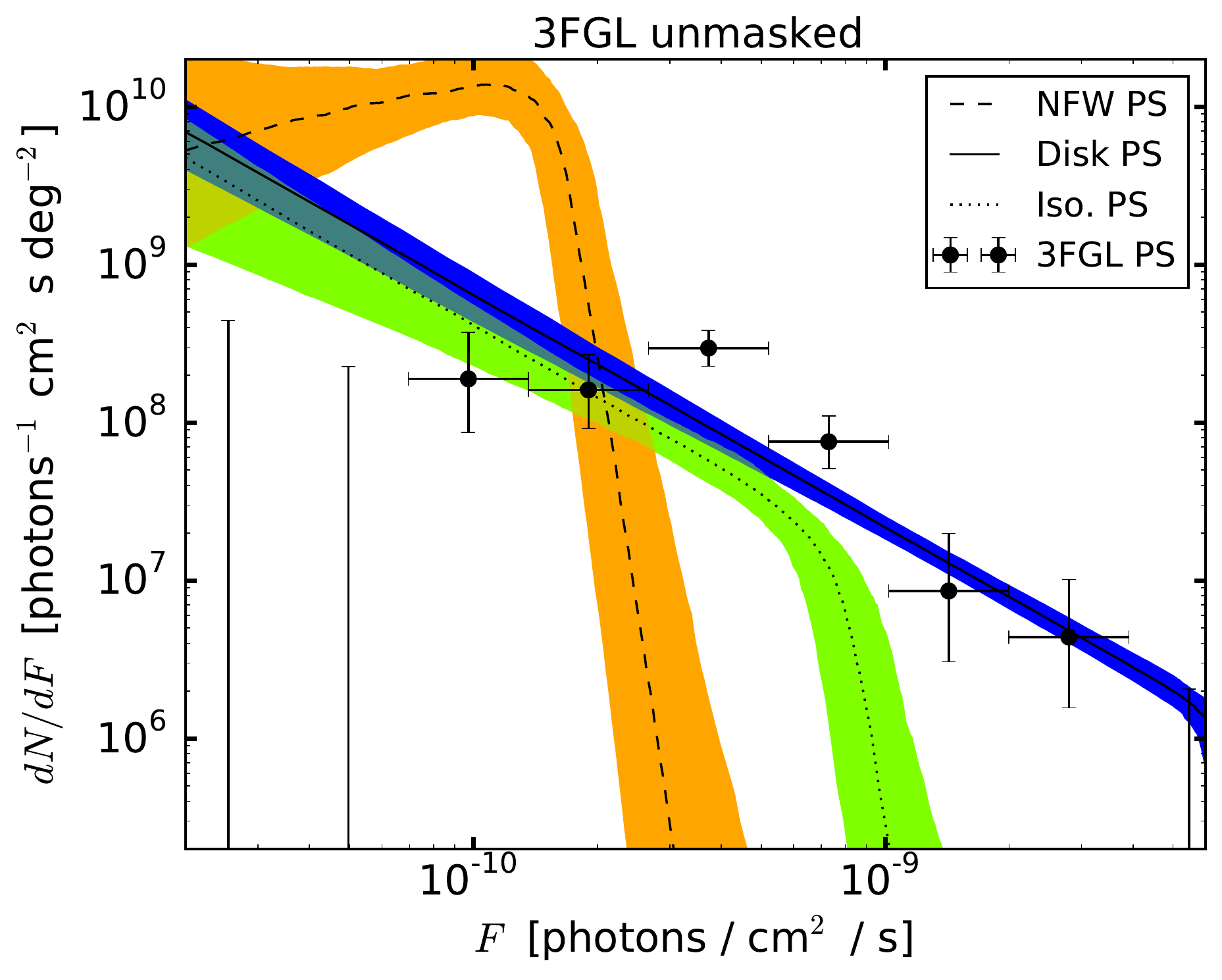}} 
	\end{array}$
	\end{center}
	\vspace{-.50cm}
	\caption{Same as the left panel of Fig.~\ref{Fig: IG_dnds_unmasked}, except setting the PSF parameter $\sigma=0.198^\circ$ (left) and $\sigma = 0.0492^\circ$ (right).  When masking identified 3FGL sources, all pixels within $5\times0.198^\circ$ are excluded.}      
	\vspace{-0.15in}
	\label{Fig: PSFhighlow}
\end{figure}

An accurate PSF is a crucial component of the NPTF procedure, as an incorrect parametrization can lead to over- or under-estimation of the PS component.  The instrument response functions for the \emph{Fermi} CTBCORE-cut data were made available by~\cite{1406.0507}.  
For a given energy range, the authors of~\cite{1406.0507} provide the 68\% and 95\% containment radii, $R_{68}$ and $R_{95}$.  

We model the PSF as a two-dimensional Gaussian 
\es{eq: PSF}{
P(x, \sigma) = \frac{1}{2 \pi \sigma^2} \exp\left[ \frac{-x^2}{2 \sigma^2} \right] \, ,
}
setting $\sigma$ such that the 68\% containment radius of~\eqref{eq: PSF} is equal to $R_{68}$.  The value of $\sigma$ varies from $\sim$0.198$^\circ$ at the lowest end to $\sim$0.0492$^\circ$ at the highest end of the energy bin (1.893-11.943~GeV).  As described earlier, the isotropic, \emph{Fermi} bubbles, and NFW DM templates are smeared with this PSF.  (The diffuse-background template is smeared with the exact energy-dependent PSF using the \emph{Fermi} Science Tools.)  The PSF must also be properly accounted for in the generating-function formalism that is used to obtain the non-Poissonian likelihood function (see~\cite{Lee:2014mza} for further details).  
  
It is known that the real PSF has power-law tails that are not captured by~\eqref{eq: PSF}~\cite{1406.0507}.
One might rightfully be concerned that ignoring these power-law tails can lead to mischaracterization of the PS population.  To illustrate the effect of mis-modeling the PSF, Fig.~\ref{Fig: PSFhighlow} shows the best-fit source-count function for the IG analysis assuming extreme values for the Gaussian-PSF parameter $\sigma$: $\sigma=0.198^\circ~(0.0492^\circ)$ in the left (right) panel.  When the assumed PSF is too wide, photons in neighboring pixels arising from diffuse emission or unresolved sources may instead be erroneously interpreted as photons from a single PS.  Thus, using a PSF that is wider than the true PSF will underestimate the number of unresolved PSs and overestimate the number of observed sources, relative to the true source-count distribution.  This effectively shifts the source-count function to large flux values, as observed in the left panel of Fig.~\ref{Fig: PSFhighlow}.

In comparison, using a PSF with a width that is too small underestimates the number of bright PSs.  For example, in the extreme limit, assuming a delta-function PSF would require all photons from a given PS to be contained in a single pixel, while the photons from bright PSs are, in actuality, smeared over several pixels by the true PSF.  
In this case, as may be seen in the right panel of Fig.~\ref{Fig: PSFhighlow},  the fitting procedure predicts fewer high-flux PSs, while at the same time predicting a larger population of unresolved sources.  
 
 The fact that the energy-averaged PSF used in Fig.~\ref{Fig: IG_dnds_unmasked} results in a source-count function that matches the high-flux observed PSs reasonably well gives us confidence in our treatment of the PSF.  Additionally, we have illustrated here that extreme variations to the PSF-parameter $\sigma$ do not significantly affect the results.  In Sec.~\ref{sec: sim}, we further verify our treatment of the PSF using simulated data. 

\subsection{Radial Distribution Profile}

\begin{figure}[b]
	\leavevmode
	\begin{center}$
	\begin{array}{cc}
	\scalebox{0.40}{\includegraphics{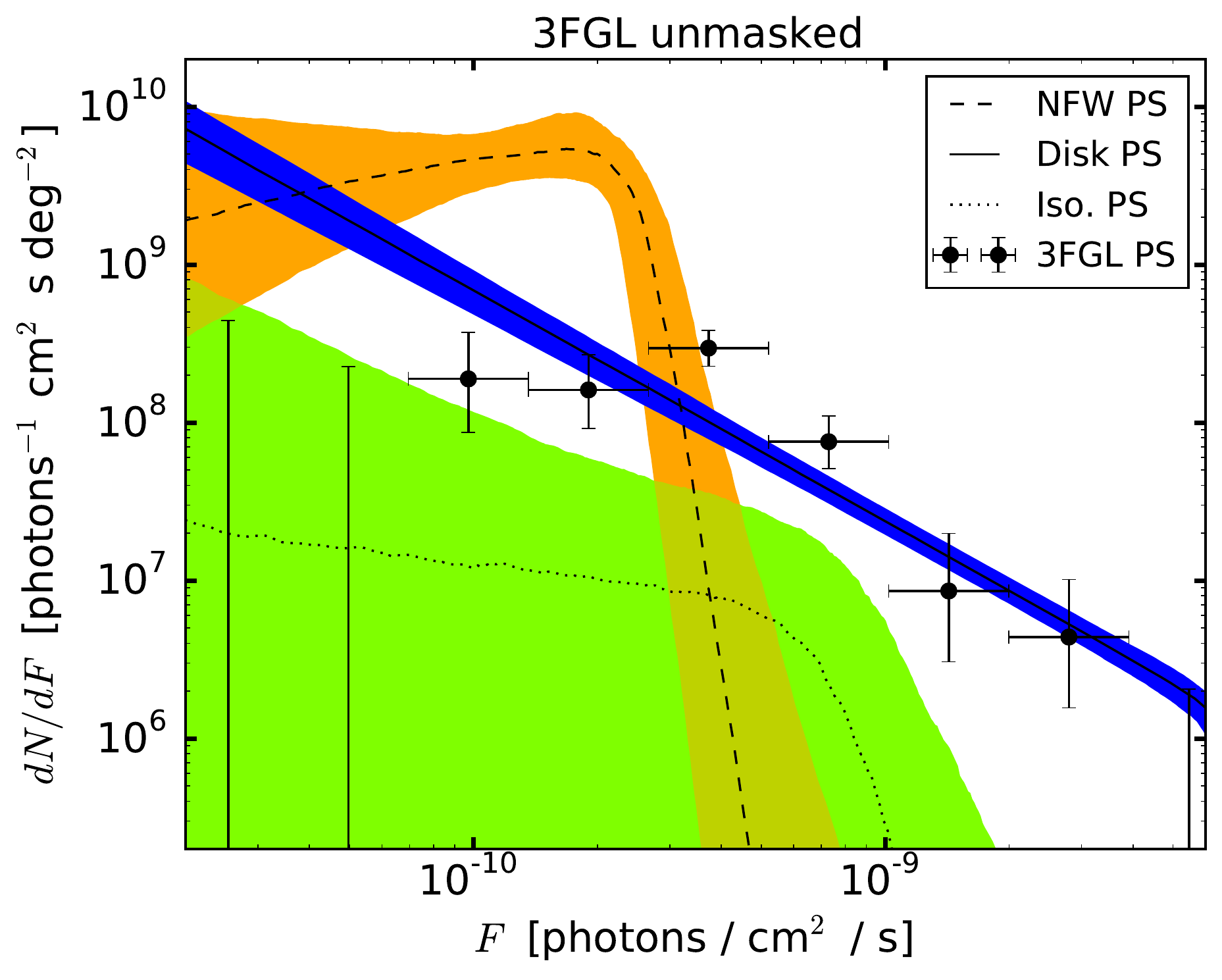}} & \scalebox{0.40}{\includegraphics{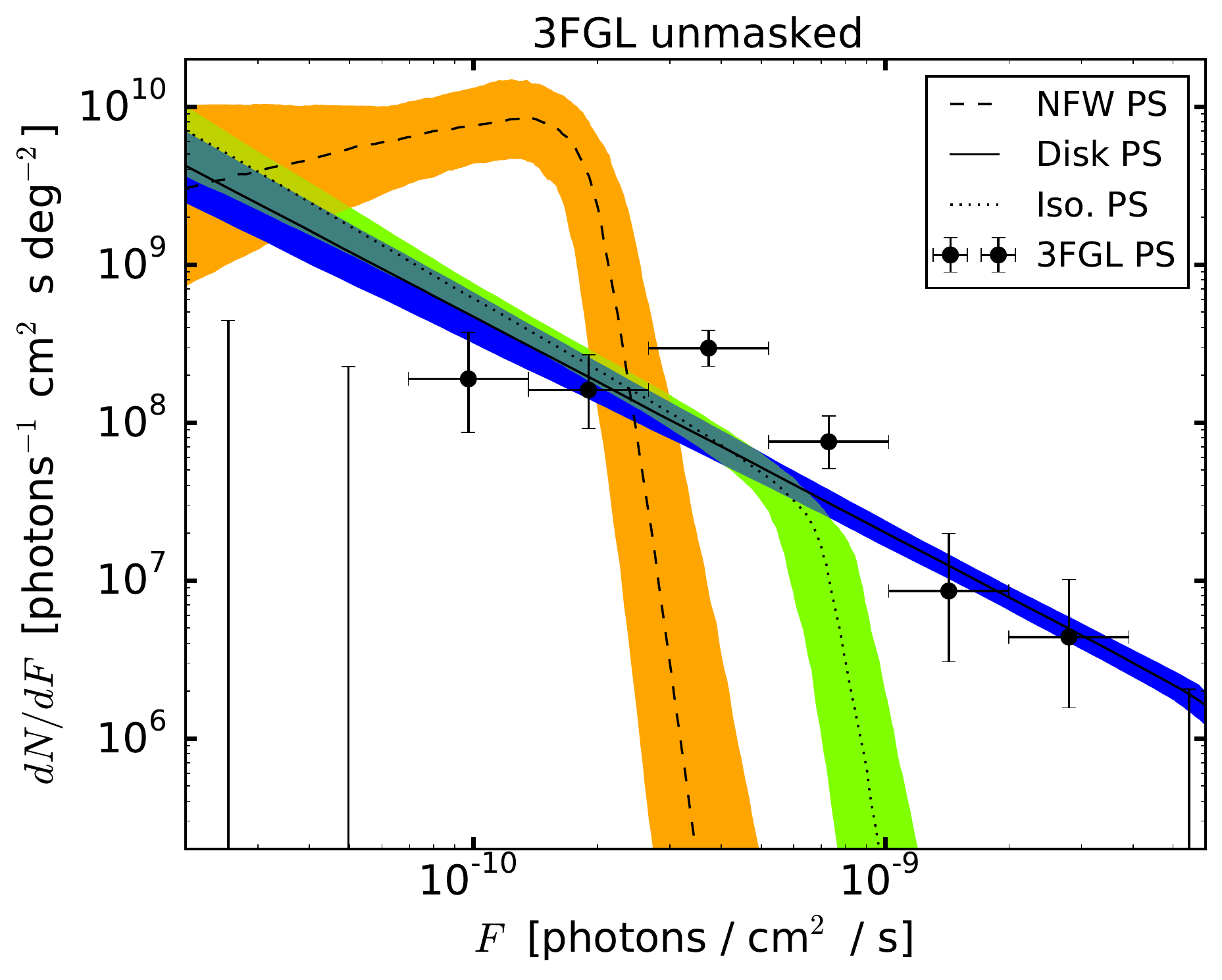}}
	\end{array}$
	\end{center}
	\vspace{-.50cm}
	\caption{Same as the left panel of Fig.~\ref{Fig: IG_dnds_unmasked}, except using $\gamma=1.1$ (left) and $\gamma=1.4$ (right) for the generalized NFW distribution. }     
	\vspace{-0.15in}
	\label{Fig: gammavariation}
\end{figure}

The analyses presented thus far have fixed the inner slope of the generalized NFW density profile to be $\gamma = 1.25$ for the DM and PS components.  Due to computational limitations, we do not scan over this parameter.  Previous studies of the GeV excess using standard template-fitting methods, such as~\cite{1402.6703, 1409.0042}, have scanned over this parameter and found best-fit values in the range between  $\gamma \approx 1.1$ and $1.4$.  Variations in $\gamma$ have negligible effects on our final result.  Fig.~\ref{Fig: gammavariation} shows the best-fit source-count function in the inner 10$^\circ$ with $|b|>2^\circ$, assuming $\gamma = 1.1$ and $1.4$ (left and right panels, respectively).  As is evident, the source-count function for NFW PSs is not sensitive to the value of the inner slope in this range.  The posterior probabilities for the flux fractions (not shown) are also not significantly changed from Fig.~\ref{Fig: IG_dnds_unmasked}.

\subsection{Diffuse-Correlated Point-Source Template}
\begin{figure}[t]
	\leavevmode
	\begin{center}$
	\begin{array}{c}
	\scalebox{0.40}{\includegraphics{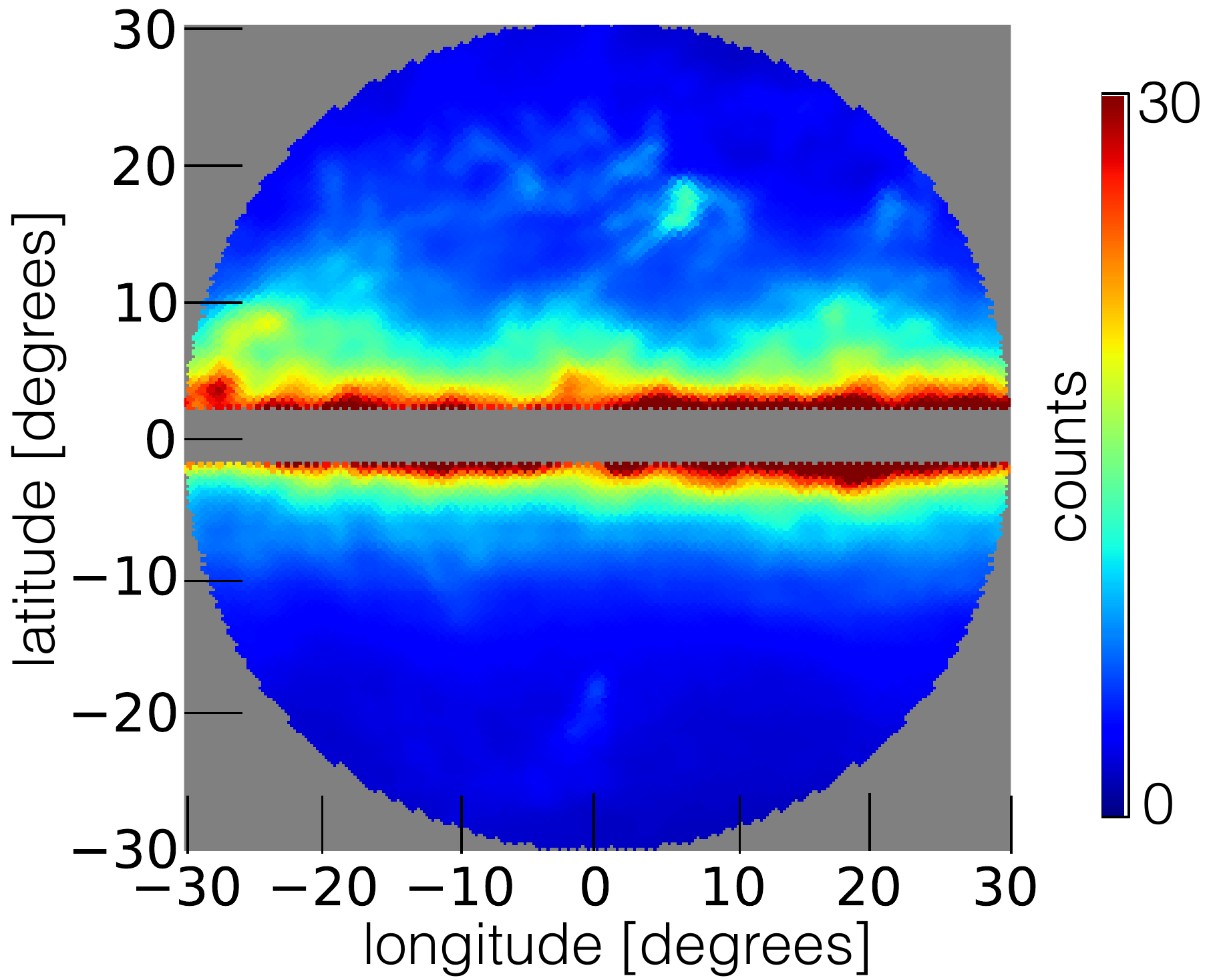}}
	\end{array}$
	\end{center}
	\vspace{-.50cm}
	\caption{The \emph{Fermi} \texttt{p6v11} diffuse background in the IG region (smoothed using  Fermi Science Tools), with $|b|<2^\circ$ masked.  Counts are clipped at 30.}
	\vspace{-0.15in}
	\label{Fig: diffuseBkgd}
\end{figure}

The conclusions described in this Letter provide evidence that the GeV excess can be accounted for by a population of unresolved PSs.  One possible systematic issue, however, is that the preference for unresolved PSs is driven by localized structure in the diffuse gamma-ray background that is not captured by our background model.  As a test of this hypothesis, we repeat the NPTF in the IG adding a PS template that traces the diffuse model.  Fig.~\ref{Fig: diffuseBkgd} shows the diffuse background in the ROI.  Because the flux from the diffuse emission is larger closer to the plane, a diffuse-correlated (diff-corr) PS template is not only sensitive to localized structure in the diffuse model, but also to the presence of a disk-like population of PSs.  
For example, if the unresolved PSs are preferentially located close to the plane, then they may be absorbed by the diff-corr PS template. 
 Breaking the degeneracy between these two interpretations requires more careful study, which we postpone to future work.  However, the preliminary results are illuminating, so we share them  here.

\begin{figure}[b]
	\leavevmode
	\begin{center}$
	\begin{array}{cc}
	\scalebox{0.40}{\includegraphics{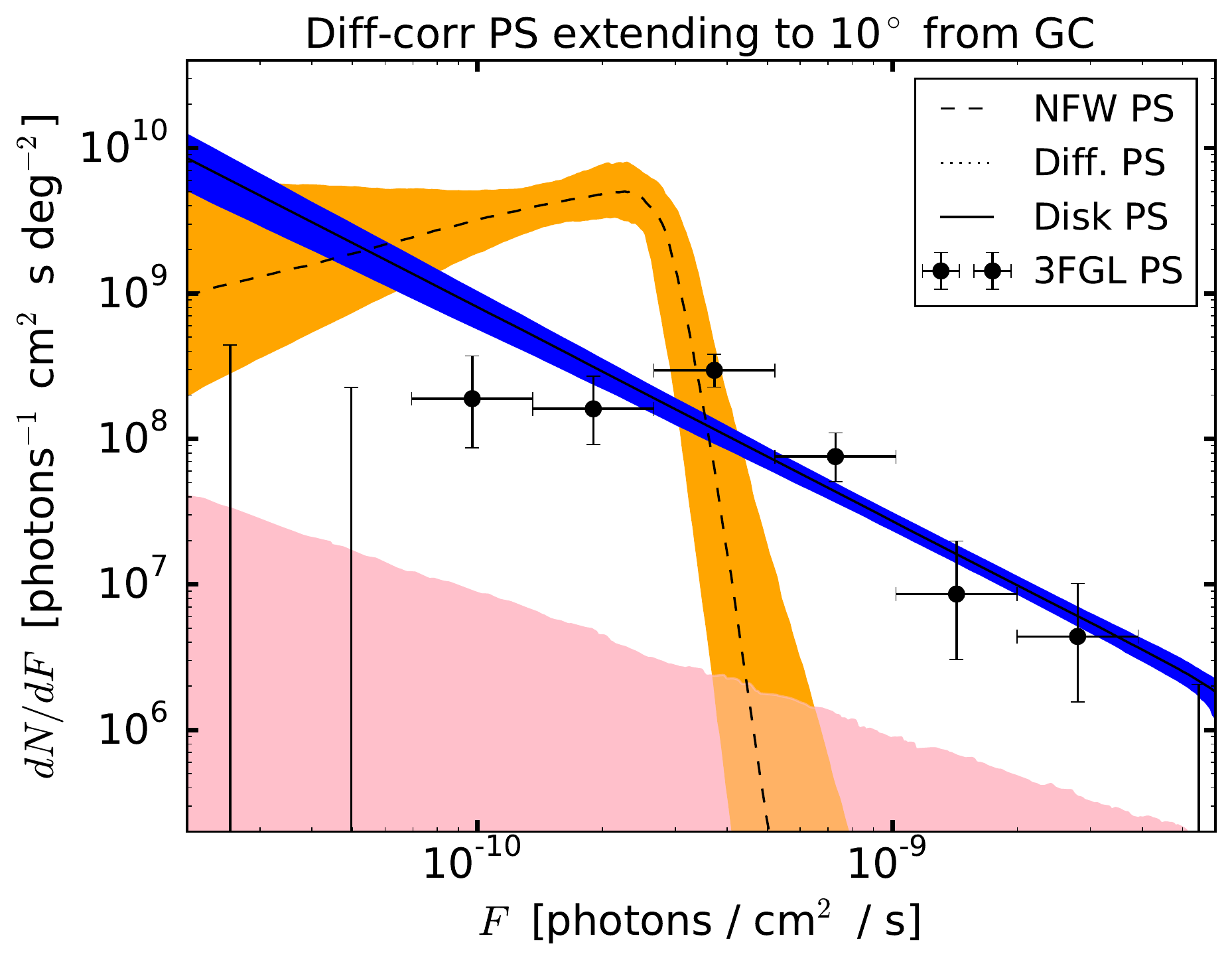}} & \scalebox{0.40}{\includegraphics{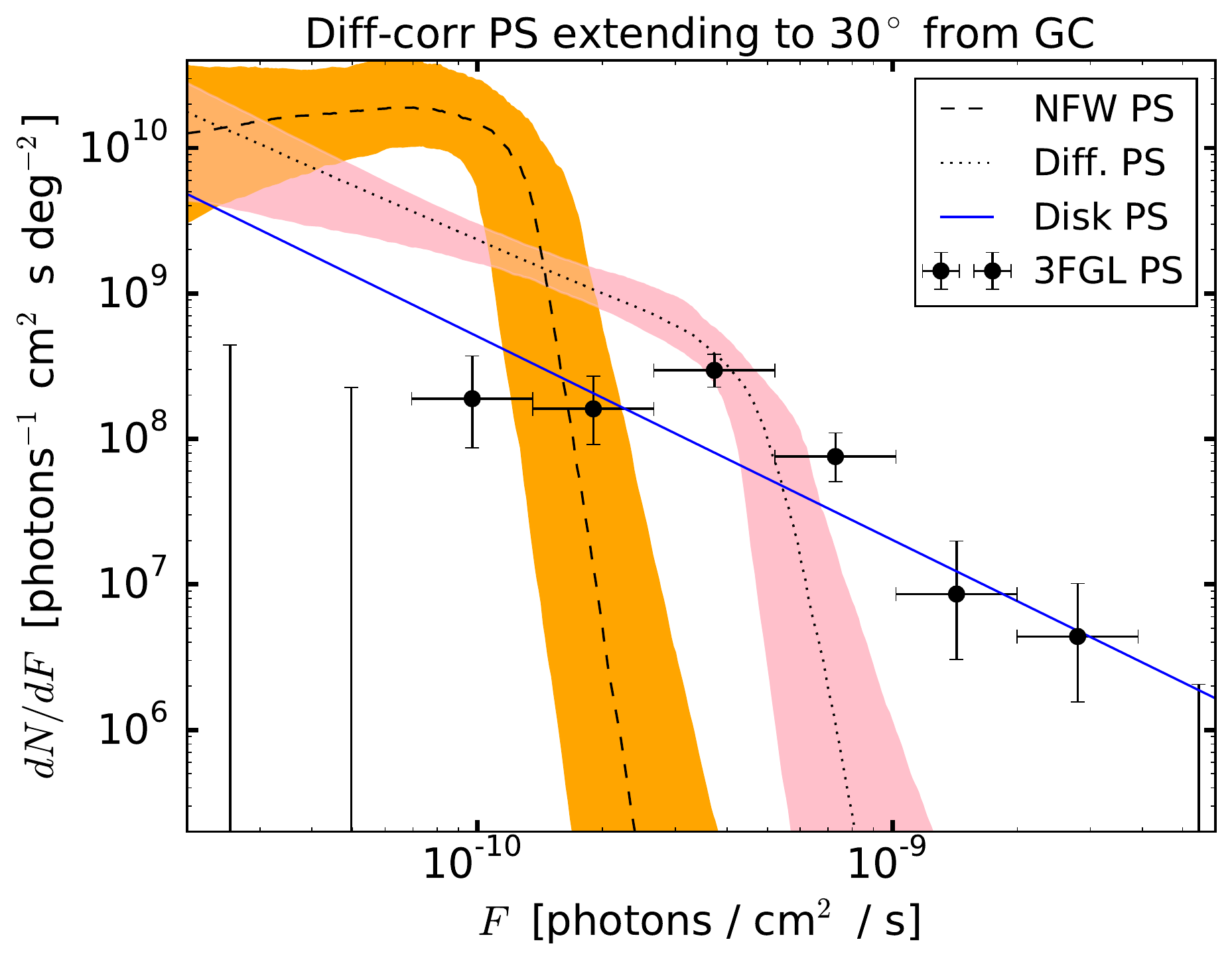}} 
	\end{array}$
	\end{center}
	\vspace{-.50cm}
	\caption{Best-fit source-count functions for PSs within $10^\circ$ of the GC and $|b| \geq 2^\circ$, with the 3FGL sources unmasked, for models with NFW PS (dashed, orange), diffuse-correlated PS (dotted, pink), and thin-disk PS (solid, blue) components.  For this analysis, the NPTF includes an additional template corresponding to diffuse-correlated PSs.  This new template is taken to have support either in the inner $10^\circ$ (left) or over the full ROI ($30^\circ$ from the GC with $|b| \geq 2^\circ$)  (right).  For the latter case, parameters for the thin-disk PS component are fixed to the best-fit values found in the standard unmasked IG analysis.
}
	\vspace{-0.15in}
	\label{Fig: diffusePSsourcecount}
\end{figure}

The left panel of Fig.~\ref{Fig: diffusePSsourcecount} shows the best-fit source-count functions obtained when doing the NPTF in the IG region (without masking the 3FGL sources), including a diff-corr PS template in addition to disk-correlated and NFW PS templates, along with the standard Poissonian templates.  
  Here, the diff-corr PS template only extends to $10^\circ$ from the GC.  
  To simplify the analysis, we do not include the subdominant isotropic PS template.  
  The best-fit flux normalization for the diffuse background is consistent with that obtained from the high-latitude analysis.  In addition, the best-fit NFW DM normalization is consistent with zero, and the recovered source-count function parameters for disk and NFW PSs are consistent with those found when not including the diff-corr PS template, since the diff-corr PS template does not absorb a significant fraction of the flux.

As a point of comparison, we repeat the procedure letting the diff-corr PS template have support over the full IG region.  Now, there is a potential degeneracy between the diff-corr PS template, the disk-correlated PS template, and the diffuse template.  To break some of this degeneracy, we fix the disk-correlated PS parameters to their best-fit values from the scan including disk-correlated, isotropic, and NFW PSs.  These results are shown in the right panel of Fig.~\ref{Fig: diffusePSsourcecount}.  Again, the best-fit NFW DM normalization is consistent with zero, however the source-count function for the NFW PSs changes.  
In particular, the source-count function for NFW PSs is shifted to lower flux, potentially suggesting that some of the near-threshold sources could either be more disk-like in morphology or associated with mis-modeling the diffuse background.
However, the preference for NFW PSs remains high, with the model including NFW PSs preferred over that without by a Bayes factor $\sim$$10^4$.  
Unlike the previous analysis that used a truncated diff-corr PS template, here the best-fit normalization of the diffuse-background template is lower than its best-fit value at high latitudes.  When the NFW PS template is not included, the normalization can be  shifted down by as much as 20\%;  the best-fit normalization of the diffuse template is still shifted down by $\sim$10\% when the NFW PS template is included.  

We caution the reader that these results are subject to considerable uncertainty due to the large number of parameters in the fitting procedure and the potential degeneracies between them. 
In addition, this analysis does not appear to be robust to changing the size of the ROI; as the ROI is increased, the best-fit normalization of the diffuse background approaches its value from high latitudes, and the flux absorbed by the diff-corr PS template decreases.

\subsection{Binned Source-Count Functions}
\label{sec:binned}

\begin{figure}[b]
	\leavevmode
	\begin{center}$
	\begin{array}{c}
	\scalebox{0.40}{\includegraphics{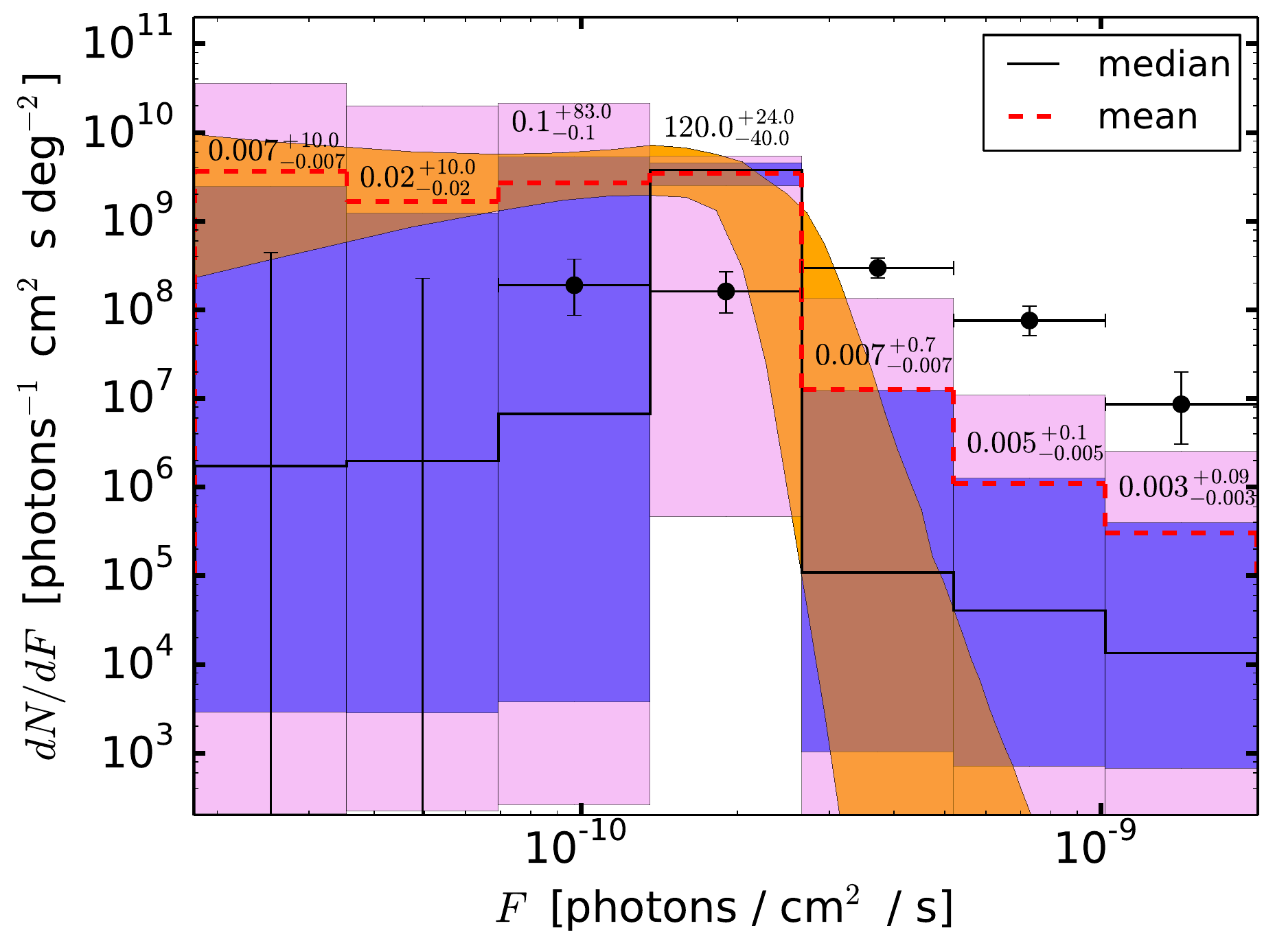}} 
	\end{array}$
	\end{center}
	\vspace{-.50cm}
	\caption{Best-fit source-count function for NFW PSs within $10^\circ$ of the GC and $|b| \geq 2^\circ$, with the 3FGL sources masked, where the number of sources in each flux bin is allowed to float freely. Blue (pink) regions indicate $68\%$ ($95\%$) confidence intervals.  The orange band is the 68\% confidence interval from the 3FGL-masked NPTF with a broken power-law source-count function. The median number of sources attributed to each bin is indicated, along with the 68\% confidence range.}
	\vspace{-0.15in}
	\label{fig:binned}
\end{figure}

The imposition of a broken power law for the source-count function might be over-constraining in some cases.  For example, it could lead to the apparent exclusion of a DM component because the extrapolation to low flux of the source-count function from high-flux sources is already too large.
 Furthermore, the broken-power law prescription may yield misleading results for the true uncertainty on the source-count function at low flux. 
 To address these concerns, we present preliminary results from an
 alternate analysis where the number of sources in each flux bin is allowed to float independently; we use the seven logarithmically-spaced flux bins shown in Fig. \ref{fig:binned}.  Within each bin, $dN/dF$ is constant as a function of $F$.  The source-count function model parameters are seven normalization parameters, one for each bin, which are taken to have log-flat priors over the range indicated for $\log_{10} A_\text{PS}$ in Tab.~\ref{tab:priors}.
 
Using the binned source-count function, we perform the NPTF on the 3FGL-masked IG ROI.  For simplicity, we leave out isotropic and disk-correlated PS templates from this analysis.  This is justified by the fact that leaving out these two templates from the broken power-law analyses does not qualitatively affect the results for the NFW PS and NFW DM components.  
We find that the DM flux fraction is consistent with zero, while the NFW PS template absorbs the excess flux.  

As shown in Fig. \ref{fig:binned}, we recover a source-count function broadly consistent with the original analyses assuming broken power laws.  The blue (pink) bands indicate the 68\% (95\%) confidence intervals for the source-count function in each bin, while the solid black (dashed red) line shows the median (mean) of the distribution.  The orange band shows the 68\% confidence interval from the masked NPTF using the broken power-law formalism.

At low flux, the mean and median of the binned result differ by multiple orders of magnitude, reflecting the fact that the posterior distributions for these parameters are skewed.  This is related to the fact that the low-flux bins are not well-constrained by the fit, so the posterior distributions for these parameters are heavily influenced by the log-flat prior distributions.  
In the broken power-law fit, only the overall normalization parameter $A_\text{PS}$ was taken to have a log-flat prior range, while in the binned fit all PS parameters have log-flat prior ranges.  This point, combined with the fact that the broken power-law is more constrained than the binned source-count function, leads to larger uncertainties at low (and, to some extent, high) flux in the binned analysis than in the broken power-law analysis.  

An additional challenge with this method is that the source counts in neighboring bins are generally highly correlated, leading to large errors in individual bins; however, the total flux is still well constrained.  For example, we find that $5.13_{-0.82}^{+0.99}$\% of the flux (in the inner 10$^\circ$ region with $\abs{b} \geq 2^\circ$) is associated with the NFW PS template.  These results are consistent with the broken power-law results, within uncertainties.  Similarly, the binned fit prefers a large number of additional unresolved sources with fluxes only slightly below the PS-detection threshold, but the exact bin in which these sources appear is more uncertain.  In future work, we plan to refine the binned analysis to make it more robust to changes in bin size and prior assumptions.  

\section{Comparison with Luminosity Functions in the Literature}
\label{sec: comp}

\begin{figure}[b]
	\leavevmode
	\begin{center}$
	\begin{array}{c}
	\scalebox{0.40}{\includegraphics{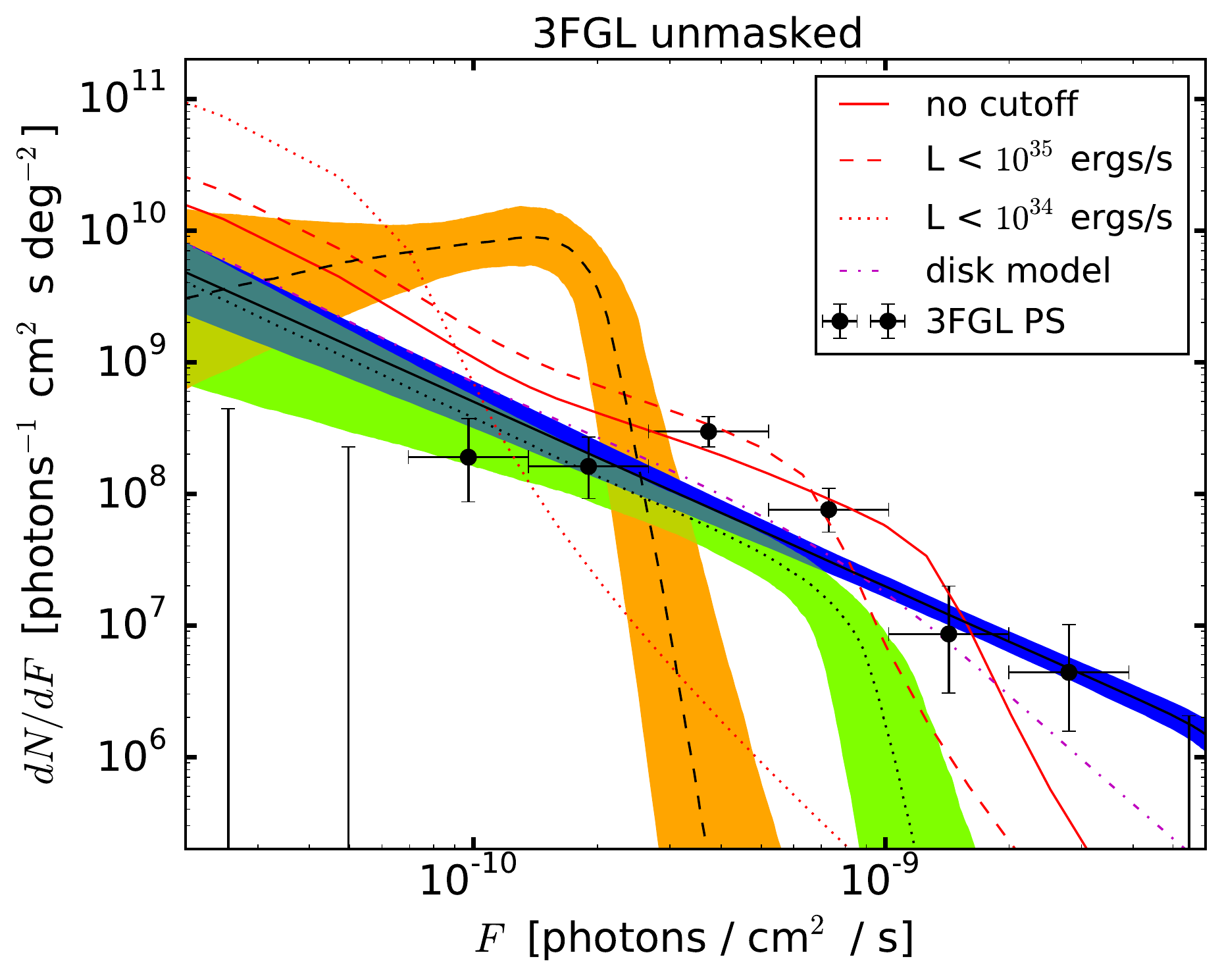}} 
	\end{array}$
	\end{center}
	\vspace{-.50cm}
	\caption{Same as Fig.~\ref{Fig: IG_dnds_unmasked}, but including comparisons to example source-count functions motivated by MSPs.  In particular, the red lines assume an NFW spatial distribution and a gamma-ray luminosity function consistent with that of observed MSPs in the Milky Way~\cite{1407.5583}.  Three scenarios are considered: a maximum luminosity cutoff of  $10^{34}$~erg/s (dotted red) or $10^{35}$~erg/s (dashed red), as well as no cutoff (solid red).  These curves are normalized to the flux of the excess in the ROI. The purple line (dot-dashed) shows the expected source-count function assuming the same MSP-motivated luminosity function, but a thick-disk spatial distribution; it is normalized to match the number of bright $F_{E > 1 \text{ GeV}} > 2.5\times10^{-9}$ photons/cm$^2$/s MSPs and MSP candidates at high latitudes ($|b| > 10^\circ$).}
	\vspace{-0.15in}
	\label{Fig:LumFigure}
\end{figure}

The best-fit source-count function recovered by the NPTF for the unresolved NFW-correlated sources is quite different to those previously considered in the literature---see \emph{e.g.},~\cite{Lee:2014mza,1407.5583,1407.5625,1411.2980,1305.0830}.  With our source-count function, both the number of sources and the flux are dominated by sources within a factor $\sim$2 of the break.
For sources $\sim$8.5~kpc from the Earth and with the spectrum of the excess, this 
break corresponds to a luminosity above 0.1 GeV of $L_\gamma \sim 2$--$3 \times 10^{34}$ erg/s.

We construct models of the expected source-count function using a luminosity function derived from observed MSPs in the nearby field of the Milky Way \cite{1407.5583}, for both NFW-distributed sources and the thick-disk model described by~\eqref{eq: diskymodel} with scale radius and height of 5~kpc and 1~kpc, respectively.  For the former case, we also consider the possibility that the luminosity function possesses a cutoff at $L_\gamma = 10^{34}$ or $10^{35}$ erg/s, following \cite{1407.5625}. The thick-disk model has been normalized to produce the correct number of high-latitude ($|b| > 10^\circ$) bright MSPs and MSP candidates, with $F_{E > 1 \text{ GeV}} > 2.5\times10^{-9}$ photons/cm$^2$/s assuming the average pulsar spectrum, as presented in \cite{1305.0830}. Because this model has the wrong spatial morphology to explain the excess, the purpose of showing this curve is to illustrate the likely contribution from a disk population of MSPs within the ROI. The models with NFW-distributed sources have instead been normalized to match the flux attributed to the NFW PS template in our analysis.

Fig.~\ref{Fig:LumFigure} shows the expected source-count functions, averaged over our standard ROI, for these scenarios, for the 3FGL-unmasked fit (Fig.~\ref{Fig: IG_dnds_unmasked}). In agreement with \cite{1407.5625}, we find that for NFW-distributed sources with the luminosity function of \cite{1407.5583} and the correct normalization to explain the excess, too many sources are predicted above the \emph{Fermi} detection threshold when there is no luminosity cutoff (solid red) or when $L_\gamma <10^{35}$ erg/s (dashed red). The case with a cutoff at $L_\gamma = 10^{34}$ erg/s (dotted red) evades this constraint, as expected, but requires $\mathcal{O}(10^4)$ new sources to explain the excess.  Using the luminosity functions of~\cite{1407.5583}, we find good agreement with the number of sources required to fit the excess as stated in \cite{1407.5625}.

The purple dot-dashed line in Fig.~\ref{Fig:LumFigure} shows the predicted source-count function for the thick-disk distribution. It is remarkably similar to the best-fit source-count function for the thin-disk PS model extracted from the data.  In particular, we estimate the slope of the purple dot-dashed line at low flux in Fig.~\ref{Fig:LumFigure} to be $\sim$$1.43$, while the NPTF predicts the slope of the thin-disk PS source-count function below the break to be $n_2  = 1.40_{-0.15}^{+0.12}$.   In agreement with \cite{1305.0830}, the unresolved sources associated with such a population should not produce enough photons to explain the excess (as may be seen by comparison to the red lines, which have the correct total flux to explain the excess).

The source-count function for the NFW PSs rises sharply above the red curves at fluxes \mbox{$F \sim 1$--$3 \times 10^{-10}$ }photons/cm$^2$/s.  As this source-count function is very shallow below the flux threshold, the total number of sources is dominated by this relatively high-flux region, as is the total flux. For this reason, only $\mathcal{O}(10^2)$ sources are needed to account for the excess in the ROI, in contrast to the $\mathcal{O}(10^3$--$10^4)$ quoted in \cite{1407.5625}.\footnote{The ROI employed in that work was also larger; for the NFW profile we employ, this corresponds to roughly a factor of $3$ difference in the expected number of sources.}  With that said, we emphasize that estimates of the total number of NFW-distributed PSs based on the NPTF are highly uncertain and subject to large systematic uncertainties since the low-flux PSs are hard to constrain with our method.   

\section{Survival Function}
\label{Sec: SF}

As a further cross-check that the PS identification is working self-consistently, we attempt to identify pixels that are likely to contain unresolved sources in the 3FGL-masked sky.  To do so, we perform a standard template fit in the ROI (excluding the PS templates) and determine the best-fit reference model by taking the median values of the posterior distributions for the appropriate fit parameters.  Using this reference model, the expected mean number of counts, $\mu_p$, in a pixel $p$ can be obtained.  From the observed counts map one can then determine the Poissonian cumulative probability to observe the real count, $n_p$, in that pixel given the expectation of the reference model.  The survival function for pixel $p$ is defined as 
\es{eq: epsilon}{
\epsilon_p \equiv 1- \text{CDF}\left[\mu_p, n_p\right] \, ,
}
where $\text{CDF}\left[\mu_p, n_p \right]$ denotes the cumulative probability of observing $n_p$ counts for a Poisson function with mean $\mu_p$.  For example, if the best-fit reference model predicts a total of $\mu_p=2$ photons in a given pixel and the observed number of counts is $n_p=6$, then $\epsilon_p \approx 5\times10^{-3}$.  In general, the smaller the value of $\epsilon_p$ in a pixel, the more likely it is that it contains an unresolved PS.  

\begin{figure}[b]
	\leavevmode
	\begin{center}$
	\begin{array}{cc}
	\scalebox{0.60}{\includegraphics{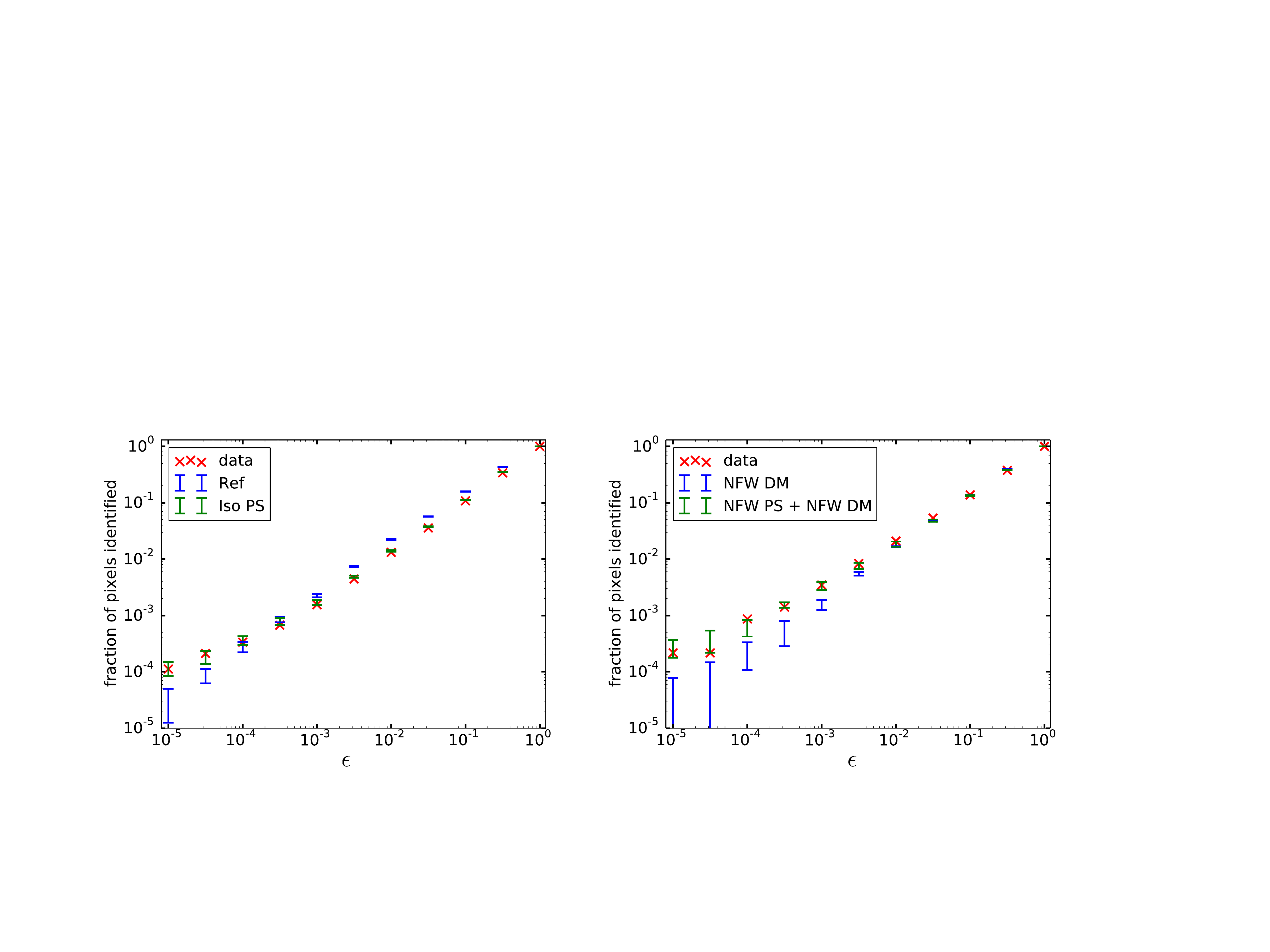}} & \scalebox{0.60}{\includegraphics{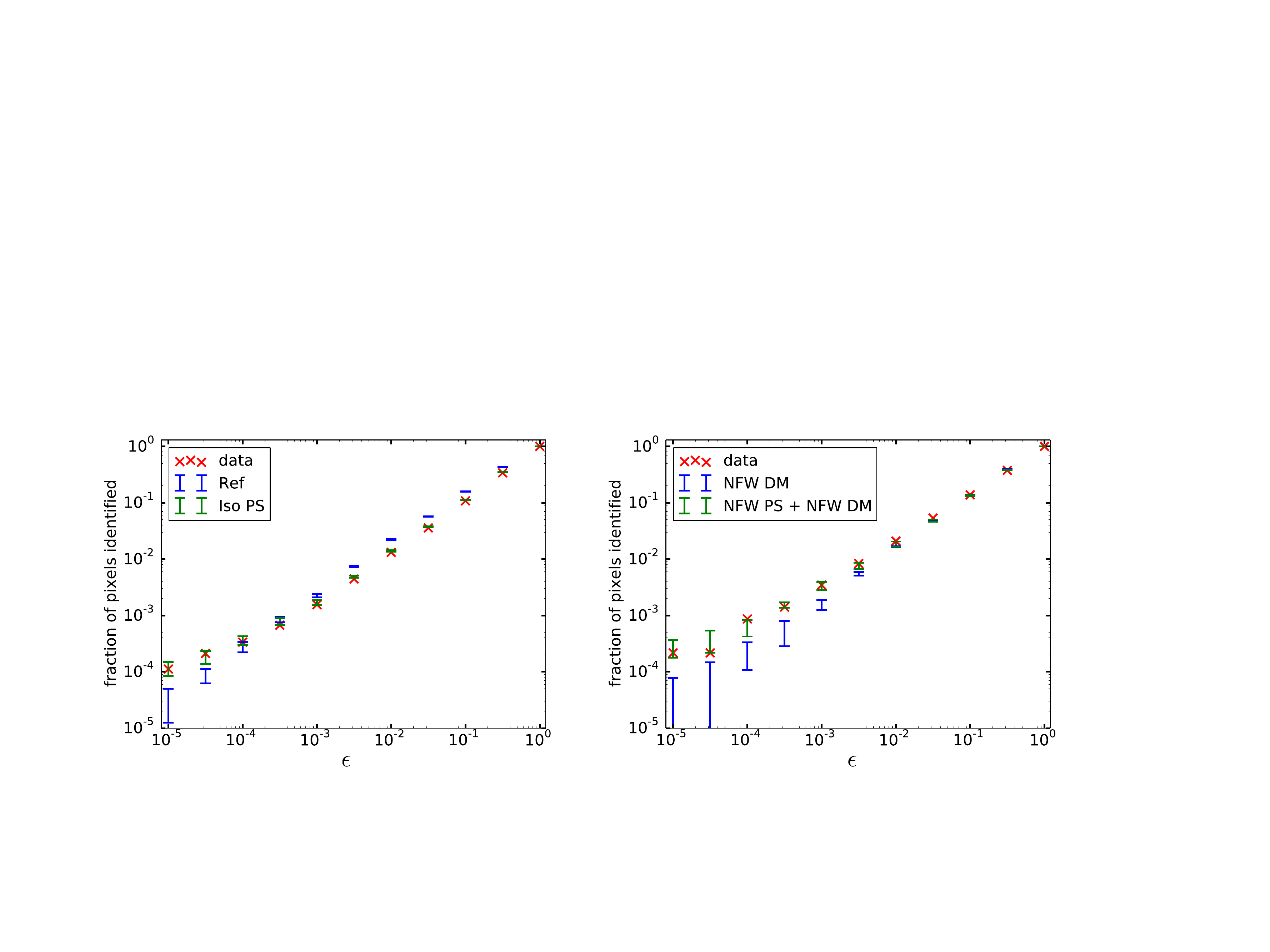}} 
	\end{array}$
	\end{center}
	\vspace{-.50cm}
	\caption{The fraction of pixels with $\epsilon_p < \epsilon$, with $\epsilon_p$ defined in~\eqref{eq: epsilon}.  (Left) The result for the high-latitude analysis, with the 3FGL sources masked.  The red crosses show the survival-function distribution for the observed data set, given the reference background model (diffuse+isotropic+bubbles).  The bars indicate the 68\% confidence intervals for simulated data sets generated from the best-fit background model (blue) and the isotropic PS model (green).  (Right) The result for the IG analysis, with the 3FGL sources masked.  This time, the reference background model includes diffuse+isotropic+bubbles+NFW DM contributions.  The survival-function distribution for the  \emph{Fermi} data set (red crosses) is shown, as well as the confidence intervals for simulated-data studies generated from the model including isotropic PSs (blue) and isotropic+NFW PSs (green).  }      
	\vspace{-0.15in}
	\label{Fig: cpc highL}
\end{figure}

\begin{figure}[t]
	\leavevmode
	\begin{center}$
	\begin{array}{cc}
	\scalebox{0.40}{\includegraphics{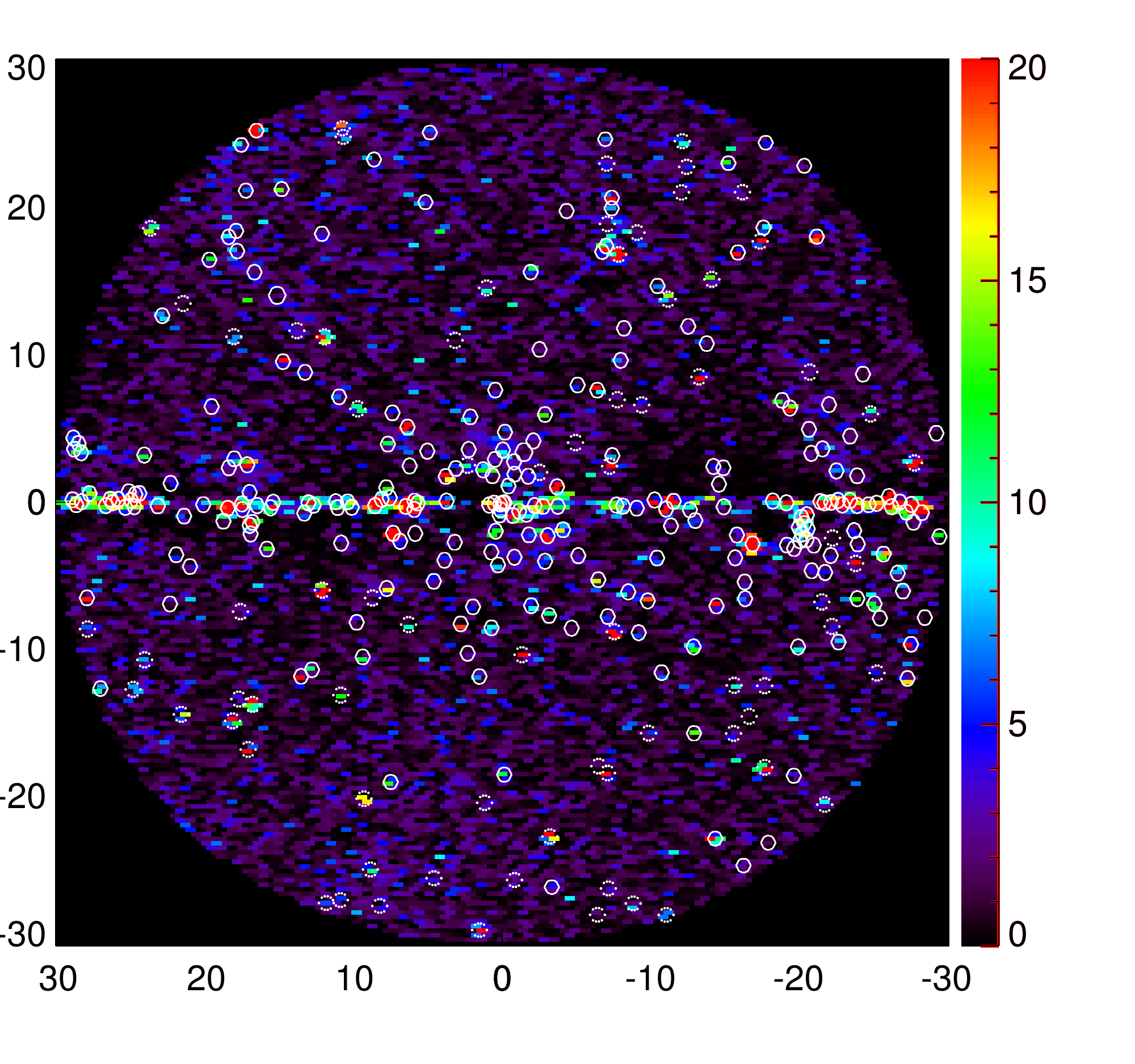}}  
	\end{array}$
	\end{center}
	\vspace{-.50cm}
	\caption{Map of the IG region (plane unmasked) showing the value of $\log \epsilon_p^{-1}$ in each pixel; the larger the value of $\log \epsilon_p^{-1}$, the more likely that the pixel contains a PS.  The circles show the locations of known 3FGL sources; a dashed border indicates an extragalactic source, while all other sources are indicated by a solid border.  The radius of each circle is set to either $0.5^\circ$ or to the semi-major axis of the 95\% confidence ellipse for the source (as provided by \emph{Fermi}), whichever is larger.  The scale is clipped at 20.  The data for this figure is plotted as additional supplementary material. }     
	\vspace{-0.15in}
	\label{Fig: CDFmap}
\end{figure}

We begin by considering the high-latitude region, where the reference model includes isotropic, diffuse, and bubbles contributions.  The reference model is built using the parameters obtained by fitting these three templates to the 3FGL-masked \emph{Fermi} data.  The total contribution from the isotropic, diffuse, and bubbles components are then compared to the observed counts to obtain a value $\epsilon_p$ in each pixel.  The left panel of Fig.~\ref{Fig: cpc highL} shows the fraction of pixels (red crosses) at high latitudes for which $\epsilon_p < \epsilon$, plotted as a function of $\epsilon$.

To better understand the observed $\epsilon$ distribution, we compare it to the results obtained using  simulated data maps.  To simulate a data map with only isotropic+diffuse+bubbles components, we draw Poisson counts from the sum of the reference templates.  Then, we treat the simulated data just as we do the real data and calculate the fraction of pixels for which $\epsilon_p < \epsilon$, as a function of $\epsilon$.  This process is repeated with $\sim$200 simulated data maps.  The result is illustrated by the blue points in the left panel of Fig.~\ref{Fig: cpc highL}; the vertical bars indicate the 68\% confidence intervals for the fraction of pixels with $\epsilon_p < \epsilon$.  The fact that the blue points  underpredict (overpredict) the fraction of pixels with small (large) $\epsilon_p$ indicates that a simulated data map containing contributions from isotropic, diffuse, and bubble emission alone does not have the same photon-count statistics as the data; the real data has both more bright and more dim pixels.  Indeed, this is a sign that the data has residual unresolved PSs, even though the known PSs are masked~\cite{1104.0010,Lee:2014mza}.

We repeat this procedure for simulated data maps that contain an isotropic PS component.  The simulated data is created using the best-fit values from the NPTF including diffuse+isotropic+bubbles components in addition to an isotropic PS component.  The appropriate number of PSs are drawn from the source-count function and placed randomly over the masked sky, following the procedure outlined in~\cite{Lee:2014mza}.  The PSs are then smeared with the energy-averaged PSF, and Poisson counts are drawn from the resulting map to create the simulated data.  Note that the PS population is independently redrawn from the source-count function for each simulated map.
The 68\% confidence interval for the survival-function distribution is shown with green error bars in the left panel of Fig.~\ref{Fig: cpc highL}.  The survival-function distribution of the simulated PS data is consistent with that of the real data, within statistical uncertainty.  This is strong evidence that the model with unresolved PSs is a better fit to the data than the model without.

Having illustrated the survival function at high latitudes, we repeat the analysis in the IG (within $30^\circ$ of the GC, with $\abs{b} \geq 2^\circ$); the results are shown in the right panel of Fig.~\ref{Fig: cpc highL}.  The best-fit reference model is obtained by doing a standard template fit with the following components: isotropic, diffuse, bubble, and NFW DM emission.  The red crosses show the survival-function distribution for the observed \emph{Fermi} data map.

The observed survival function in the IG can be compared to the expected distributions determined from two different sets of simulated data maps.  The first set of simulated data (68\% confidence intervals in blue) was generated assuming the best-fit values for the NPTF with diffuse+isotropic+bubbles+NFW DM templates in addition to isotropic PS contributions.  The second set of simulated data maps (68\% confidence intervals in green) was generated by also adding an NFW PS contribution.  Notice that the NFW PS contribution is required to explain the survival function observed by \emph{Fermi}.

Figure~\ref{Fig: CDFmap} shows a map of $\log\epsilon_p^{-1}$ for all pixels within $30^\circ$ of the GC.  Notice that the brightest pixels (\emph{i.e.}, those with small $\epsilon_p$) do not exhibit any distinctive spatial features, other than the fact that there are more such pixels in regions of higher flux closer to the plane.  The white circles indicate the locations of known 3FGL sources.  The brightest pixels are well-correlated with the locations of these sources, as one would expect.  There are also bright pixels that are not in the 3FGL sources.  In a followup paper, we plan to provide locations of these likely PS candidates to potentially guide future dedicated multi-wavelength studies.  It would also be interesting to correlate the $\log\epsilon_p^{-1}$ maps with other catalogs and data sets.

\section{Simulated data}
\label{sec: sim}

We can check the validity of the NPTF using simulated data generated by Monte Carlo.  In Fig.~\ref{Fig: sim_data_1}, we showed the results of two analyses of simulated data in the IG  including (1) isotropic and disk-correlated PSs, in addition to NFW DM, and (2) isotropic, disk-correlated, and NFW PSs, without NFW DM.  We showed that the Bayes factors and source-count functions recovered from those fits behave as expected.  In particular, when analyzing the simulated data including NFW PSs, we find results consistent with those from the real data.  When, instead, the simulated data has NFW DM instead of NFW PSs, we do not find a preference for NFW PSs when performing the NPTF.  In this section, we give more details for how we generated the simulated data and we also summarize additional simulated-data tests that we have performed.     

The simulated data is generated through the following procedure.  First, we follow the procedure outlined in Sec.~\ref{Sec: SF} to generate the population of PSs for each PS component and place them on the sky.  Second, we smooth the PSs using a PSF described, in each energy bin, by a King function~\cite{1406.0507}.  We divide the full energy range of our analysis into 80 smaller energy bins.  To divide the photons for each PS into the different energy bins, we must make assumptions about the energy spectrum of the PSs.  For the NFW PSs, we take the energy spectrum to be that of NFW DM, over this energy range, as computed in the standard ROI with a standard Poissonian template fit.  For the isotropic PSs, we take the spectrum to be that of the high-latitude isotropic emission, again measured from a standard Poissonian template fit.  For the disk-correlated PSs, we take the spectrum to be the average of the Galactic and unassociated 3FGL PSs over the ROI.  

Small variations to these spectra and to the form of the PSF do not significantly affect our results.  For example, we have checked that analyzing simulated data where we smooth the PSs using a single Gaussian PSF, with 68\% containment radius matching the energy-averaged value we assume for the NPTF, produces results consistent within uncertainties with those obtained analyzing the simulated data that incorporated spectral information and used the more accurate King-function approximation to the PSF.  This may be seen as an additional justification for our treatment of the PSF in the NPTF.       

We may additionally use simulated data to study the effect of mismodeling the diffuse background on the results of the NPTF.  For these studies we follow the simplified procedure described in Sec.~\ref{Sec: SF} for constructing simulated data, which is less computationally intensive.  
 \begin{figure}[tb]
	\leavevmode
	\begin{center}$
	\begin{array}{cc}
\scalebox{0.40}{\includegraphics{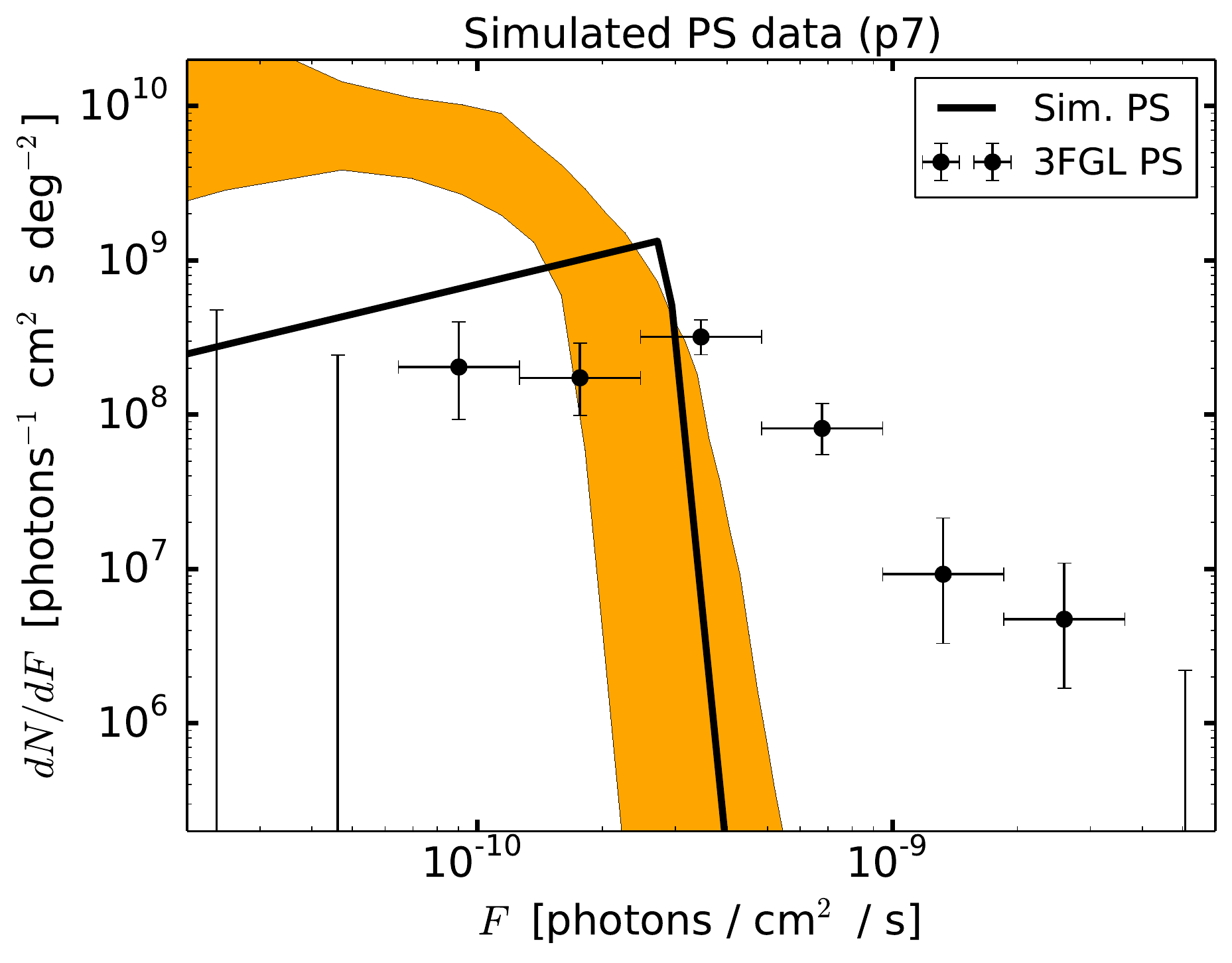}}  &  \scalebox{0.40}{\includegraphics{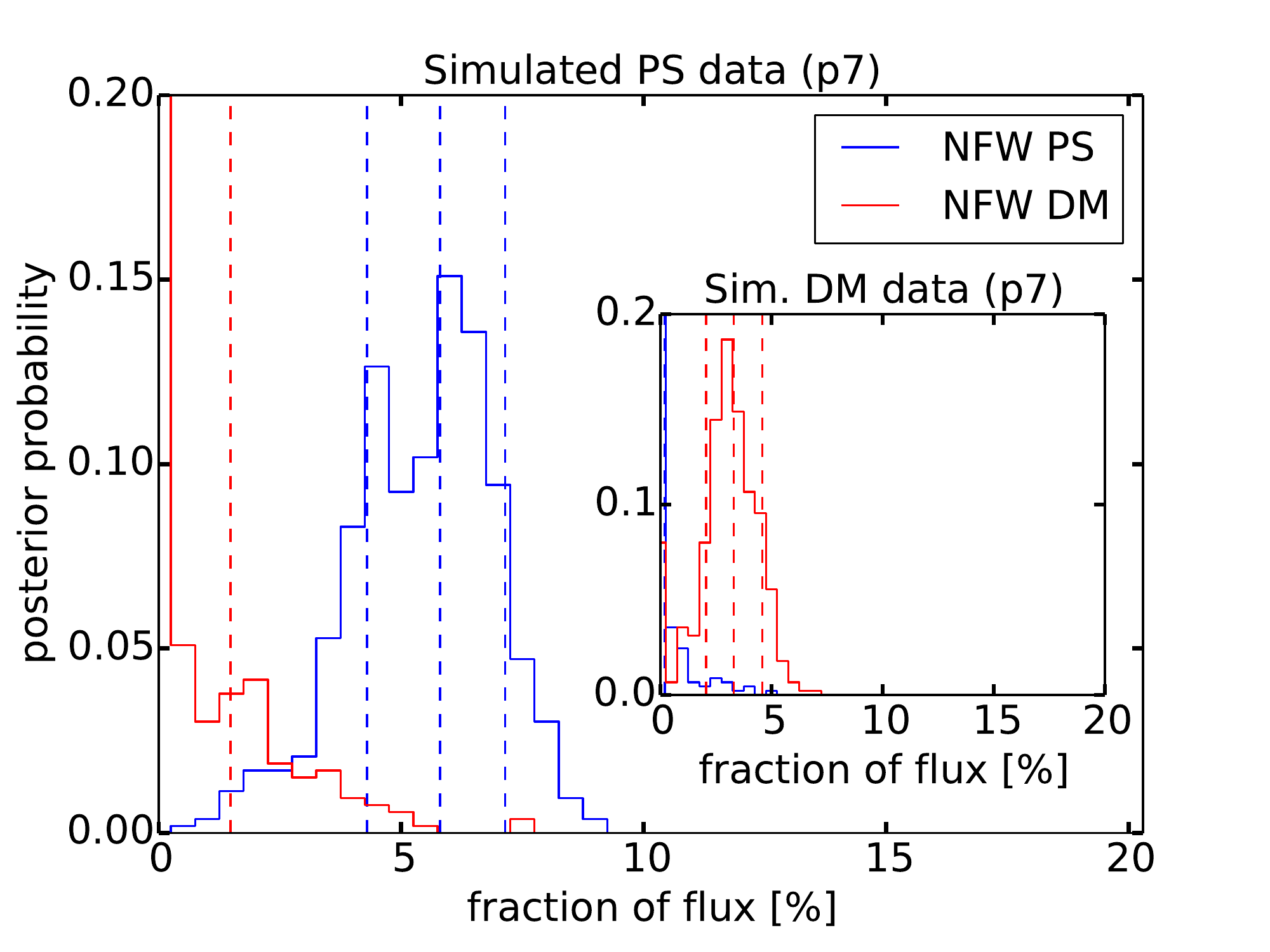}}  
	\end{array}$
	\end{center}
	\vspace{-.50cm}
	\caption{
	Results obtained by applying the NPTF to simulated data.  
	The simulated data is generated within the 3FGL-masked region using the \emph{Fermi} \texttt{p7v6} diffuse background and assuming the best-fit parameters from the masked analysis described in Sec.~\ref{sec:varydiffuse}. 
	 For simplicity, we leave out the sub-dominant disk-correlated and isotropic PS components and only include NFW PSs.  We use the \emph{Fermi} \texttt{p6v11} diffuse model in the NPTF analyzing the simulated data.   The left panel shows the source-count function for the NFW-distributed PSs used to generate the simulated data in solid black; the 68\% confidence interval for the inferred posterior source-count function is shown by the orange bands.  The right panel shows the fraction of flux from the NFW PS and NFW DM templates.  The inset plot shows the results of analyzing simulated data with NFW DM but without NFW-distributed PSs.          All plots are relative to the region within 10$^\circ$ of the GC with $|b| \geq 2^\circ$.  For the flux-fraction plots, the fractions are computed relative to the total number of counts observed in the real data.  }
	\vspace{-0.15in}
	\label{Fig: sim_data}
\end{figure}
For simplicity, we consider the 3FGL-masked IG ROI and we only include NFW PSs in the simulations.  We generate the simulated data using the best-fit values found on the real data with the \texttt{p7v6} diffuse model and a NFW PS template.  However, when analyzing the simulated data using the NPTF, we use the \emph{Fermi} \texttt{p6v11} diffuse model for the diffuse-background template.
This exercise is meant to probe the effect of errors in the diffuse model, and the possibility that such errors could bias the NPTF in favor of PSs.  The results of this analysis are shown in Fig.~\ref{Fig: sim_data}.  
From the right panel, we see that when NFW-distributed PSs are included in the simulated data, we find evidence for their presence; from the inset, we see that when  NFW-distributed PSs are not included in the simulated data, we find no evidence for their presence, but correctly find evidence for the presence of DM.  When performing the analysis on the simulated data containing NFW-distributed PSs, we find that the model with a PS template is preferred over that without by a Bayes factor $\sim$$10^{5}$; the Bayes factor is $\sim$$10^{-1}$ when using the simulated data that does not include NFW PSs.    

The source-count function we recover from the analysis differs from that used to generate the data at low flux in this example.  This may be seen in the left panel of Fig.~\ref{Fig: sim_data}, and should be taken as an indication that the source-count functions we recover should not be trusted too far below the PS-detection threshold.    

The conclusions are unaffected when using any of the other thirteen diffuse models described in Sec.~\ref{sec:varydiffuse} to generate the simulated data.  For all cases, the NPTF is able to adequately identify an NFW PS contribution (with Bayes factors $\sim$10$^6$--10$^9$).  When the data set is generated with a DM component rather than a PS component, the NPTF no longer finds evidence in favor of the PS model, with the Bayes factors typically $\lesssim 1$ and at most $\mathcal{O}(10)$.  
Note that for these tests, the NPTF uses a \texttt{p6v11} template, even if the simulated data set is generated with a different diffuse model. 

We have also considered simulated data sets for which  the excess is split between a DM component and an NFW PS component.  In this case, we find it is difficult to recover the correct source-count function; for some simulated data sets, the NPTF overpredicts the number of low-flux PSs, while for other data sets it underpredicts the number of PSs.  However, the NPTFs do consistently recover Bayes factors $\mathcal{O}(100)$ in favor of the model containing NFW PSs.
While these Bayes factors are considerably smaller than the Bayes factors recovered from both the fit to the real data and the fits to simulated data sets with only an NFW PS component, they are also larger than the Bayes factors in the analyses of simulated data that have no PS component.  This suggests that mixed DM and PS models may be distinguishable using Bayesian techniques, even if the correct source-count function cannot be recovered.

\end{document}